A CATALOG OF DIFFUSE INTERSTELLAR BANDS IN THE SPECTRUM OF HD 204827

L. M. Hobbs[1], D. G. York[2,3], T. P. Snow[4], T. Oka[2,3], J. A. Thorburn[1], M. Bishof[2], S. D. Friedman[5], B. J. McCall[6], B. Rachford[7], P. Sonnentrucker[8], and D. E. Welty[2]

ABSTRACT

Echelle spectra of the double-lined spectroscopic binary HD 204827 were obtained on five nights, at a resolving power R = 38,000 and with a S/N ratio = 750 near 6000 Å in the final, combined spectrum. The stars show E(B-V) = 1.11 and spectral types near O9.5V and B0.5III. A catalog is presented of 380 diffuse interstellar bands (DIBs) measured between 3900 and 8100 Å in the stars' spectrum. The central wavelengths, the widths (FWHM), and the equivalent widths of nearly all of the bands are tabulated, along with the minimum uncertainties in the latter. The reliable removal of very weak stellar lines from the catalog, and of some stellar lines from the less severe blends with DIBs, is made generally easy by the highly variable radial velocities of both stars. The principal result of this investigation is that the great majority of the bands in the catalog are very weak and relatively narrow. Typical equivalent widths amount to a few mÅ, and the band widths (FWHM) are most often near 0.55 Å. Therefore, most of these DIBs can be detected only in spectra obtained at a resolving power and a S/N ratio at least comparable to those used here. In addition, the anomalous interstellar reddening and the very high value of the ratio $N(C_2)/E(B-V)$ seen toward HD 204827 indicate that the physical conditions in one or more of the several interstellar clouds seen in this direction differ significantly from those found toward the prototypical DIB target HD 183143, for example. Probably primarily for these reasons, 113 of the 380 bands, or 30%, were not detected in four previous, modern surveys of the DIBs seen in the spectra of stars other than HD 204827. No preferred wavenumber spacings among the 380 bands are reliably identified which could provide clues to the identities of the large molecules thought to cause the DIBs. The very numerous weak DIBs may be vital in eventually achieving convincing identifications of the absorbers. Both the tabulated data for the DIBs and a plot of our spectrum of HD 204827 are available online at `http://dibdata.org`.

Subject Headings:   ISM: lines and bands --- ISM: molecules --- (stars:) binaries: spectroscopic --- stars: individual (HD 204827)

1) University of Chicago, Yerkes Observatory, Williams Bay, WI 53191
2) University of Chicago, Department of Astronomy and Astrophysics, 5640 South Ellis Avenue, Chicago, IL 60637
3) Also at Enrico Fermi Institute, University of Chicago, Chicago, IL 60637
4) University of Colorado, CASA-Campus Box 389, Boulder, CO 80309
5) Space Telescope Science Institute, 3700 San Martin Drive, Baltimore, MD 21218
6) University of Illinois, Departments of Chemistry and Astronomy, 600 South Matthews Avenue, Urbana, IL 61801
7) Embry-Riddle Aeronautical University, Department of Physics, 3700 Willow Creek Road, Prescott, AZ 86301
8) Johns Hopkins University, Department of Physics and Astronomy, 34th and Charles Streets, Baltimore, MD21218

1. INTRODUCTION

The diffuse interstellar bands (DIBs) are widely assumed to be caused by large molecules which have not yet been conclusively identified (Herbig 1995; Sarre 2006). Under interstellar conditions, and in the wavelength range from about 0.4 to 1.0 μm, several classes of molecules considered to be possible DIB absorbers may produce a few strong bands along with a much larger array of appreciably weaker bands (e.g. Hudgins & Allamandola 1999; Le Page, Snow, &Bierbaum 2003; Webster 2004; Cordiner & Sarre 2007). Based on their large numbers alone, these weak bands may eventually be decisive in achieving convincing identifications of the various DIB absorbers, so the most complete census possible of the weaker bands is desirable. In part in an attempt to advance such efforts, we have acquired echelle spectra of very high quality of more than 30 early-type stars with E(B-V) > 0.50.

Among the potentially most useful stars for this purpose in our survey is HD 204827. The star shows a spectral type of approximately B0V, a large color excess E(B-V) = 1.11, and a relatively bright visual magnitude V = 7.94. The star's spectrum shows a number of weak, narrow DIBs which are not seen in a spectrum of HD 183143 that shows comparable detection limits and was obtained with the same telescope and spectrograph (Thorburn et al 2003, hereafter Paper I). HD 183143 is a still more heavily reddened B7Iae star whose spectrum has often served as a prototype for the detection of DIBs (e.g. Jenniskens & Desert 1994; Herbig 1995). In addition, the column densities of diatomic and triatomic carbon molecules observed toward HD 204827 are, respectively, the largest values found among the 24 stars surveyed for this purpose by Oka et al (2003). Like the group of weak, narrow DIBs just noted, the $C_2$ lines easily detected in the spectrum of HD 204827 are also absent from our spectrum of HD 183143. In Paper I, we therefore referred to the group of DIBs in question as the "$C_2$ DIBs". Observations of HD 204827 clearly promise to add to a census of the DIB population a set of bands formed under physical conditions that must differ significantly from those found along light paths such as that toward HD 183143 (Valencic et al 2003; Sonnentrucker et al 2007).

Our data show that HD 204827 is a double-lined spectroscopic binary (SB2). The observations of the system extended over five nights, and the radial velocity of each of the two stars varied by 40 km s$^{-1}$ more over this period. The resulting, very clear, kinematic separation of interstellar lines and bands from the photospheric lines of both stars is extremely useful in our present study. The central absorption depths of many features of both classes seen in our spectra amount to 1% or less of the stellar continuum and thus cannot be considered to have been previously well catalogued at the required precision.

2. THE DATA

2.1. Observations and Data Reduction

The principal features of the overall DIBs observing program and of the reduction of the resulting spectra have been detailed previously in Paper I. Therefore, those aspects of the present program will be described here in a brief summary form only.

The spectra were obtained with the 3.5m telescope and the ARC echelle spectrograph (ARCES) at Apache Point Observatory. Each exposure provides nearly complete spectral coverage from about 3,700 to 10,000 Å at a resolving power R = 38,000 (Wang et al 2003). The absolute sensitivity of the instrument (including the effects of atmospheric transmission) varies substantially over this wide wavelength range, reaching a broad maximum near 7000 Å for HD 204827. Exposures of the star's spectrum were obtained on each of four nights from 2001 September 5 through 2001 October 29 and on 2003 October 25. In all, 21 untrailed exposures lasting 30 minutes each were obtained and were subsequently combined during data reduction. The total observing time utilized was 10.5 hours. The reduction of the many individual exposures to a single, combined, final spectrum followed standard methods, except for several modifications needed for ARCES data (Paper I). Observations were also obtained of 10 Lac, a sharp-lined, lightly reddened, MK standard star of spectral type O9V. This comparison spectrum provides an independent estimate of the stellar lines likely to be present in the spectrum of HD 204827, although the evidence provided on this point by the kinematic information is generally more decisive. In part because HD 204827 is an SB2, the match of its spectrum to that of 10 Lac (and to those of other spectral standard stars) is only approximate. Observations were also obtained of several bright, broad-lined, early-type stars, to allow removal of telluric lines by division of the two types of spectra. At the level of precision desired here, better than 1% of the continuum, this removal is generally not fully successful for the stronger telluric lines. The resulting, weak, residual telluric lines present in the final spectrum can be readily recognized, as usual, by their fixed geocentric wavelength from night to night.

2.2. The S/N Ratio

Empirical estimates of the S/N ratio, per pixel, achieved as a function of wavelength in the final spectrum of HD 204827 are presented in column 2 of Table 1. The peak value, S/N = 750, is attained near 6000 Å. The corresponding S/N ratios per resolution element are larger by a factor of $\sqrt{2.3} = 1.5$. Three independent methods were used to determine or to corroborate these estimates.

First, the S/N ratio in the continuum was measured directly in various wavelength segments of the spectrum which appear to be as free as possible from all types of absorption lines (interstellar, stellar, and residual telluric), to an absorption depth of about 1% of the continuum. Few such wavelength segments can be found, especially at $\lambda < 4730$ Å, where a host of stellar lines is present. The various segments used extend over ranges between 8 and 90 Å, and each segment was broken into several smaller ones separated by short gaps caused by isolated interfering lines. As a function of wavelength, the satisfactorily concordant results of this process were fit by a quadratic function, from which the entries in column 2 of Table 1 were interpolated.

The second method began with the choice of a number of unblended, relatively narrow DIBs of intermediate strength. Representative parameters of these DIBs include band widths at half depth of $0.4 < \text{FWHM} < 0.7$ Å, equivalent widths $10 < W_\lambda < 20$ mÅ, and central absorption depths averaging about 2.5%. For each band chosen, the equivalent width was measured separately in each of the five averaged spectra respectively obtained on the various nights of our observations, and the resulting mean for all nights and the corresponding rms deviation were then calculated. This rms error, $\Delta W_\lambda$, was converted to an effective S/N ratio by the formula $\Delta W_\lambda = 1.064 \times \text{FWHM} / (S/N)$, which expresses the equivalent width of a line of fractional central depth $\sigma = 1/(S/N)$ and of Gaussian profile. (Many DIBs have irregular, asymmetric profiles, but most of those used for this particular purpose indeed are of crudely Gaussian form.) As a function of wavelength, and after fitting by a quadratic function, the S/N ratio deduced from these DIBs showed satisfactory agreement with the results previously determined directly from the continuum. The scatter about the fitting function was larger for this second method than for the first, however, in part owing to the paucity of DIBs at $\lambda < 4900$ Å. With one exception noted below, this second method was therefore used only as a consistency check on the first.

Thirdly, from the measured exposure levels cumulatively achieved in the combined, final spectrum of HD 204827, we calculated the S/N ratio expected theoretically from the photon shot noise alone. As a function of wavelength, the resulting function shows an amplitude and a shape quite different from those exhibited by the empirical results given in Table 1. Reached near 7000 Å, the maximum S/N ratio set by photon noise alone proves to be about 2400, a value well in excess of the maximum S/N ratio actually achieved. As a further test, the second method discussed above was extended to broader DIBs. To within the uncertainties, the resulting, empirical detection limit, $\Delta W_\lambda$, proves to be almost linearly proportional to FWHM (a result already used above), rather than to the square root of FWHM as would be expected if photon noise primarily limits the S/N ratio. These two facts together indicate that the errors arising in flat fielding, continuum placement, recognition of weak blending lines, and other systematic effects primarily establish the detection limits encountered in our final spectrum. Although not negligible for the narrowest DIBs, for which many of the systematic errors are minimized, the effects of photon noise are of secondary importance here for all of the DIBs. These questions will be considered further in section 3.4.

The formula for $\Delta W_\lambda$ used above provides a practical measure of the minimum equivalent width that is detectable in our final spectrum. The result applies only to a line with a Gaussian profile, but it will be used here as a simple, convenient model for estimating the detection limit $\Delta W_\lambda$ for all DIBs. The corresponding formula for a triangular line profile, for example, differs only in that the numerical coefficient becomes unity, which is a minor difference in practice. An overview of the resulting 1σ detection limits, $\Delta W_\lambda$, expected as a function of wavelength is presented in columns 3 and 4 of Table 1, for two illustrative values of the band width.

3. DATA ANALYSIS AND RESULTS

3.1 Background

Our primary goal is to compile a list of the DIBs that are present in the spectrum of HD 204827 above a detection limit near an equivalent width of a few mÅ. Our aim is to exclude from this list all features that are not genuine DIBs, while not being so restrictive as to exclude at the same time a significant number of very weak DIBs that are actually detectable in our spectra. The method used consists of two steps. First, the respectively averaged spectra from each of the five nights were intercompared by eye, in an attempt to identify all genuine DIBs that are detectable in our spectra. The two primary attributes required of the DIB candidates were an invariant heliocentric radial velocity for all five nights, along with a satisfactory repeatability of the band's apparent strength and profile. The spectrum of 10 Lac was also used as a guide to the stellar lines that are likely to be present in the spectrum of HD 204827. The S/N ratios of the respectively combined spectra from the five nights of observations were roughly similar, so that this visual comparison could be carried out fairly simply and directly. An example of the data is shown in Figure 1. Any important blends with stellar lines, residual telluric lines, or other DIBs were also noted. The second step then consisted of measuring the central wavelength, the band width (FWHM), and the equivalent width as well as its uncertainty for each of these bands, in the combined, final spectrum from all five nights.

Over the wavelength range $3900 < \lambda < 8100$ Å, the results of these measurements are listed in Table 2, and the combined, final spectrum of HD 204827 is displayed in Appendix A[9]. For each of the 380 DIBs reported here, columns 5 and 6 of Table 2 give the central wavelength and the band width (FWHM), respectively; columns 7 and 8 give the equivalent width and its uncertainty, respectively; and column 9 indicates the presence of significant blended features or other information. These various measurements will be described in more detail below. Column 10 provides sequential numbers for the DIBs, for convenient reference. Columns 1 through 4 of Table 2 present the central wavelengths, as reported by the previous authors, of those DIBs detected in our spectra that had already been noted in one or more of the DIB atlases compiled by Jenniskens & Desert (1994), Galazutdinov et al (2000), Tuairisg et al (2000), or Weselak, Schmidt, & Krelowski (2000). The first two digits of those previous wavelengths are redundant (with those in column 5) and have been suppressed. The essential features of these four atlases for our purposes are briefly summarized in Appendix B.

We emphasize that none of these four previous studies included observations of HD 204827. Therefore, the presence in Table 2 of, in particular, a weak DIB not reported in the previous investigations (or, conversely, the absence of one that was previously reported) does not necessarily indicate a discrepancy between the two studies, for at least two reasons. First, the relative strengths of some DIBs vary significantly from one line of sight to another. Second, when the spectral types of two background stars differ appreciably, the blending with stellar lines may affect very differently the detectability of a given DIB in the two spectra. A considerable number of generally weak DIBs reported in one or more of the four previous surveys indeed do not appear in Table 2. Some of those bands may actually be present in our spectra of HD 204827 too, but are so weak in this line of sight that they do not satisfy our criterion of repeatability from night to night well enough to merit inclusion in Table 2.

3.2. Central Wavelengths and Band Widths

The zero-point of the wavelength scale adopted here is set by assigning the laboratory wavelength of 7698.9645 Å (Morton 2003) to the interstellar K I line recorded in our spectra (Figure 2). That is, our wavelengths are those that would be measured in the rest frame of the effectively unresolved group of interstellar clouds revealed by K I at the ARCES resolution of 8 km s$^{-1}$. These unresolved clouds show a heliocentric radial velocity of -17.4 ± 0.4 km s$^{-1}$, or an LSR velocity of –3.2 ± 0.4 km s$^{-1}$. Because laboratory wavelengths are currently unknown for all DIBs, no direct comparison of them with the K I wavelength is possible. Therefore, various, small, physically real offsets probably exist between the correct, but unknown, laboratory values of the DIB wavelengths and those deduced here on the basis of K I atoms. Other previous studies of DIB wavelengths are necessarily subject to the same limitation as well.

---
[9] These tabular and graphical data are publicly available at http://dibdata.org.

Furthermore, at higher instrumental resolution, the interstellar K I line is resolved into at least four components (Pan et al 2004), the two strongest of which are separated by about 5 km s$^{-1}$, or 0.13 Å at the λ7699 line, for example. We might have assigned the laboratory K I wavelength to any one of these several resolved clouds instead, thereby changing all of the absolute wavelengths reported here by a uniform fractional shift Δλ/λ. Differences of these magnitudes are substantially larger than the random errors (described below) that arise in measuring the wavelengths of most DIBs in our catalog. Depending upon such currently arbitrary choices made for wavelength zero-points, significant systematic differences are consequently expected among the absolute wavelengths reported in various investigations. For this reason, specifically systematic differences among our wavelengths and the varied values found in columns 1 through 4 of Table 2 may be of little physical significance, in some cases.

Another operational definition may partially account for differences among the wavelengths reported in different surveys. Many DIBs show irregular profiles, so precise definitions of the central wavelength and of the band width are necessary. In the case of asymmetric DIBs, the results determined in the various surveys will reflect, in part, the distinct definitions adopted for these quantities, and both the absolute and the relative wavelengths of the asymmetric DIBs may be affected by these choices. The definitions used here are as follows. Within a DIB's profile, the midpoint along the vertical line drawn at fixed wavelength between the continuum and the point of deepest absorption is located. The horizontal line drawn through that midpoint intersects the shortest-wavelength and the longest-wavelength wings of the profile at wavelengths denoted by $\lambda_1$ and $\lambda_2$, respectively. Then, $\lambda_c = (\lambda_1 + \lambda_2) / 2$ and FWHM = $\lambda_2 - \lambda_1$.

The smallest random errors of measurement that are attained here for both $\lambda_c$ and FWHM typically amount to ± 0.03 Å for the narrower DIBs in Table 2, with widths near 0.4 Å. This uncertainty increases roughly in proportion to FWHM for the broader bands. The values of FWHM reported here are the measured values without any correction for instrumental broadening. The FWHM of the ARCES instrumental profile is about 0.16 Å at 6000 Å, for example, and the narrowest DIBs in Table 2 show measured widths that exceed the instrumental resolution by an approximate factor of only 2.5. For the narrow DIBs that are only partially resolved, the intrinsic widths of the features must be smaller than the measured values given in Table 2 by as much as 10%. For example, after quadratic subtraction of the instrumental width from the observed FWHM of 0.30 ± 0.02 Å, the intrinsic FWHM of the DIB at 4817.64 Å is about 0.27 ± 0.02 Å.

One of the most difficult problems in analyzing the subset of strongly asymmetric DIBs stems from the unknown identities of the absorbing molecules and therefore of their spectra. Thus, no reliable method is available for deciding whether an irregular DIB profile with two distinct absorption maxima reveals one intrinsically asymmetric band or a blend of two narrower, more symmetric bands. The effectively arbitrary choices made here in such cases can be ascertained only from the numerical results reported in Table 2. For example, the profile of the band at 6376.08 Å is shown in Figure 3, and Table 2 shows that this asymmetric feature has been treated as a single DIB in our compilation. In two of the previous surveys cited in the table, this DIB was instead regarded as a blend of two distinct DIBs, a clearly plausible choice. To emphasize this difference in method, the notation "blend" has been entered in column 2 in place of the pair of distinct wavelengths reported by Galazutdinov et al (2000). The corresponding comments apply to column 3. The notation "blend" is used throughout Table 2 to indicate that a feature treated here as one, asymmetric DIB was regarded in the previous investigation in question as two (or more) distinct, blended DIBs in the spectrum of a star other than HD 204827. This correspondence is sometimes ambiguous, and the set of "blend" entries indicated in Table 2 may occasionally be incomplete. If needed, the pertinent wavelengths of the separate DIBs are available in the previous papers. A final comment in this connection is that having data for several lines of sight will not necessarily be conclusive in accounting for differences in such band profiles, in the absence of molecular identifications and knowledge of their spectra. It may be difficult to distinguish between different excitation conditions for a single molecule and different relative column densities of two molecules.

3.3. Equivalent Widths and Their Errors

The equivalent widths presented in column 7 of Table 2 were calculated by integration over the DIB profiles in the combined, final spectrum formed from the 21 individual spectra obtained on all five nights. The corresponding $1\sigma$ errors listed in column 8 are estimates of the *minimum* errors. These errors were calculated from the formula $\Delta W_\lambda = 1.064 \times \text{FWHM} / (S/N)$ of section 2, using the values of the FWHM listed in Table 2 and the values of the continuum S/N ratio interpolated from Table 1.

The actual uncertainties in the equivalent widths must sometimes exceed the minimum values calculated routinely via the formula. The principal known source of additional uncertainty consists of blends with residual telluric lines, stellar lines, or other DIBs. A warning is provided in column 9 of Table 2 when such a blend significantly increases the uncertainties in measuring the wavelength, the width, and/or the equivalent width. A flag "d", "s", "t", or "det" in column 9 identifies a blend with another DIB, a stellar line, a telluric line, or a detector artifact, respectively. In these flagged cases, a crude deblending, by means of either Gaussian or polynomial fitting, has been carried out, if a definite value of the equivalent width has been specified for the DIB. If deblending instead appeared infeasible, only an upper limit is reported, although the DIB is definitely present. If the blending, usually with a strong stellar line, is so severe as to prevent useful measurement entirely, the notation "p" is entered in column 7 to indicate that the DIB is present. That is, we consider that all of the bands listed in Table 2 are definitely present in the spectrum of HD 204827.

Another general limitation is that sufficiently weak cosmic-ray impacts on the CCD detector generally are not distinguishable in the 21 individual exposures. Each of those spectra shows a S/N ratio lower than the values given in Table 1 by a factor amounting crudely to $\sqrt{21} = 4.6$. To an unknown extent, such unrecognized impacts must also distort the profile of the typical, very shallow DIB investigated here.

Among the 380 DIBs in Table 2, one (at 6809.51 Å) constitutes a $1.5\sigma$ detection, two (at 6825.80 and 6814.20 Å) are $2\sigma$ detections, and 12 others are detected at less than the $2.5\sigma$ level. These 15 formally most uncertain bands have been scrutinized with care and do appear to meet the requirements of stationarity and repeatability for inclusion here. The total number of detections in our catalog at a level of 3σ or higher is 344.

Each of the well known DIBs at λ5780 and λ5797 shows a relatively deep, narrow absorption core superimposed on a much shallower, broader absorption feature with a complex, composite profile. Owing in part to the difficulty in reliably defining the continuum for the broader features, the data for these two bands given in Table 2 refer to the respective narrow cores alone, as indicated by the "nc" flag listed in column 9. In these two cases alone, the equivalent widths reported in Paper I have been adopted here without a remeasurement.

For the very broad DIBs exemplified by λ4428, an additional error enters into the data reduction. The wings of such broad bands extend over several orders of the echelle spectrum, and the process of correctly locating the continuum within this extended profile then becomes much more difficult. In such cases, the continuum level was interpolated smoothly across the affected gap, between the many other, undistorted orders (Paper I). The accuracy of the entries in Table 2 for the broad DIBs such as those at 4428.29, 5450.62, 5525.48, and 6590.42 Å may be substantially reduced by this additional uncertainty.

Preliminary values of the central wavelengths, the widths, and the equivalent widths were reported in Paper I (in Tables 2 and 3) for 32 of the bands listed in Table 2; the independent results presented here supersede the earlier ones. The definitions of $\lambda_c$ and FWHM described in section 3.2 were not applied in the earlier measurements, and some blends have been treated differently in the two cases. For example, the DIBs at 5705.13 and 6203.08 Å in Paper I correspond approximately to the combinations of bands in Table 2 at 5705.08, 5706.51, and 5707.77 Å and at 6203.05 and 6204.49 Å, respectively. Except in cases where such procedural differences are important, the two independent measurements of the same data generally agree satisfactorily.

3.4. Selection Effects

The most sensitive detection limits are attained in our spectra within the approximate range 5100 < λ < 6865 Å (Table 1). At shorter wavelengths, the poorer detection limits are attributable primarily to interference by the plethora of stellar lines present there and to progressively increasing photon noise. At longer wavelengths, the poorer sensitivity stems principally from three distinct difficulties: the many strong telluric lines present in much of the region; the progressively stronger interference fringes arising in the CCD detector, which degrade the accuracy of the flat-fielding and of the continuum definition; and the steadily increasing photon noise at λ > 7000 Å. In order to preserve a crudely uniform fractional completeness of detection, we have therefore limited the compilation in Table 2 to 3900 < λ < 8100 Å. Nevertheless, an appreciable fraction of the region included in Table 2 at λ > 6850 Å unavoidably allows only much poorer detection sensitivity than at the shorter wavelengths, owing to the frequent presence of strong telluric absorption bands. Even when the observational selection effects introduced by these various impediments are taken into account, it is clear that the intrinsic spectral density of detectable DIBs decreases fairly sharply at λ < 4900 Å and at λ > 7100 Å. For completeness, we note that two DIBs, at 8439.44 and 8530.09 Å, are reliably detected at λ > 8100 Å in our spectrum of HD 204827 (Wallerstein, Sandstrom, & Gredel 2007), and that none is found at λ < 3900 Å.

An additional selection effect arises in connection with the DIB widths. Instrumental broadening in our spectra probably is effectively unimportant for most of the bands listed in Table 2. Therefore, at any fixed equivalent width, a relatively broad band also must show relatively shallow absorption, on average. Both of these effects increase the difficulty of interpolating the continuum correctly across the DIB profile, so that the resulting systematic errors in determining the wavelength, width, and equivalent width of a band will increase with increasing FWHM (Table 1). Therefore, at a given equivalent width, the fractional error in the equivalent width will also increase correspondingly, reducing the likelihood that progressively broader bands will be detected. In practice, this effect will be very large for the broadest bands. For example, the estimated 1σ detection limit in our spectrum for the λ4428 DIB, with FWHM = 22.5 Å, is 40 mÅ (Table 2).

3.5. The HD 204827 Binary

HD 204827 is a probable member of the Cep OB2 association, which is quite extended and complex (Patel et al 1998; de Zeeuw et al 1999). In addition, the star has often been considered to be an outlying member of Trumpler 37, a very young open cluster embedded in Cep OB2 (Simonson 1968; Garrison & Kormendy 1976; Marschall & van Altena 1987). The most luminous member of Trumpler 37, HD 206267, is the Trapezium-like, exciting star of the large, well known HII region IC 1397. Both the unusual reddening of HD 204827, characterized by $R_V \approx 2.65$ (Morbidelli et al 1997; Valencic et al 2003), and the star's binarity must be taken into account when estimating the distance to the star. The resulting spectroscopic distance is consistent with the independently determined, more accurately known value of about 950 pc to Trumpler 37 (Garrison & Kormendy 1976).

The member of the HD 204827 spectroscopic-binary system that shows the much broader spectral lines will be referred to here as HD 204827 A, and the one with the narrower lines as HD 204827 B. The contrast in the line widths is illustrated in Figure 4. The width of the weak Si III line at 4552.62 Å, for example, can be measured fairly accurately for both stars. The line shows observed widths (FWHM) of 124 and 17 km s$^{-1}$ for components A and B, respectively; the intrinsic width of the latter is 15 km s$^{-1}$, after correction for instrumental broadening. Both stars display spectral types very close to O9.5V or B0V, values that have been previously assigned to the composite spectrum (Simonson 1968; Garrison & Kormendy 1976). For example, both spectra show the lines of He I, and the spectrum of component A also shows well developed lines of He II. Any contribution of the narrow-lined component B to the intrinsically broad He II lines probably is very minor, as judged from the radial velocities observed for those lines (Figure 4). This conclusion is amplified below. Based on various lines of H I, He I, He II, C IV, N II, N III, O II, and Si III, the spectral types of components A and B appear to be approximately O9.5V and B0.5III, respectively. The luminosity class assigned to the cooler component is based primarily on the Balmer lines, whose systematically variable profiles presumably betray the expected blend of lines from both binary components. However, the radial velocities derived from the Balmer lines agree roughly with those derived from He II, and hence of component A. Because component B shows the slightly cooler temperature, its apparently relatively minor contribution to the composite Balmer lines may perhaps be accounted for by the brighter luminosity class.

The orbital period of the system appears to be unknown at present (Petrie & Pearce 1961; Gies 1987). The heliocentric radial velocities of both stars measured on five nights are presented in

Table 3. The results for HD 204827 B were derived from the narrow components of lines of Si III and O II, while the results for HD 204827 A were deduced from lines of He II and C IV. Laboratory wavelengths for the He II lines were taken from Garcia & Mack (1965). The He II and the C IV lines yield results in good mutual agreement and show no evidence of a blended, narrow component arising from HD 204827 B. This evidence is strong in the case of C IV, where the lines are not intrinsically broad, as can be seen in the spectrum of 10 Lac. A crude upper limit on the period appears to be about 10 days, when the velocities in Table 3 are combined with those derived from additional ARCES spectra acquired on three nights within an interval of 12 days in 2007 October. (The latter spectra show relatively low S/N ratios and are not otherwise considered or used here.) Much more complete temporal coverage of the orbit, perhaps at intervals of about a day, is needed. The star is known to be a variable of low amplitude (T. Gandet, private communication), and accurate photometry of the system might reveal the period more easily. If the combined mass of the SB2 is taken to be 30 Mo, a period $P < 10$ days requires an orbital semi-major axis $a < 0.3$ AU.

Figure 5 compares the measured radial velocities of the members of the SB2. If HD 204827 effectively consists only of two simple, point sources viewed from an arbitrary angle, a curve fitted through these data should be monotonically decreasing and single-valued. The apparent contradiction of this hypothesis revealed by the radial velocities observed on 2001 Sep 05 suggests that HD 204827 is a more complex system.

Speckle interferometry of HD 204827 has revealed a possible third member of the stellar system, at an angular separation of about 0.093 arcsec, or 88 AU at 950 pc (Mason et al 1998). No evidence is found in our data for the line spectrum of this distant, apparently fainter (and thus, presumably less massive) companion to the SB2. If the third star is a physical member of the HD 204827 system, its orbital period about the compact pair must exceed 150 yr, unless the companion happened to be near apastron in a substantially eccentric orbit when the speckle observations were made in 1994 and 1996. The presumably very much longer period and the relatively great distance of the speckle companion from the SB2 suggest that the speckle companion is not the cause of the kinematic irregularity indicated in Figure 5, even if this companion does prove to be a member of the system.

## 4. DISCUSSION

### 4.1. New DIBs

By "new" DIBs, we shall mean bands that were not reported in any of the four previous surveys cited in Table 2, which are used here as a convenient basis for a comparison. Thus, in some cases, these "new" DIBs may actually have been reported previously, elsewhere in the literature. The total number of DIBs we have detected in the spectrum of HD 204827 is 380, of which 113, or 30%, are new by this definition (Table 2). In the region $3900 < \lambda < 5300$ Å, 35 of the 50 DIBs detected, or 70%, did not appear in those previous surveys, in a spectral region where the average spacing between adjacent pairs of these 50 features is 28.0 Å (Table 4). The corresponding values for the region $6417 < \lambda < 6694$ Å reveal a much lower discovery fraction amounting to 7 of 50, or 14%, in a region with a much smaller average spacing of 5.5 Å between adjacent DIBs. Thus, by a wide margin, our discovery fraction happens to be highest where the spectral density of DIBs is lowest. Similar statistical results are collected in Table 4 for the full spectral range included in this study. The table is organized not by uniform wavelength intervals, but rather by successive groups of 50 DIBs each, in order of increasing wavelength.

The principal cause of this result almost certainly is not the confusion potentially introduced by overlapping DIBs, however. A perusal of Table 2 shows that most of the new DIBs found at $\lambda <$ 5300 Å are relatively weak and narrow, and many of these probably are examples of the $C_2$ DIBs defined in Paper I. The detection of such bands generally requires the availability of a spectrum of high S/N ratio and high resolution, as well as a light path with the highest possible value of $N(C_2)/E(B-V)$. As one pertinent comparison, the latter ratio toward HD 204827 exceeds that toward HD 183143 by a factor of at least 80. As therefore illustrated specifically by the case of HD 204827, the $C_2$ DIBs appear to be concentrated primarily toward the blue end of the visible spectral region. The comparatively high values of the S/N ratio, the instrumental resolving power, and the ratio $N(C_2)/E(B-V)$ toward HD 204827 which characterize the present observing program therefore seem primarily to account for our high discovery fraction at $\lambda < 5300$ Å, a region relatively sparsely populated by the classical DIBs. We note incidentally that cases of the opposite kind are also fairly common. For example, the DIB at 8038 Å is not reliably detectable in our spectrum of HD 204827, at an upper limit $W_\lambda < 14$ mÅ, but it does appear as a very strong, broad, nearly symmetric feature with $W_\lambda = 221$ mÅ in our spectrum of HD 183143 (Cordiner & Sarre 2007).

A quantitative test confirms the impression that a very small fraction of all DIBs which are actually present is hidden by their mutual overlapping. For this purpose, the various wavenumber splittings between the 379 pairs of immediately adjacent DIBS in Table 2 are analyzed. Differences in wavenumber, rather than in wavelegth, are used for this purpose, because the former are proportional to the corresponding energy differences, which are directly related to the presumed molecular structure. As a function of wavenumber splitting, the histogram of Figure 6 plots the logarithmic numbers of splittings found in twelve bins. All of these bins are 10 cm$^{-1}$ wide, their centers are separated by 10 cm$^{-1}$, and the first of these bins is centered at 5 cm$^{-1}$. At $\lambda = 5000$ Å, an interval $\Delta\sigma = 10$ cm$^{-1}$ corresponds to $\Delta\lambda = 2.5$ Å. A cutoff in the histogram at large splittings has been introduced owing to the nearly vanishing populations of those bins. The approximately linear relation seen in Figure 6 reveals that the number of these splittings declines nearly exponentially with increasing splitting. This result is expected if the wavenumber differences are randomly distributed and if there is negligible overlap between adjacent members of this sample (Welty, Hobbs, & Kulkarni 1994; see their Figure 7). The central result for the present purpose is that no relative falloff in numbers is seen in the bins containing the smallest splittings. This histogram further suggests that the wavenumber splittings are distributed randomly, so that no preferred splitting established by molecular structure is evident in this sample, a point that will be discussed further in section 4.3.

4.2. The Observed Distributions of Wavelengths, Widths, and Equivalent Widths

The histogram of Figure 7 shows the distribution by wavelength of all 380 DIBs in Table 2. The first populated bin spans the range from 4100 to 4300 Å and contains only the moderately broad, shallow DIB at 4259.01 Å. The bins containing the highest number of DIBs are centered at 6000 and 6200 Å. Each of these bins contains 46 DIBs, and the corresponding average spacing between adjacent pairs is 4.4 Å in these regions of maximum crowding.

Figure 8 shows the distribution by observed band width of the 376 DIBs in Table 2 for which definite values were measured. The first populated bin is centered at 0.25 Å and extends from 0.2 to 0.3 Å; it contains only the DIB at 4817.64 Å, with FWHM = 0.30 ± 0.02 Å. The latter corresponds to an intrinsic FWHM = 0.27 ± 0.02 Å, after approximate correction for instrumental broadening. The most highly populated bin is seen to be centered at 0.55 Å. The falloff in numbers at FWHM < 0.55 Å results from instrumental broadening and, presumably, from an unknown lower limit on the intrinsic widths of most DIBs. In any case, the most important result of Figure 8 is the steeply decreasing number of bands that show progressively greater widths at FWHM ≥ 0.55 Å. These results demonstrate that, when narrow DIBs can be efficiently detected, their numbers strongly dominate this distribution.

Figure 9 shows the distribution by equivalent width of the 360 DIBs in Table 2 for which definite values were measured. The first bin extends from 0 to 2 mÅ and contains 15 DIBs. An example of this group is the DIB at 5137.07 Å, with $W_\lambda$ = 1.9 ± 0.6 mÅ. The sharp falloff in the distribution seen in the bin centered at $W_\lambda$ = 1 mÅ almost certainly reflects the approximate detection limit for the narrowest bands in our study. For the broader DIBs, this detection limit must occur at proportionally larger equivalent widths (section 3.4). The most important overall result is the steeply decreasing number of progressively stronger DIBs at $W_\lambda$ ≥ 3 mÅ. When very weak DIBs can be efficiently detected, their numbers also strongly dominate the corresponding distribution.

4.3. On the DIB Molecules

The structures of the molecules that are presumed to give rise to the DIBs may introduce some recognizable statistical signatures into the pattern of the wavelengths in our relatively large sample of DIBs. As was discussed in section 4.1, the histogram of Figure 6 provided no such evidence, but it was based on the wavenumber splittings among the nearest-neighbor DIBs alone. The much larger sample of wavenumber splittings available from all possible pairs of DIBs listed in Table 2 permits a further test (Herbig 1995).

The total number of all possible pairs of DIBs available from Table 2 is 72,010. A conveniently smaller sample can be isolated by considering only those DIB pairs whose wavenumber splittings do not exceed some specified upper limit. As an initial example, an upper limit of 400 cm$^{-1}$ was chosen, which yields 8,698 DIB pairs. The corresponding wavenumber splittings are distributed as shown in the histogram of Figure 10, where the splittings have been grouped into 40 bins. All of these bins are 10 cm$^{-1}$ wide, their centers are separated by 10 cm$^{-1}$, and the first of these bins is centered at 5 cm$^{-1}$. The observed distribution is relatively flat, declining from an average of about 240 cases per bin at the smaller splittings to about 200 cases per bin at the larger splittings, with no statistically significant excesses or deficiencies in any of the individual bins. The numbers in the two adjacent groups centered at 65 and 75 cm$^{-1}$ stand above those in neighboring bins, but only by amounts comparable to the square roots of the numbers in question. The intrinsic scatter in this distribution presumably is established by the absorbing molecules and therefore does not assume a Gaussian form. Nevertheless, if the two adjacent peaks are treated as if they constitute an emission line superimposed on a sloping continuum in the presence of photon shot noise, the peaks represent a feature with only 2.5σ significance. Once again, the pertinent molecular energy-level differences appear to be effectively distributed randomly in our sample, without any reliably identified, preferred values.

As an additional check, the pattern of the binning was not changed, but the upper limit imposed on the wavenumber splittings was extended to 900 cm$^{-1}$, a choice which yields 19,058 pairs. Still another check consisted of narrowing the bin widths to 2 cm$^{-1}$, while a splitting upper limit of 400 cm$^{-1}$ was maintained. In both of these variations of the original test, the final conclusion again is that no preferred energy-level differences are convincingly evident in the resulting histograms corresponding to Figure 10. In the case of the narrower bins, the population of the first bin, centered at a splitting $\Delta\sigma = 1$ cm$^{-1}$, or $\Delta\lambda = 0.25$ Å at 5000 Å, becomes nearly negligible. This effect is expected, because such splittings only slightly exceed the instrumental resolution itself, so that such DIB pairs would be effectively unresolved in our data.

## 5. SUMMARY

In high-quality echelle spectra of HD 204827 obtained on five different nights, 380 diffuse interstellar bands are measured between 3900 and 8100 Å. The principal conclusions derived from these data are the following.

1) The primary DIB population seen toward HD 204827 consists of bands that are very shallow (< 1% fractional absorption), relatively narrow (FWHM near 0.55 Å), and therefore very weak ($W_\lambda$ near a few mÅ). The detection and measurement of this majority population requires spectra with a resolving power and a S/N ratio at least comparable to those achieved in this study.
2) As was previously known, the set of interstellar clouds toward HD 204827 shows anomalous reddening and, as underlined in Paper I, a very high value of $N(C_2)/E(B-V)$. Therefore, the physical conditions in one or more of these several clouds must differ significantly from those found toward the prototypical DIB target HD 183143, for example.
3) HD 204827 is a double-lined spectroscopic binary, and both stars show variations in radial velocity that exceed 40 km s$^{-1}$ over the five nights for which we have data. This variability generally allows the reliable removal of lines arising from both components of the binary, including very weak stellar lines, from the list of the DIBs and from the less severe blends with DIBs.
4) Among the 380 bands measured here, 113 (or 30%) were not detected in four previous, modern surveys of the DIBs in the spectra of stars other than HD 204827. The three previous conclusions listed here may primarily account for this relatively large fraction of new DIBs.
5) The central wavelengths, the widths, and the equivalent widths of nearly all of these bands are presented in Table 2, along with the minimum errors for the latter values. The random and the systematic uncertainties in the wavelengths and in the widths of the bands are discussed in sections 3.2 and 3.4. For the narrowest DIBs, the systematic, zero-point uncertainty in all of the absolute wavelengths generally exceeds substantially the random measurement errors, which amount to ± 0.03 Å. This systematic uncertainty arises from the unknown identities and laboratory spectra of the presumed absorbing molecules, and from the presence of multiple clouds along the light path. In higher-resolution spectra, the two strongest, partially resolved components of the interstellar K I line are separated by about 5 km s$^{-1}$, or 0.13 Å at the λ7699 line, for example. .
6) Over 200 Å ranges, the average spacing between adjacent DIBs found toward HD 204827 decreases to a minimum of 4.4 Å near 6200 Å. However, the distribution of the wavenumber gaps between adjacent bands appears to be both effectively random and substantially unaffected by mutual overlapping at R = 38,000. In particular, among adjacent DIBs, no preferred wavenumber interval is evident that might be identified with the molecular structure of a specific DIB absorber.
7) Similarly, no preferred wavenumber splitting is convincingly evident from a limited investigation of the wavenumber splittings among all possible pairs of DIBs in the catalog.

It is a pleasure to thank Joe Reader for providing valuable information about laboratory wavelengths of He II and C IV lines, and John Hutchings and Hal McAlister for discussions of the velocity curves of the binary system. Financial support for this work was provided by grants NNG04GL34G, NNX06AB14G, and NAG-12279 from the National Aeronautics and Space Administration to the University of Colorado (T.S.); by grant PHY 03-54200 from the National Science Foundation to the University of Chicago (T.O.); and by LSTA grant NAG-13114 from the National Aeronautics and Space Administration to the Johns Hopkins University (P.S.).

APPENDIX A

A general overview of the final, combined spectrum of HD 204827 and of the spectrum of 10 Lac is provided in Figure 11, where the DIBs present in the spectrum of HD 204827 are marked by vertical lines. (Some of the stronger DIBs are also present in the spectrum of 10 Lac, for which E(B-V) = 0.11.) The vertical lines are numbered in order of wavelength, at intervals of five bands, for easy comparison with the entries in Table 2. In contrast to the cases of Figures 1 and 4, the wavelength scale in Figure 11 is insufficiently expanded to allow scrutiny of the many relatively narrow, weak DIBs, but their more important characteristics are listed in Table 2. In addition, Paper I presented preliminary plots at a more expanded scale of our spectra of both HD 204827 and HD 183143. The plots display fifteen selected wavelength regions, each 50 Å wide, located at $4350 < \lambda < 6750$ Å.

The averaged stellar line profiles of HD 204827 presented in Figure 11 are substantially distorted by the orbital motion of the binary, especially in the case of the narrow-lined component B (cf. Figure 1). Because the spectra obtained on the five different nights are not shown separately in Figure 11, many of the weaker stellar lines in the combined spectrum of HD 204827 cannot be readily recognized as such. The slight, but important, mismatch in spectral type between 10 Lac and both components of HD 204827 is also especially evident in the comparative strengths of some of the weaker stellar lines. The imperfect removal of telluric lines is often obvious, especially in the region at $\lambda > 6865$ Å. Almost all of the apparent emission lines in both spectra in that region result from erroneous telluric corrections and are spurious.

A number of atomic and (diatomic or triatomic) molecular interstellar lines are also present in this line of sight (Adamkovics et al 2003; Pan et al 2004; Sonnentrucker et al 2007). The atomic lines and their strengths are reported in Table 5, while the corresponding data for the molecular lines appear in Table 6. Upper limits listed in the latter table indicate blends of molecular lines with stellar or telluric lines. The lines listed in Tables 5 and 6 are identified in Figure 11 by asterisks placed above the spectrum of 10 Lac, although the asterisks refer to the spectrum of HD 204827. The Fe I line at 3859.91 Å and the lines of the (2-0) Phillips band of $C_2$ near 8760 Å lie outside the wavelength limits adopted for the catalog of DIBs and for Figure 11. However, the very narrow lines of Fe I and $C_2$ can be measured reliably in our final spectrum and are therefore included in the tables as well. All of the nearly unsaturated lines in Tables 5 and 6 show FWHM $\leq 14$ km s$^{-1}$, or intrinsic breadths of FWHM $\leq 11.5$ km s$^{-1}$. These lines are narrower than all of the DIBs, and the detection limits for these lines are correspondingly lower (cf. Table 1). Nevertheless, the reality of the five lines in Table 6 with $W_\lambda \leq 1.5$ mÅ should be regarded as uncertain, although all appear nominally to be real.

APPENDIX B

As the basis for their atlas of DIBs, Jenniskens & Désert (1994) show the spectra of one unreddened star and of four reddened stars with a range in E(B-V) from 0.30 to 1.28, over the wavelength range $3800 < \lambda < 8680$ Å. They denote the DIBs in their atlas with a "+" to indicate a certain detection; an "o" for a probable detection; and a "–" to indicate a possible detection. No quantitative uncertainties are given. Among the 131 DIBs in common between their atlas and our Table 2, the numbers of certain, probable, and possible detections indicated by Jenniskens & Désert are 101, 27, and 3, respectively.

Galazutdinov et al (2000) used o Per (HD 23180, E(B-V) = 0.32) as their primary DIBs standard, and they co-added the spectra of seven other heavily reddened stars for comparison, over the range $4460 < \lambda < 8800$ Å. Since o Per is a short-period spectroscopic binary, the authors were generally able to distinguish the interstellar from the stellar features, in identifying 271 DIBs in its spectrum. Galazutdinov et al also used synthetic stellar spectra as a further means of identifying blends of DIBs with stellar lines. On these grounds, the authors consider all of their DIBs as positive detections. No equivalent widths are reported, because the primary emphasis of this study is on accurate wavelengths for the DIBs.

The DIB atlas of Tuairisg et al (2000) is based on data for three highly reddened stars with a range in E(B-V) from 1.01 to 2.00, over the range $3906 < \lambda < 6812$ Å Synthetic spectra were used for comparisons. The authors' Table 4 lists 226 DIBs and $1\sigma$ uncertainties for the equivalent widths.

Finally, in the atlas of Weselak et al (2000), the data reduction and the uncertainties are handled in much the same way as by Galazutdinov et al (2000). Weselak et al co-added at least three spectra in each case, and also compared the results with synthetic spectra, in order to identify and eliminate stellar lines. These authors subdivided their 14 stars (with a range in E(B-V) frin 0.16 to 0.52) into separate "zeta" and "sigma" categories before co-addition, in order to assess the DIBs' behavior in the two groups, which were defined on the basis of the 5780/5797 strength ratio (Krelowski & Walker 1987). Quantitative uncertainties are not provided, but all listed DIBs appear in all of the reddened spectra. No discussion of possible blends with specific stellar lines is presented, but the authors indicate that the broad range in stellar spectral types should allow the interstellar bands to be distinguished from stellar lines. Weselak et al detected 62 DIBs, plus 27 others that were less certain, in the range $5650 < \lambda < 6865$ Å. For the present survey, we compare DIBs only within this spectral range that was specifically searched for DIBs by Weselak et al, although they included in their list some other DIBs known throughout the visible region that were taken from other surveys

# REFERENCES

Adamkovics, M., Blake, G. A., & McCall, B. J. 2003, ApJ, 595, 235
Cordiner, M. A., & Sarre, P. J. 2007, A&A, 472, 537
De Zeeuw, P. T., Hoogerwerf, R., de Bruijne, J. H. J., Brown, A. G. A., & Blaauw, A. 1999, AJ, 117, 354
Galazutdinov, G. A., Musaev, F. A., Krelowski, J. & Walker, G. A. H. 2000, PASP, 112, 648
Garcia, J. D., & Mack, J. E. 1965, J. Opt. Soc. Am., 55, 654
Garrison, R. F., & Kormendy, J. 1976, PASP, 88, 865
Gies, D. R. 1987, ApJS, 64, 545
Herbig, G. H. 1995, ARA&A, 33, 19
Hudgins, D. M., & Allamandola, L. J. 1999, ApJ, 513, L69
Jenniskens, P., & Desert, F-X. 1994, A&AS, 106, 39
Krelowski, J., & Walker, G. A. H. 1987. ApJ 312, 860
Le Page, V., Snow, T. P., & Bierbaum, V. M. 2003, ApJS, 584, 316
Marschall, L. A., &  van Altena, W. F. 1987, AJ, 94, 71
Mason, B. D., Gies, D. R., Hartkopf, W. I., Bagnuolo, W. G., ten Brummelaar, T., & McAlister, H. A. 1998, AJ, 115, 821
Morbidelli, L., Patriarchi, P., Perinotto, M., Barbaro, G., & Di Bartolomeo, A. 1997, A&A, 327, 125
Morton, D. C. 2003, ApJS, 149, 205
Oka, T., Thorburn, J. A., McCall, B. J., Friedman, S. D., Hobbs, L. M., Sonnentrucker, P., Welty, D. E., & York, D. G. 2003, ApJ, 582, 823
Pan, K., Federman, S. R., Cunha, K., Smith, V. V., & Welty, D. E. 2004, ApJS, 151, 313
Patel, N. A., Goldsmith, P. F., Heyer, M. H., Snell, R. L. , & Pratap, P. 1998, ApJ, 507, 241
Petrie, R. M., & Pearce, J. A. 1961, Publ. Dom. Astrophys. Obs. Victoria, 12, 1
Sarre, P. J. 2006, J. Mol. Spectr. 238, 1
Simonson, S. C. 1968, ApJ, 154, 923
Sonnentrucker, P., Welty, D. E., Thorburn, J. A., & York, D. G. 2007, ApJS, 168, 58
Thorburn, J. A., et al 2003, ApJ, 584, 339
Tuairisg, S. O., Cami, J., Foing, B. H., Sonnentrucker, P., & Ehrenfreund, P. 2000, A&AS, 142, 225
Valencic, L. A., Clayton, G. C., Gordon, K. D., & Smith, T. L. 2003, ApJ, 598, 369
Wallerstein, G., Sandstrom, K., & Gredel, R. 2007, PASP (in press)
Wang, S., et al 2003, Proc. SPIE, 4841, 1145
Webster, A. 2004, MNRAS, 349, 263
Welty, D. E., Hobbs, L. M., & Kulkarni, V. 1994, ApJ, 436, 152
Weselak, T., Schmidt, M., & Krelowski, J. 2000, A&AS, 142, 239
17

Table 1

1σ Detection Limits

| $\lambda$(Å) | S/N[a] | $\Delta W_\lambda$(mÅ) FWHM = 0.5 Å | $\Delta W_\lambda$(mÅ) FWHM = 2.0 Å |
|---|---|---|---|
| 4000 | 503 | 1.1 | 4.2 |
| 4500 | 613 | 0.9 | 3.5 |
| 5000 | 690 | 0.8 | 3.1 |
| 5500 | 736 | 0.7 | 2.9 |
| 6000 | 750 | 0.7 | 2.8 |
| 6500 | 731 | 0.7 | 2.9 |
| 7000 | 681 | 0.8 | 3.1 |
| 7500 | 599 | 0.9 | 3.6 |
| 8000 | 485 | 1.1 | 4.4 |

[a] Per pixel.

Table 2

DIB Parameters

| JD 91 $\lambda_c$(Å) | GM 00 $\lambda_c$(Å) | TC 00 $\lambda_c$(Å) | WS 00 $\lambda_c$(Å) | $\lambda_c$(Å) | FWHM (Å) | $W_\lambda$ (mÅ) | $\Delta W_\lambda$ (mÅ) | note | number |
|---|---|---|---|---|---|---|---|---|---|
| | | | | 4259.01 | 1.05 | 21.5 | 2.0 | s | 1 |
| | | | | 4363.86 | 0.46 | 14.7 | 0.8 | | 2 |
| 28.88 | | 27.96 | | 4428.19 | 22.50 | 1221.0 | 40.0 | s | 3 |
| 1.80 | 1.80 | 1.65 | | 4501.79 | 2.05 | 31.5 | 3.6 | s | 4 |
| | | | | 4659.82 | 0.45 | 5.8 | 0.7 | | 5 |
| | | | | 4668.65 | 0.67 | 16.1 | 1.1 | s | 6 |
| | | | | 4680.20 | 0.65 | 9.2 | 1.1 | | 7 |
| | | | | 4683.03 | 0.43 | 20.3 | 0.7 | s | 8 |
| | | | | 4688.89 | 0.48 | 8.3 | 0.8 | s | 9 |
| blend | 26.27 | blend | | 4726.83 | 2.74 | 283.8 | 4.2 | | 10 |
| | | | | 4734.79 | 0.42 | 15.9 | 0.7 | | 11 |
| 62.57 | 62.67 | 62.57 | | 4762.61 | 1.00 | 50.8 | 1.6 | s? | 12 |
| 80.09 | 80.04 | 80.10 | | 4780.02 | <1.67 | <41.3 | 2.7 | s | 13 |
| | | | | 4817.64 | 0.30 | 3.9 | 0.5 | | 14 |
| 80.35 | | | | 4879.96 | 1.58 | 10.8 | 2.5 | | 15 |
| | | | | 4887.14 | 0.71 | 4.5 | 1.1 | | 16 |
| | | | | 4947.38 | 0.43 | 2.2 | 0.7 | | 17 |
| | | 51.05 | | 4951.12 | 0.61 | 10.3 | 0.9 | | 18 |
| | | | | 4959.63 | 0.68 | 3.4 | 1.1 | | 19 |
| | | | | 4961.95 | 0.51 | 5.6 | 0.8 | | 20 |
| 63.96 | 63.90 | 63.89 | | 4963.88 | 0.62 | 53.4 | 1.0 | | 21 |
| | | | | 4965.22 | 0.62 | 4.3 | 1.0 | | 22 |
| | | | | 4966.01 | 0.51 | 4.3 | 0.8 | | 23 |
| | | | | 4969.14 | 0.76 | 14.7 | 1.1 | | 24 |
| | | 79.28 | | 4979.61 | 0.64 | 13.5 | 1.0 | | 25 |
| | | | | 4982.14 | 0.41 | 3.2 | 0.6 | | 26 |
| 84.73 | 84.81 | 84.78 | | 4984.79 | 0.50 | 31.1 | 0.8 | | 27 |
| | | | | 4987.42 | 1.88 | 30.4 | 2.9 | | 28 |
| | | | | 5003.58 | 0.58 | 14.6 | 0.9 | s | 29 |
| | | | | 5027.47 | 0.62 | 11.6 | 0.9 | | 30 |
| | | | | 5054.84 | 0.56 | 11.2 | 0.9 | | 31 |
| | | 61.56 | | 5061.49 | 0.51 | 16.6 | 0.8 | | 32 |
| | | | | 5074.47 | 0.48 | 24.6 | 0.7 | | 33 |
| | | | | 5092.09 | 0.43 | 4.6 | 0.7 | d, t | 34 |
| | | | | 5100.95 | 0.57 | 4.8 | 0.9 | | 35 |
| 9.70 | | 10.81 | | 5110.89 | 1.53 | 6.3 | 2.3 | | 36 |
| | | | | 5117.62 | 0.75 | 5.6 | 1.1 | | 37 |
| | | | | 5133.14 | <0.94 | <10.6 | 1.4 | s | 38 |
| | | | | 5137.07 | 0.43 | 1.9 | 0.6 | | 39 |
| | | 70.69 | | 5170.49 | 0.47 | 12.1 | 0.7 | | 40 |

|       |       |       |         |       |        |      |     |    |
|-------|-------|-------|---------|-------|--------|------|-----|----|
|       |       | 76.00 | 5176.04 | 0.62  | 35.6   | 0.9  | s   | 41 |
|       |       |       | 5178.10 | 0.48  | 2.0    | 0.7  |     | 42 |
|       |       |       | 5217.85 | 0.41  | 5.8    | 0.6  |     | 43 |
|       |       |       | 5229.76 | 0.50  | 3.3    | 0.7  |     | 44 |
|       |       | 36.34 | 5236.18 | 1.66  | <25.7  | 2.5  | s   | 45 |
|       |       |       | 5245.48 | 0.64  | 10.1   | 0.9  |     | 46 |
|       |       |       | 5251.78 | 0.41  | 2.7    | 0.6  |     | 47 |
|       |       |       | 5257.47 | 0.82  | 13.6   | 1.2  |     | 48 |
|       |       |       | 5262.48 | 0.54  | 6.2    | 0.8  |     | 49 |
|       |       |       | 5297.97 | 0.72  | 4.1    | 1.1  |     | 50 |
|       |       |       | 5304.25 | 0.55  | 5.8    | 0.8  |     | 51 |
|       |       |       | 5340.38 | 0.76  | 9.7    | 1.1  |     | 52 |
|       |       |       | 5342.54 | 0.60  | 3.0    | 0.9  |     | 53 |
|       |       | 59.58 | 5358.75 | 0.71  | 2.6    | 1.0  |     | 54 |
| 63.60 | 63.60 | 63.78 | 5363.70 | 1.24  | 10.8   | 1.8  |     | 55 |
|       |       | 70.36 | 5371.28 | 1.77  | 8.5    | 2.6  |     | 56 |
|       |       |       | 5384.19 | 0.81  | 3.2    | 1.2  |     | 57 |
|       |       |       | 5390.85 | 0.61  | 2.1    | 0.9  |     | 58 |
|       |       |       | 5395.64 | 1.11  | 6.6    | 1.6  |     | 59 |
| 4.56  | 4.50  | 4.52  | 5404.56 | 0.91  | 14.9   | 1.3  |     | 60 |
|       | 18.90 | blend | 5418.87 | 0.76  | 49.1   | 1.1  |     | 61 |
|       |       |       | 5424.10 | 0.65  | 3.9    | 0.9  |     | 62 |
|       |       |       | 5433.50 | 0.45  | 3.0    | 0.7  |     | 63 |
| 49.63 |       | blend | 5450.62 | 11.06 | <136.5 | 16.0 | s,t | 64 |
|       |       |       | 5480.79 | 0.54  | 6.1    | 0.8  |     | 65 |
| 87.43 | 87.67 | 87.54 | 5487.69 | 5.20  | 73.8   | 7.5  | s?  | 66 |
| 94.14 | 94.10 | 94.10 | 5494.10 | 0.50  | 24.2   | 0.7  |     | 67 |
|       |       | 97.00 | 5497.08 | 2.61  | 21.1   | 3.8  | s   | 68 |
|       |       | 3.16  | 5502.91 | 1.27  | 7.7    | 1.8  |     | 69 |
|       |       |       | 5504.31 | 0.41  | 3.3    | 0.6  |     | 70 |
|       |       | 6.06  | 5506.28 | 0.92  | 8.0    | 1.3  | d   | 71 |
| 8.35  | 8.35  | 8.03  | 5508.12 | 2.37  | 84.1   | 3.4  | s   | 72 |
|       | 12.64 | 12.66 | 5512.68 | 0.48  | 20.8   | 0.7  |     | 73 |
|       |       | 15.97 | 5515.99 | 1.07  | 8.7    | 1.5  |     | 74 |
|       |       | 24.47 | 5524.98 | 1.65  | 10.6   | 2.4  | d   | 75 |
| 24.89 |       | 25.48 | 5525.48 | 10.56 | 112.8  | 15.2 |     | 76 |
|       |       |       | 5527.55 | 0.52  | 4.7    | 0.7  | d   | 77 |
|       |       |       | 5530.07 | 0.61  | 3.5    | 0.9  | d   | 78 |
| 35.68 |       | 35.26 | 5535.20 |       | p      |      | s   | 79 |
| 37.00 |       | 37.51 | 5537.27 | 0.54  | 2.3    | 0.8  |     | 80 |
| 40.98 | 41.62 | 41.78 | 5541.84 | 0.56  | 16.8   | 0.8  |     | 81 |
| 44.97 | 44.96 | 45.02 | 5545.06 | 0.80  | 29.8   | 1.1  |     | 82 |
|       | 46.46 | 46.52 | 5546.48 | 0.55  | 17.4   | 0.8  |     | 83 |
|       |       |       | 5547.48 | 0.36  | 2.5    | 0.5  |     | 84 |
|       |       |       | 5551.07 | 0.46  | 3.0    | 0.7  |     | 85 |
|       |       |       | 5553.98 | 0.73  | 4.8    | 1.0  |     | 86 |
|       |       | 56.27 | 5556.44 | 1.28  | 10.5   | 1.8  |     | 87 |
|       |       | 59.93 | 5560.03 | 1.34  | 7.9    | 1.9  |     | 88 |
|       |       |       | 5566.11 | 1.41  | 6.2    | 2.0  |     | 89 |

|  |  |  |  |  |  |  |  |  |  |
|---|---|---|---|---|---|---|---|---|---|
|  |  |  |  | 5580.82 | 0.53 | 4.2 | 0.8 |  | 90 |
|  |  |  |  | 5585.56 | 0.64 | 4.0 | 0.9 |  | 91 |
|  | 94.59 | 94.54 |  | 5594.60 | 0.45 | 8.1 | 0.6 | s | 92 |
|  |  |  |  | 5599.70 | 0.61 | 4.2 | 0.8 |  | 93 |
|  |  | 0.49 |  | 5600.85 | 0.91 | 4.9 | 1.3 |  | 94 |
| 9.96 | 9.73 | 9.96 |  | 5609.78 | 0.63 | 7.1 | 0.9 |  | 95 |
|  |  | 34.73 |  | 5634.98 | 1.12 | 6.5 | 1.6 |  | 96 |
|  |  | 45.43 |  | 5645.49 | 0.68 | 4.1 | 1.0 | t | 97 |
|  |  |  |  | 5669.33 | 1.47 | 9.7 | 2.1 |  | 98 |
| blend | 5.20 | 5.10 | 5.43 | 5705.08 | 2.58 | 41.6 | 3.7 |  | 99 |
|  |  |  |  | 5706.51 | 0.49 | 4.8 | 0.7 | d | 100 |
|  |  |  |  | 5707.77 | 0.58 | 5.0 | 0.8 | d | 101 |
|  |  |  |  | 5711.45 | 0.50 | <19.5 | 0.7 | s | 102 |
|  |  |  |  | 5716.32 | 0.51 | 3.3 | 0.7 |  | 103 |
| 19.43 | 19.30 | 19.68 | 19.40 | 5719.48 | 0.69 | 16.7 | 1.0 |  | 104 |
|  |  |  |  | 5734.97 | 0.58 | 3.5 | 0.8 |  | 105 |
|  |  |  |  | 5735.89 | 0.35 | 2.0 | 0.5 |  | 106 |
|  |  |  |  | 5753.47 | 0.48 | 2.2 | 0.7 |  | 107 |
|  |  |  | 56.07 | 5756.12 | 0.64 | 3.4 | 0.9 |  | 108 |
|  |  |  |  | 5758.90 | 0.66 | 2.1 | 0.9 |  | 109 |
|  | 60.40 | 60.64 | 60.44 | 5760.53 | 0.65 | 7.6 | 0.9 |  | 110 |
| 62.50 | 62.70 | 62.80 | 62.69 | 5762.74 | 0.51 | 11.8 | 0.7 |  | 111 |
| 66.25 | 66.16 | 66.17 | 66.15 | 5766.10 | 0.98 | 29.8 | 0.9 | det | 112 |
| 69.10 | 69.09 | 69.32 | 69.03 | 5769.09 | 0.54 | 19.4 | 0.8 |  | 113 |
|  |  |  |  | 5769.92 | 0.52 | 3.9 | 0.7 |  | 114 |
| 72.60 | 72.60 | 72.67 | 72.53 | 5772.61 |  | p |  | s | 115 |
| 76.08 | 75.78 | 76.21 | 75.75 | 5775.90 | 1.18 | 7.8 | 1.7 |  | 116 |
| 80.59 | 80.37 | 80.55 | 80.50 | 5780.48 | 2.11 | 257.0 | 3.0 | nc | 117 |
| 84.90 | 85.05 | 84.86 | 85.11 | 5785.04 | 0.89 | 10.0 | 1.3 |  | 118 |
|  | 93.22 | 93.13 | 93.19 | 5793.21 | 0.79 | 19.1 | 1.1 |  | 119 |
| 95.23 | 95.16 |  | 95.20 | 5795.20 | 0.93 | 5.4 | 1.3 | d | 120 |
| 97.11 | 96.96 | 97.08 | 96.97 | 5797.06 | 0.77 | 199.0 | 1.1 | nc | 121 |
|  | 6.68 | 6.52 |  | 5806.62 | 0.37 | 1.8 | 0.5 |  | 122 |
| blend | 9.24 | 9.53 | 9.22 | 5809.23 | 1.09 | 7.8 | 1.6 |  | 123 |
| 14.50 |  |  | 14.21 | 5814.26 | 0.44 | 2.7 | 0.6 | s | 124 |
|  | 15.71 | 15.80 | 15.78 | 5815.74 | 0.58 | 5.2 | 0.8 |  | 125 |
| 18.85 | 18.75 | 18.47 | 18.69 | 5818.72 | 0.60 | 9.6 | 0.9 |  | 126 |
|  | 21.23 |  | 21.22 | 5821.17 | 0.39 | 2.7 | 0.6 |  | 127 |
| 28.40 | 28.46 | 28.56 | 28.52 | 5828.47 | 0.86 | 18.3 | 1.2 | s | 128 |
|  | 38.00 | 37.92 | 38.08 | 5838.02 | 0.49 | 9.4 | 0.7 |  | 129 |
|  |  |  |  | 5838.97 | 0.48 | 2.4 | 0.7 |  | 130 |
|  | 40.65 | 40.62 | 40.72 | 5840.66 | 0.58 | 9.8 | 0.8 |  | 131 |
|  | 42.23 | blend | blend | 5842.93 | 2.13 | 13.8 | 3.0 | d, t | 132 |
| 44.19 | 44.80 | 44.95 | 44.80 | 5844.92 | 0.50 | 6.2 | 0.7 |  | 133 |
| 49.78 | 49.80 | 49.81 | 49.85 | 5849.81 | 0.82 | 95.6 | 1.2 |  | 134 |
|  | 54.50 |  | 54.54 | 5854.54 | 0.42 | 11.2 | 0.6 | d | 135 |
|  | 55.63 |  | 55.72 | 5855.68 | 0.41 | 8.8 | 0.6 | d | 136 |
|  |  |  | 59.00 | 5859.05 | 0.52 | 2.4 | 0.8 |  | 137 |
|  |  |  |  | 5885.37 | 0.58 | 5.7 | 0.8 |  | 138 |

| | | | | | | | | |
|---|---|---|---|---|---|---|---|---|
| | | | | 5888.75 | 0.42 | 2.1 | 0.6 | | 139 |
| | | | | 5893.54 | 0.71 | 10.9 | 1.0 | | 140 |
| | 0.40 | 0.58 | 0.56 | 5900.58 | 0.59 | 6.8 | 0.8 | | 141 |
| | 4.60 | | 4.52 | 5904.63 | 0.85 | 4.6 | 1.2 | | 142 |
| | 10.54 | 10.54 | 10.40 | 5910.57 | 0.74 | 21.4 | 1.0 | | 143 |
| | | | | 5913.74 | 0.53 | 3.0 | 0.8 | | 144 |
| | | | | 5914.79 | 0.38 | 2.1 | 0.5 | | 145 |
| | | | 17.51 | 5917.05 | 1.39 | 6.6 | 2.0 | | 146 |
| | 22.25 | | 22.31 | 5922.32 | 0.50 | 6.8 | 0.7 | | 147 |
| | 23.40 | 23.51 | 23.39 | 5923.47 | 0.59 | 13.9 | 0.8 | | 148 |
| | 25.90 | 25.81 | 25.85 | 5925.94 | 0.81 | 11.3 | 1.1 | | 149 |
| | 27.68 | | 27.70 | 5927.63 | 0.59 | 8.1 | 0.8 | | 150 |
| | | | 28.96 | 5928.86 | 0.44 | 2.6 | 0.6 | s | 151 |
| | | | | 5934.60 | 0.43 | 2.0 | 0.6 | | 152 |
| | 45.47 | 45.53 | 45.44 | 5945.54 | 0.49 | 9.9 | 0.7 | | 153 |
| | 47.29 | 47.29 | 47.25 | 5947.32 | 0.44 | 8.9 | 0.6 | | 154 |
| | 48.86 | 48.87 | 48.88 | 5948.88 | 0.47 | 6.1 | 0.7 | | 155 |
| | | | | 5952.26 | 0.45 | 4.7 | 0.6 | | 156 |
| | | | | 5954.25 | 0.34 | 2.6 | 0.5 | | 157 |
| | 58.90 | 58.89 | 58.89 | 5958.91 | 1.15 | 19.7 | 1.6 | | 158 |
| | | | | 5962.89 | 0.96 | 6.3 | 1.4 | | 159 |
| | 73.75 | 73.78 | 73.78 | 5973.81 | 0.38 | 3.7 | 0.5 | | 160 |
| | 75.74 | 75.66 | 75.58 | 5975.75 | 0.41 | 7.0 | 0.6 | | 161 |
| 82.00 | 82.93 | 82.77 | 82.83 | 5982.67 | 0.54 | 4.0 | 0.8 | | 162 |
| | 86.61 | 86.60 | 86.43 | 5986.55 | 0.49 | 2.4 | 0.7 | | 163 |
| | 88.08 | 88.04 | 88.03 | 5988.11 | 0.73 | 14.2 | 1.0 | | 164 |
| | 89.44 | 89.51 | 89.48 | 5989.46 | 0.57 | 6.3 | 0.8 | | 165 |
| | 95.75 | 95.73 | 95.73 | 5995.83 | 0.79 | 5.1 | 1.1 | | 166 |
| | 99.63 | | 99.18 | 5999.54 | 0.62 | 2.8 | 0.9 | | 167 |
| 4.55 | 5.03 | 4.80 | 4.93 | 6004.89 | 2.65 | 21.1 | 3.8 | d | 168 |
| 10.58 | 10.65 | 10.48 | 10.80 | 6010.75 | 3.27 | 31.5 | 4.6 | d | 169 |
| | | | | 6014.81 | 0.73 | 3.4 | 1.0 | | 170 |
| 19.34 | 19.36 | 19.45 | 19.23 | 6019.32 | 0.79 | 4.2 | 1.1 | s | 171 |
| 27.39 | 27.48 | 27.09 | 27.48 | 6027.68 | 1.99 | 18.5 | 2.8 | | 172 |
| | 30.40 | | | 6030.49 | 0.62 | 2.6 | 0.9 | | 173 |
| 37.56 | 37.61 | 37.47 | 37.38 | 6037.63 | 1.71 | 14.8 | 2.4 | | 174 |
| | | | | 6048.60 | 0.77 | 4.6 | 1.1 | | 175 |
| | | | | 6054.50 | 0.98 | 4.2 | 1.4 | | 176 |
| | | | 57.58 | 6057.52 | 0.59 | 2.7 | 0.8 | | 177 |
| | 59.67 | 59.88 | 59.28 | 6059.26 | 0.54 | 12.5 | 0.8 | | 178 |
| 65.38 | 65.20 | 65.19 | 65.23 | 6065.28 | 0.54 | 7.3 | 0.8 | | 179 |
| 70.53 | | 71.08 | 70.98 | 6071.31 | 0.91 | 4.0 | 1.3 | | 180 |
| | | | | 6081.10 | 0.52 | 1.9 | 0.7 | s | 181 |
| | | | 82.18 | 6082.33 | 0.88 | 3.6 | 1.3 | s | 182 |
| | 84.75 | 84.94 | 84.78 | 6084.94 | 0.82 | 6.8 | 1.2 | | 183 |
| | | | | 6087.48 | 0.96 | 4.6 | 1.4 | | 184 |
| 89.80 | 89.78 | 89.79 | 89.78 | 6089.85 | 0.54 | 28.0 | 0.8 | | 185 |
| | | | 93.18 | 6093.40 | 0.95 | 5.9 | 1.4 | | 186 |
| | 2.38 | | 2.33 | 6102.72 | 1.22 | 6.1 | 1.7 | s? | 187 |
|  |  |  |  | 6106.36 | 0.48 | 2.5 | 0.7 | d | 188 |
|  |  |  |  | 6107.16 | 0.45 | 1.5 | 0.6 | d | 189 |
| 8.21 | 8.05 | 8.14 | 8.13 | 6108.05 | 0.37 | 5.6 | 0.5 |  | 190 |
|  |  |  | 9.98 | 6109.88 | 0.50 | 4.2 | 0.7 |  | 191 |
| 13.22 | 13.20 | 13.20 | 13.18 | 6113.18 | 0.68 | 24.3 | 1.0 |  | 192 |
| 16.65 | 16.80 | 16.74 | 16.83 | 6116.84 | 0.81 | 13.9 | 1.2 |  | 193 |
|  | 18.68 | 18.66 | 18.63 | 6118.63 | 0.56 | 5.4 | 0.8 |  | 194 |
|  |  |  |  | 6136.08 | 1.55 | 5.5 | 2.2 |  | 195 |
| 39.77 | 39.94 | 39.94 | 40.03 | 6139.98 | 0.61 | 15.5 | 0.9 |  | 196 |
|  | 41.91 |  |  | 6142.09 | 0.68 | 4.9 | 1.0 | s | 197 |
|  | 45.69 | 45.41 | 45.50 | 6145.61 | 0.59 | 3.6 | 0.8 |  | 198 |
|  |  | 47.02 |  | 6148.40 | 0.84 | 5.9 | 1.2 |  | 199 |
|  |  | 51.15 |  | 6151.14 | 1.01 | 5.6 | 1.4 | d? | 200 |
|  | 58.54 |  | 58.53 | 6158.57 | 0.70 | 15.4 | 1.0 |  | 201 |
|  | 61.93 | 61.83 | 61.96 | 6161.88 | 0.46 | 9.7 | 0.7 |  | 202 |
|  |  |  |  | 6163.49 | 0.39 | 5.0 | 0.6 |  | 203 |
|  | 65.97 | 65.72 |  | 6165.68 | 1.48 | 12.3 | 2.1 |  | 204 |
|  |  | 70.71 |  | 6170.48 | 1.67 | 10.0 | 2.4 |  | 205 |
|  |  |  |  | 6174.80 | 0.55 | 2.1 | 0.8 |  | 206 |
| 77.27 |  | 77.72 |  | 6177.63 | 1.25 | 4.1 | 1.8 |  | 207 |
|  |  |  | 82.60 | 6182.58 | 1.04 | 6.4 | 1.5 |  | 208 |
|  | 85.81 | 85.89 | 85.98 | 6185.81 | 0.42 | 5.7 | 0.6 |  | 209 |
|  |  | 87.19 | 87.58 | 6187.22 | 0.63 | 3.3 | 0.9 |  | 210 |
| 89.53 |  | 89.31 | 89.47 | 6189.52 | 0.86 | 4.8 | 1.2 |  | 211 |
| 94.87 | 94.73 | blend | 94.77 | 6194.74 | 0.38 | 7.9 | 0.5 |  | 212 |
| 96.19 | 95.96 | 95.99 | blend | 6195.98 | 0.42 | 37.8 | 0.6 |  | 213 |
| 99.21 | 98.87 | 98.82 | 99.04 | 6198.97 | 0.40 | 2.5 | 0.6 |  | 214 |
| 3.19 | 3.08 | 3.06 | 3.05 | 6203.05 | 1.21 | 57.1 | 1.7 | d | 215 |
| 4.33 | 4.66 | blend | 4.86 | 6204.49 | 4.87 | 71.6 | 6.9 | s, d | 216 |
| 12.19 | 11.67 | 11.74 | 11.73 | 6211.69 | 0.59 | 9.2 | 0.8 |  | 217 |
|  | 12.90 | 12.95 | 12.87 | 6212.91 | 0.71 | 5.3 | 1.0 |  | 218 |
| 15.71 | 15.79 | 15.80 |  | 6215.81 | 1.03 | 6.7 | 1.5 |  | 219 |
| 23.65 | 23.56 | 23.53 | 23.63 | 6223.67 | 0.56 | 4.9 | 0.8 |  | 220 |
|  |  |  |  | 6225.24 | 0.48 | 2.4 | 0.7 |  | 221 |
|  | 26.30 | 26.02 | 26.08 | 6226.27 | 0.57 | 6.8 | 0.8 |  | 222 |
| 34.27 | 34.03 | 34.01 | 34.05 | 6234.01 | 0.78 | 25.3 | 1.1 |  | 223 |
| 36.58 | 36.67 | 36.71 |  | 6236.92 | 0.68 | 6.0 | 1.0 |  | 224 |
|  |  | blend | 45.47 | 6244.46 | 1.72 | 25.4 | 2.5 | s | 225 |
|  | 50.84 | 50.77 | 50.83 | 6250.88 | 0.64 | 5.1 | 0.9 |  | 226 |
|  |  |  |  | 6252.43 | 0.47 | 1.6 | 0.7 |  | 227 |
|  |  |  |  | 6259.70 | 0.69 | 4.0 | 1.0 |  | 228 |
| 70.06 | 69.75 | blend | 69.86 | 6269.85 | 1.17 | 77.0 | 1.7 |  | 229 |
|  |  |  |  | 6275.58 | 0.38 | 3.4 | 0.5 |  | 230 |
| 84.31 | 83.85 | blend | 83.80 | 6283.84 | 4.77 | 459.7 | 6.9 | t | 231 |
| 87.12 | 87.47 | 87.57 | 87.54 | 6287.59 | 0.51 | 13.9 | 0.7 | d | 232 |
| 8.93 | 9.10 | 8.92 |  | 6308.80 | 5.04 | 54.4 | 7.2 |  | 233 |
| 17.58 | 17.06 | 17.75 |  | 6317.86 | 5.18 | 46.6 | 7.4 |  | 234 |
| 25.10 | 24.80 | 24.81 |  | 6324.45 | 1.60 | 13.6 | 2.3 |  | 235 |
| 30.42 | 29.97 | 30.14 |  | 6329.98 | 0.63 | 13.8 | 0.9 | d? | 236 |
|  |  |  |  | 6338.00 | 0.71 | 3.1 | 1.0 |  | 237 |
| 53.49 | 53.34 | 53.18 | 53.25 | 6353.38 | 1.90 | 17.7 | 2.7 |  | 238 |
|  |  |  |  | 6355.45 | 0.83 | 4.8 | 1.2 |  | 239 |
|  |  | 58.54 | 58.35 | 6358.14 | 1.29 | 9.8 | 1.9 |  | 240 |
| 62.35 | 62.30 | 62.23 | 62.50 | 6362.15 | 1.95 | 13.3 | 2.8 |  | 241 |
| 67.22 | 67.25 | 67.28 | 67.41 | 6367.34 | 0.46 | 11.8 | 0.7 |  | 242 |
|  |  |  |  | 6374.66 | 0.68 | 2.6 | 1.0 |  | 243 |
| 76.07 | blend | blend | 76.10 | 6376.08 | 0.75 | 44.6 | 1.1 |  | 244 |
| 79.27 | 79.29 | 79.22 | 79.46 | 6379.32 | 0.58 | 94.9 | 0.8 |  | 245 |
|  |  |  |  | 6384.97 | 0.51 | 2.8 | 0.7 |  | 246 |
| 97.39 | 97.39 | 96.95 | 96.63 | 6397.01 | 1.21 | 25.0 | 1.7 |  | 247 |
|  | 0.30 | 0.37 | 0.25 | 6400.46 | 0.81 | 7.7 | 1.2 |  | 248 |
|  | 10.18 | 10.08 |  | 6410.20 | 0.60 | 6.4 | 0.9 |  | 249 |
| blend | 13.93 | blend | 14.18 | 6414.67 | 1.99 | 14.9 | 2.9 |  | 250 |
|  |  | 17.27 |  | 6417.03 | 1.12 | 6.1 | 1.6 |  | 251 |
|  |  | 18.54 |  | 6418.61 | 0.71 | 5.9 | 1.0 |  | 252 |
| 25.72 | 25.70 | 25.61 | 25.90 | 6425.66 | 0.76 | 16.5 | 1.1 | t | 253 |
| 39.34 | 39.50 | 39.42 | 39.50 | 6439.48 | 0.75 | 25.4 | 1.1 |  | 254 |
| 45.53 | 45.20 | 45.25 | 45.25 | 6445.28 | 0.46 | 35.7 | 0.7 |  | 255 |
| 49.13 | 49.14 | 49.30 | 49.28 | 6449.22 | 0.76 | 29.4 | 1.1 | t | 256 |
|  |  |  |  | 6452.12 | 0.47 | 2.2 | 0.7 |  | 257 |
|  |  | 56.08 | 55.86 | 6456.01 | 0.89 | 10.1 | 1.3 |  | 258 |
| 60.29 | 60.00 | 60.31 | 60.28 | 6460.44 | 0.51 | 4.1 | 0.7 |  | 259 |
|  | 63.61 | 63.63 | 63.41 | 6463.68 | 0.98 | 16.7 | 1.4 |  | 260 |
|  | 65.48 |  | 65.44 | 6465.37 | 0.65 | 2.6 | 0.9 |  | 261 |
|  | 66.74 | 66.95 | 66.65 | 6466.86 | 0.57 | 8.9 | 0.8 |  | 262 |
|  | 68.70 |  |  | 6468.77 | 0.93 | 8.2 | 1.3 |  | 263 |
|  | 74.27 |  | 74.23 | 6474.24 | 0.57 | 11.3 | 0.8 | s? | 264 |
|  |  |  | 86.75 | 6485.71 | 0.59 | 2.6 | 0.9 |  | 265 |
|  | 89.38 | 89.29 | 89.62 | 6489.47 | 0.84 | 9.4 | 1.2 | d | 266 |
| 91.88 | 92.02 | 92.17 | 92.15 | 6492.01 | 0.51 | 3.9 | 0.7 | d | 267 |
| 94.17 |  | 92.92 |  | 6494.05 | 8.98 | 69.2 | 13.1 | d | 268 |
| 97.55 |  | 97.79 | 97.82 | 6498.08 | 0.66 | 4.2 | 1.0 |  | 269 |
| 20.70 | 20.56 | 20.95 |  | 6520.62 | 0.89 | 23.7 | 1.3 |  | 270 |
|  |  |  | 23.36 | 6523.28 | 0.43 | 2.8 | 0.6 |  | 271 |
| 36.44 |  | 36.86 | 36.89 | 6536.53 | 0.71 | 12.7 | 1.0 |  | 272 |
|  | 43.20 |  | 43.33 | 6543.08 | 0.52 | 8.9 | 0.8 |  | 273 |
|  | 53.82 | 53.76 | 53.89 | 6553.92 | 0.45 | 10.2 | 0.7 | s | 274 |
| 91.40 |  | 91.03 |  | 6590.42 | 7.53 | 102.2 | 11.1 | d,s | 275 |
|  | 94.13 |  |  | 6594.36 | 0.63 | 3.5 | 0.9 |  | 276 |
| 97.39 | 97.31 | 97.47 | 97.34 | 6597.31 | 0.55 | 8.9 | 0.8 |  | 277 |
|  |  |  |  | 6600.08 | 0.42 | 3.4 | 0.6 |  | 278 |
|  |  |  |  | 6607.07 | 0.48 | 1.6 | 0.7 |  | 279 |
| 13.72 | 13.56 | 13.63 | 13.52 | 6613.62 | 0.93 | 165.1 | 1.4 |  | 280 |
|  |  | 22.59 | 22.90 | 6622.82 | 0.64 | 8.1 | 0.9 |  | 281 |
|  |  |  |  | 6625.77 | 0.48 | 2.2 | 0.7 |  | 282 |
|  |  |  | 27.88 | 6628.14 | 1.02 | 5.7 | 1.5 | s | 283 |
|  |  |  |  | 6629.60 | 0.62 | 3.5 | 0.9 |  | 284 |
|  | 30.80 |  | 30.61 | 6630.77 | 0.54 | 5.6 | 0.8 |  | 285 |
|  | 31.66 |  |  | 6631.69 | 0.38 | 3.5 | 0.6 |  | 286 |
| 32.93 | 32.85 | 32.86 | 32.65 | 6633.12 | 1.30 | 9.7 | 1.9 | s | 287 |
|  |  | 35.50 | 35.44 | 6635.74 | 1.12 | 5.3 | 1.7 |  | 288 |
|  | 46.03 | 45.95 |  | 6645.95 | 0.82 | 6.6 | 1.2 |  | 289 |
|  | 54.58 |  |  | 6654.73 | 0.67 | 5.7 | 1.0 |  | 290 |
|  |  |  |  | 6657.11 | 0.41 | 2.5 | 0.6 |  | 291 |
|  |  |  |  | 6658.74 | 0.56 | 2.2 | 0.8 |  | 292 |
| 60.64 | 60.64 | 60.62 | 60.70 | 6660.71 | 0.58 | 33.0 | 0.9 |  | 293 |
|  |  | 62.25 | 62.20 | 6662.21 | 0.53 | 3.7 | 0.8 |  | 294 |
|  |  |  | 63.72 | 6663.71 | 0.88 | 3.9 | 1.3 |  | 295 |
|  | 65.15 |  | 65.15 | 6665.27 | 0.54 | 7.3 | 0.8 |  | 296 |
|  | 72.15 |  | 72.39 | 6672.27 | 0.66 | 17.1 | 1.0 |  | 297 |
|  |  | 84.83 |  | 6684.86 | 1.59 | 6.3 | 2.4 |  | 298 |
|  | 89.30 | 89.38 | 89.32 | 6689.44 | 0.70 | 9.3 | 1.0 |  | 299 |
|  | 93.35 |  |  | 6693.52 | 0.54 | 2.9 | 0.8 | d | 300 |
| 94.46 | 94.48 | 94.40 | 94.50 | 6694.54 | 0.60 | 6.9 | 0.9 | d | 301 |
| 99.37 | 99.24 | 99.28 | 99.24 | 6699.32 | 0.63 | 21.6 | 0.9 |  | 302 |
| 1.98 | 1.98 | 1.87 | 1.95 | 6702.02 | 0.63 | 13.3 | 0.9 |  | 303 |
| 9.24 | 9.39 | 9.65 | 9.44 | 6709.43 | 0.49 | 1.9 | 0.8 |  | 304 |
|  |  | 19.58 |  | 6719.18 | 0.40 | 2.3 | 0.6 |  | 305 |
|  |  |  |  | 6727.47 | 0.56 | 2.9 | 0.8 | d | 306 |
|  | 29.28 | 29.20 | 29.28 | 6729.27 | 0.53 | 18.5 | 0.8 | d | 307 |
|  |  | 33.35 |  | 6733.16 | 1.24 | 5.8 | 1.8 |  | 308 |
|  | 37.13 |  | 37.14 | 6737.28 | 0.61 | 4.0 | 0.9 |  | 309 |
| 40.96 | 40.99 |  | 40.96 | 6740.82 | 0.86 | 3.3 | 1.3 |  | 310 |
|  |  |  |  | 6747.78 | 1.13 | 5.5 | 1.7 |  | 311 |
|  |  |  |  | 6751.51 | 2.81 | 16.7 | 4.2 |  | 312 |
|  |  |  |  | 6762.05 | 0.62 | 2.3 | 0.9 |  | 313 |
|  |  | 65.39 |  | 6765.34 | 0.61 | 4.3 | 0.9 |  | 314 |
|  | 67.74 |  | 67.58 | 6767.76 | 0.70 | 3.8 | 1.1 |  | 315 |
| 70.05 | 70.05 | 70.23 | 70.12 | 6770.21 | 0.59 | 4.9 | 0.9 |  | 316 |
|  | 78.99 |  | 79.04 | 6778.98 | 0.63 | 2.9 | 0.9 |  | 317 |
|  |  |  |  | 6780.65 | 0.58 | 4.1 | 0.9 |  | 318 |
| 88.89 | 88.66 | 89.64 | 88.66 | 6788.76 | 0.62 | 2.8 | 0.9 |  | 319 |
| 92.45 | 92.52 | 92.54 | 92.42 | 6792.46 | 0.56 | 3.2 | 0.8 |  | 320 |
| 95.37 | 95.24 | 95.18 | 95.20 | 6795.21 | 0.50 | 7.4 | 0.8 |  | 321 |
| 1.41 | 1.37 | 1.39 | 1.41 | 6801.46 | 0.61 | 4.9 | 0.9 |  | 322 |
| 3.28 | 3.29 |  | 3.26 | 6803.21 | 0.46 | 1.5 | 0.7 |  | 323 |
|  |  |  |  | 6809.51 | 0.54 | 1.2 | 0.8 |  | 324 |
| 11.44 |  | 11.30 | 11.27 | 6811.15 | 0.44 | 3.1 | 0.7 | d | 325 |
| 12.82 |  | 12.39 | 12.80 | 6812.73 | 1.03 | 4.6 | 1.6 | d | 326 |
|  |  |  |  | 6814.20 | 0.56 | 1.6 | 0.8 |  | 327 |
|  | 23.30 |  | 23.38 | 6823.45 | 0.54 | 3.5 | 0.8 |  | 328 |
|  |  |  |  | 6825.80 | 0.48 | 1.4 | 0.7 | d | 329 |
| 27.28 | 27.30 |  | 27.22 | 6827.33 | 0.71 | 8.8 | 1.2 | d | 330 |
|  | 37.70 |  | 37.70 | 6837.71 | 0.61 | 3.2 | 0.9 |  | 331 |
|  |  |  | 39.05 | 6839.49 | 0.44 | 1.5 | 0.7 |  | 332 |
| 41.59 | 41.49 |  | 41.54 | 6841.52 | 0.46 | 2.6 | 0.7 |  | 333 |
| 43.44 | 43.60 |  | 43.69 | 6843.63 | 0.89 | 12.3 | 1.4 |  | 334 |

|       |       |       |         |      |       |     |     |     |
|-------|-------|-------|---------|------|-------|-----|-----|-----|
|       | 45.30 | 45.26 | 6845.34 | 0.69 | 2.3   | 1.1 |     | 335 |
|       | 46.60 | 46.53 | 6846.50 | 0.48 | 3.6   | 0.7 | s   | 336 |
| 52.90 | 52.67 | 52.91 | 6852.53 | 0.65 | 7.1   | 1.0 |     | 337 |
|       | 62.53 | 62.47 | 6862.51 | 0.63 | 3.9   | 1.0 |     | 338 |
|       | 64.65 |       | 6864.72 | 0.50 | 2.3   | 0.8 |     | 339 |
| 19.25 | 19.44 |       | 6919.21 | 1.08 | <22.2 | 1.7 | t   | 340 |
| 44.53 | 44.56 |       | 6944.60 | 0.81 | 10.8  | 1.2 | t   | 341 |
|       | 73.55 |       | 6973.76 | 0.56 | <5.6  | 0.9 | t   | 342 |
| 78.54 | 78.28 |       | 6978.40 | 0.56 | <5.8  | 0.9 | t   | 343 |
| 93.18 | 93.18 |       | 6993.13 | 0.76 | <49.5 | 1.2 | t   | 344 |
|       | 30.35 |       | 7030.29 | 0.64 | <6.5  | 1.0 | t   | 345 |
|       | 31.56 |       | 7031.52 | 0.57 | <7.2  | 0.9 | t   | 346 |
| 45.65 | 45.87 |       | 7045.89 | 0.46 | 3.7   | 0.7 |     | 347 |
| 60.81 | 61.00 |       | 7061.05 | 0.50 | <6.8  | 0.8 | t   | 348 |
| 62.70 | 62.65 |       | 7062.68 | 0.47 | <6.5  | 0.7 | t,s | 349 |
| 69.65 | 69.48 |       | 7069.55 | 0.78 | 8.1   | 1.2 | s   | 350 |
| 78.02 | 78.11 |       | 7077.86 | 1.04 | <6.4  | 1.6 | t   | 351 |
| 85.10 | 84.94 |       | 7084.65 | 2.56 | <21.2 | 4.1 | t   | 352 |
|       |       |       | 7092.67 | 1.04 | 8.7   | 1.6 |     | 353 |
|       |       |       | 7116.31 | 0.56 | <5.3  | 0.9 | t   | 354 |
| 19.94 |       |       | 7119.71 | 1.05 | <10.7 | 1.7 | t   | 355 |
|       | 59.51 |       | 7159.83 | 1.46 | <12.1 | 2.4 | t   | 356 |
|       | 62.96 |       | 7162.87 | 0.59 | 4.4   | 1.0 | t   | 357 |
|       |       |       | 7203.65 | 0.53 | 7.2   | 0.9 | t   | 358 |
| 24.18 | 24.00 |       | 7224.03 | 1.02 | 84.9  | 1.7 | t   | 359 |
| 34.33 |       |       | 7334.48 | 0.69 | 9.8   | 1.2 | t   | 360 |
| blend | 57.60 |       | 7357.56 | 1.41 | 12.5  | 2.4 | t   | 361 |
| 66.61 | 67.12 |       | 7367.13 | 0.54 | 34.7  | 0.9 |     | 362 |
| 75.90 | 75.90 |       | 7375.87 | 0.60 | 5.0   | 1.0 |     | 363 |
| 85.92 | 85.83 |       | 7385.89 | 0.52 | 4.8   | 0.9 |     | 364 |
| 5.77  |       |       | 7405.71 | 0.68 | 2.7   | 1.2 |     | 365 |
|       | 19.07 |       | 7419.05 | 0.42 | 2.5   | 0.7 |     | 366 |
|       | 70.35 |       | 7470.32 | 0.88 | 8.4   | 1.5 |     | 367 |
|       | 72.65 |       | 7472.60 | 0.65 | 2.9   | 1.1 |     | 368 |
|       | 83.02 |       | 7482.91 | 0.61 | 2.3   | 1.1 |     | 369 |
|       | 84.09 |       | 7484.17 | 0.55 | 3.2   | 1.0 |     | 370 |
|       | 94.89 |       | 7495.02 | 0.60 | 13.6  | 1.1 |     | 371 |
|       |       |       | 7520.60 | 0.48 | 3.3   | 0.9 |     | 372 |
| 58.50 | 59.35 |       | 7559.41 | 0.54 | 3.9   | 1.0 |     | 373 |
| 62.24 |       |       | 7562.25 | 1.23 | 29.2  | 2.2 |     | 374 |
|       |       |       | 7706.80 | 0.77 | 5.8   | 1.5 |     | 375 |
|       | 7.96  |       | 7708.06 | 0.60 | 3.3   | 1.1 |     | 376 |
|       |       |       | 7822.80 | 0.89 | 4.1   | 1.8 |     | 377 |
| 32.72 | 32.81 |       | 7832.89 | 0.63 | 13.4  | 1.3 | s   | 378 |
| 62.34 | 62.39 |       | 7862.43 | 0.52 | 4.0   | 1.1 | t   | 379 |
| 26.21 | 26.27 |       | 8026.25 | 0.61 | 13.4  | 1.4 | t   | 380 |

Table 3

Heliocentric Radial Velocities (km s$^{-1}$)

| UT Date | HD 204827 A[a] | HD 204827 B[b] |
|---|---|---|
| 2001 Sep 05 | +9.8 | +0.6 |
| 2001 Sep 08 | +12.4 | -22.8 |
| 2001 Sep 09 | +20.1 | -27.9 |
| 2001 Oct 29 | -48.5 | +13.7 |
| 2003 Oct 25 | -5.0 | -10.4 |

[a] A typical uncertainty is ± 3 km s$^{-1}$.
[b] A typical uncertainty is ± 0.5 km s$^{-1}$.

Table 4

Some Statistical Characteristics of the DIBs

| DIB # | $\lambda_{min}$(Å) | $\lambda_{max}$(Å) | $\Delta\lambda$(Å) | Sp[a](Å) | New[b] | Prev[c] | New Fr[d] |
|---|---|---|---|---|---|---|---|
| 1–50 | 3900.0 | 5300.0 | 1400 | 28.0 | 35 | 15 | 0.70 |
| 51–100 | 5300.0 | 5707.0 | 407 | 8.1 | 21 | 29 | 0.42 |
| 101–150 | 5707.0 | 5928.0 | 221 | 4.4 | 14 | 36 | 0.28 |
| 151–200 | 5928.0 | 6155.0 | 227 | 4.5 | 12 | 38 | 0.24 |
| 201–250 | 6155.0 | 6417.0 | 262 | 5.2 | 10 | 40 | 0.20 |
| 251–300 | 6417.0 | 6694.0 | 277 | 5.5 | 7 | 43 | 0.14 |
| 301–350 | 6694.0 | 7074.0 | 380 | 7.6 | 8 | 42 | 0.16 |
| 350–380 | 7074.0 | 8100.0 | 1026 | 34.2 | 6 | 24 | 0.20 |
| 1–380 | 3900.0 | 8100.0 | 4200 | 11.0 | 113 | 267 | 0.30 |

a The average spacing between DIBs in each wavelength range.
b The number of "new" DIBs in each wavelength range.
c The number of previously known DIBs in each wavelength range.
d The fraction of "new" DIBs in each wavelength range.

Table 5

Atomic Interstellar Lines

| Atom/Ion | $\lambda$(Å) | $W_\lambda$(mÅ) |
|---|---|---|
| Fe I  | 3859.91 | 6.0 |
| Ca II | 3933.66 | 293. |
| Ca II | 3968.47 | 198. |
| K I   | 4044.14 | 8.1 |
| K I   | 4047.21 | <4.0 |
| Ca I  | 4226.73 | 11.0 |
| Na I  | 5889.95 | 486. |
| Na I  | 5895.92 | 444. |
| Li I  | 6707.81 | 5.3 |
| K I   | 7664.91 | 391. |
| K I   | 7698.97 | 316. |
| Rb I  | 7800.27 | 4.4 |
| Rb I  | 7947.60 | --- |

Table 6

Molecular Interstellar Lines

| Molecule | Band | Line | λ(Å) | $W_\lambda$(mÅ) |
|---|---|---|---|---|
| CN | B-X (0-0) | R(1) | 3874.00 | 27. |
| | | R(0) | 3874.60 | 69. |
| | | P(1) | 3875.77 | 15. |
| | | P(2) | 3876.31 | 1.5 |
| CH | B-X (0-0) | $R_2$(1/2) | 3878.77 | 4.9 |
| | | $Q_2$(1/2) | 3886.41 | 18. |
| | | $P_2$(1/2) | 3890.21 | 10. |
| $CH^+$ | A-X (1-0) | R(0) | 3957.71 | 17. |
| $C_3$ | A-X (0-0) | band | 4051.6 | 11. |
| $CH^+$ | A-X (0-0) | R(0) | 4232.54 | 29. |
| CH | A-X (0-0) | $R_2$(1/2) | 4300.32 | 63. |
| $C_2$ | A-X (4-0) | R(6)+R(4) | 6909.37+6909.41 | 2.7 |
| | | R(8)+R(2) | 6910.31+6910.58 | 5.7 |
| | | R(0) | 6912.70 | 2.0 |
| | | Q(2) | 6914.98 | 3.5 |
| | | Q(4) | 6916.79 | 2.8 |
| | | Q(6) | 6919.66 | 1.2 |
| | | P(4) | 6922.69 | 1.0 |
| $C_2$ | A-X (3-0) | R(6)+R(4) | 7714.58+7714.94 | 8.3 |
| | | R(8) | 7715.42 | 1.1 |
| | | R(2) | 7716.53 | 7.3 |
| | | R(10) | 7717.47 | 1.7 |
| | | R(0) | 7719.33 | 7.5 |
| | | Q(2) | 7722.10 | <21. |
| | | Q(4) | 7724.22 | 11. |
| | | P(2) | 7725.82 | 3.0 |
| | | Q(6) | 7727.56 | 7.3 |
| | | P(4)+Q(8) | 7731.66+7732.12 | 5.5 |
| | | Q(10) | 7737.90 | 1.7 |
| | | P(6) | 7738.74 | 2.1 |
| CN | A-X (2-0) | $R_{21}$(0) | 7871.64 | 3.1 |
| | | $R_2$(1) | 7873.99 | 1.5 |
| | | $R_2$(0) | 7874.85 | 5.0 |
| | | $R_1$(0) | 7906.60 | 9.8 |
| $C_2$ | A-X (2-0) | R(6) | 8750.85 | 10.1 |
| | | R(8)+R(4) | 8751.49+8751.68 | 20.9 |
| | | R(10)+R(2) | 8753.58+8753.95 | 17.8 |
| | | R(0) | 8757.69 | 14.9 |
| | | Q(2) | 8761.19 | 28.2 |
| | | Q(4) | 8763.75 | 25.5 |
| | | P(2) | 8766.03 | 5.0 |

| | | |
|---|---|---|
| Q(6) | 8767.76 | 19.5 |
| Q(8)+P(4) | 8773.22+8773.43 | 17.5 |
| Q(10) | 8780.14 | <6.9 |
| P(6) | 8782.31 | 5.4 |
| Q(12) | 8788.56 | 4.2 |
| P(8) | 8792.65 | <5.2 |
| Q(14) | 8798.46 | <3.3 |

FIGURE CAPTIONS

Fig. 1.   Spectra of 10 Lac (upper plot) and of HD 204827 (the lower six plots), in the region near 5250 Å. For HD 204827, an average of the five, respective, nightly averages appears above those five. Note that the nightly averages are not plotted in entirely chronological order. The various spectra are displaced vertically by 3% of the continua. The wavelength scale used for HD 204827 throughout this paper is a heliocentric one that incorporates a zero-point offset of –17.4 km s$^{-1}$, so that interstellar absorption lines are shown at their rest wavelengths (section 3.2). In the spectrum of HD 204827, three DIBs are seen at 5245.48, 5251.78, and 5257.47 Å, along with a narrow, stellar, Fe III line with a rest wavelength of 5243.23 Å. The equivalent width of the DIB at 5251.78 Å is 2.7 ± 0.6 mÅ. The stellar line shows a variable radial velocity that smears its profile in the average spectrum. The two stellar lines present in the spectrum of 10 Lac with central depths of about 1.5% of the continuum are undetectable in the spectrum of HD 204827, owing to a slight mismatch in the spectral types.

Fig. 2.  The interstellar λ7699 line of K I in the spectrum of HD 204827, at a resolution of 8 km s$^{-1}$.

Fig. 3.  DIBs at 6376.08 Å and 6379.32 Å in the spectrum of HD 204827.

Fig. 4.   Spectra of 10 Lac (upper plot) and of HD 204827 (the lower five plots), in the region of the He II λ4541.59 and the Si III λ4552.62 lines. The broad lines of binary component A are readily seen at both transitions, and the narrow line of component B is also evident in Si III. The various spectra are displaced vertically by 12% of the continua.

Fig. 5.  Radial velocities of HD 204827 A and HD 204827 B. Typical errors of measurement are ± 3.0 and ± 0.5 km s$^{-1}$, respectively.

Fig. 6.   Distribution by wavenumber splitting of the separations between 379 pairs of adjacent DIBs. The logarithms of the numbers of splittings in each of twelve bins are shown. All bins are 10 cm$^{-1}$ wide, their centers are also separated by 10 cm$^{-1}$, and the first bin is centered at 5 cm$^{-1}$. At λ = 5000 Å, an interval Δσ = 10 cm$^{-1}$ corresponds to Δλ = 2.5 Å.

Fig. 7.  Distribution by wavelength of 380 DIBs in Table 2. The wavelength bins are 200 Å wide, their centers are also spaced by 200 Å, and the first bin is centered at 4000 Å.

Fig. 8.  Distribution by observed width of the 376 DIBs in Table 2 for which definite values were measured. The widths are the measured values, uncorrected for instrumental broadening, which amounts to 0.16 Å at 6000 Å. The bins are 0.1 Å wide, their centers are also spaced by 0.1 Å, and the first bin is centered at 0.05 Å. The bin centered at FWHM = 3.05 Å accounts for all DIBs with FWHM > 3.0 Å

Fig. 9.  Distribution by equivalent width of the 360 DIBs in Table 2 for which definite values were measured. The bins are 2 mÅ wide, their centers are also spaced by 2 mÅ, and the first bin is centered at 1 mÅ. The bin centered at W$_λ$ = 61 mÅ accounts for all DIBs with W$_λ$ > 60 mÅ.

Fig. 10. Distribution of the wavenumber splittings that arise between 8,698 pairs of DIBs in Table 2 with separations not exceeding 400 cm$^{-1}$. The bins are 10 cm$^{-1}$ wide, their centers are also spaced by 10 cm$^{-1}$, and the first bin is centered at 5 cm$^{-1}$. The error bars span $\pm \sqrt{N}$, where N is the number of splittings in each bin.

Fig. 11. Final, combined spectrum of HD 204827 in the range 3900 < λ < 8100 Å, along with the spectrum of 10 Lac. See the explanatory description in the text. The upper and lower panels on each page display the same data at different vertical scales; both scales are appreciably expanded in order to emphasize the generally weak DIBs. The DIBs present in the spectrum of HD 204827 are marked by vertical lines and are numbered in order of increasing wavelength. The stronger of these DIBs are also seen in the spectrum of 10 Lac. The atomic and the (diatomic or triatomic) molecular lines listed in Tables 5 and 6 are identified by asterisks placed above the spectrum of 10 Lac (in order to avoid crowding), although the asterisks refer to the spectrum of HD 204827. Only three illustrative pages among the total of 42 are printed here; the entire figure is publicly available at http://dibdata.org.

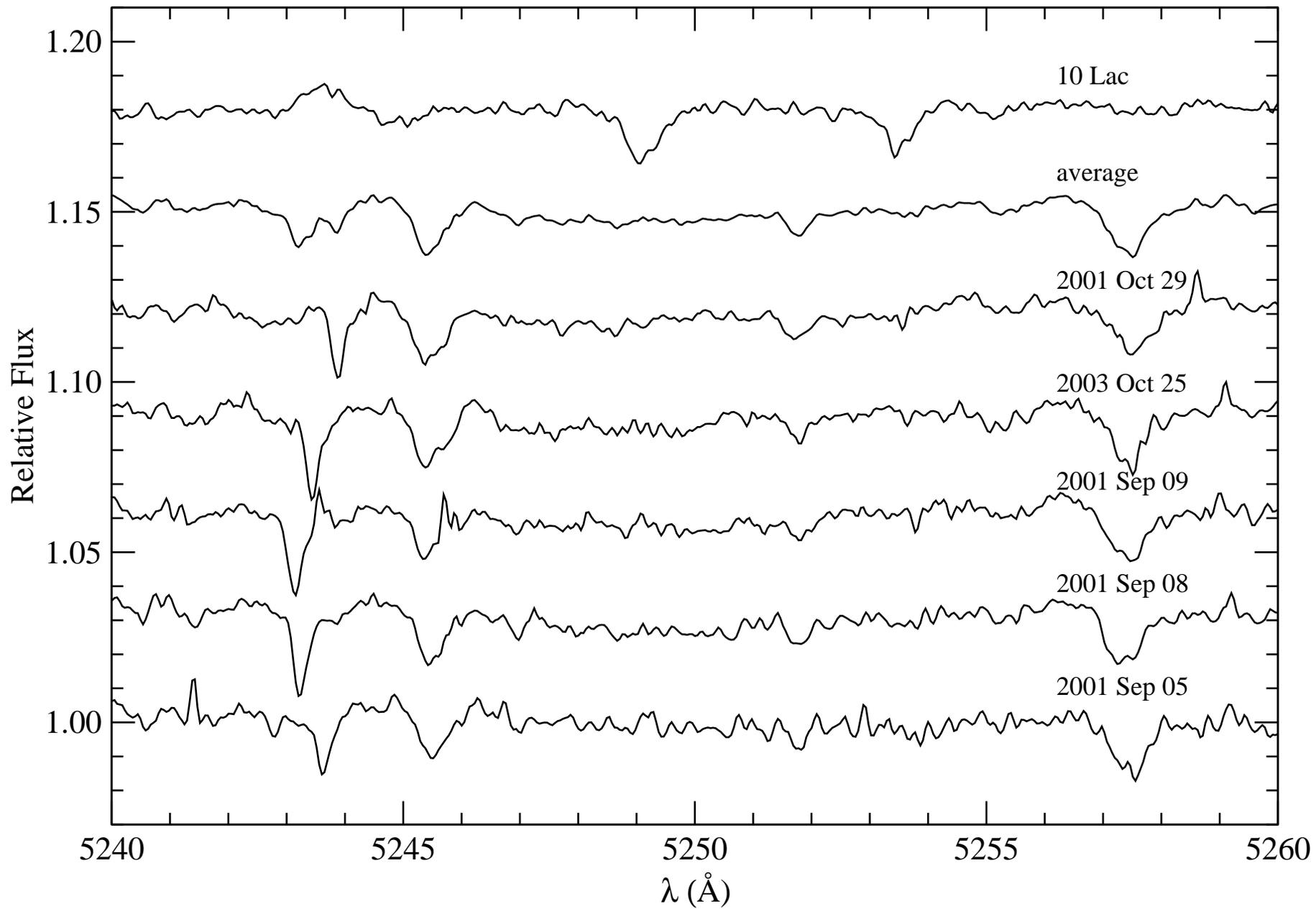

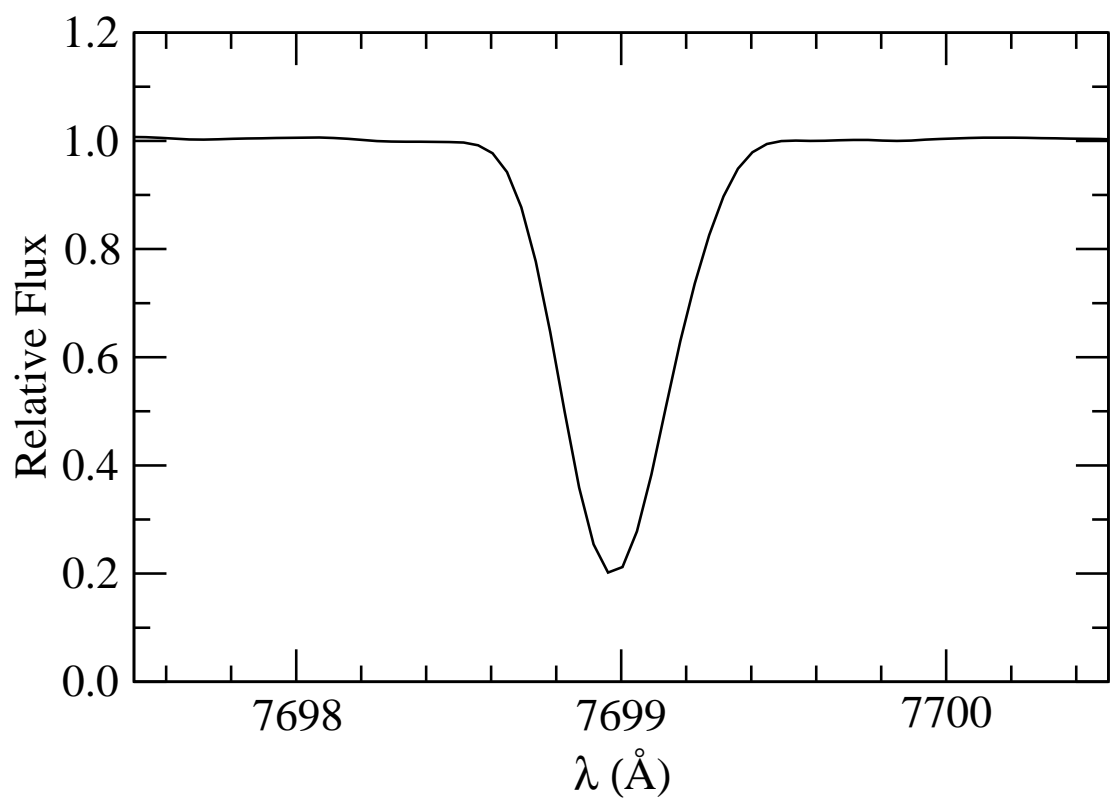

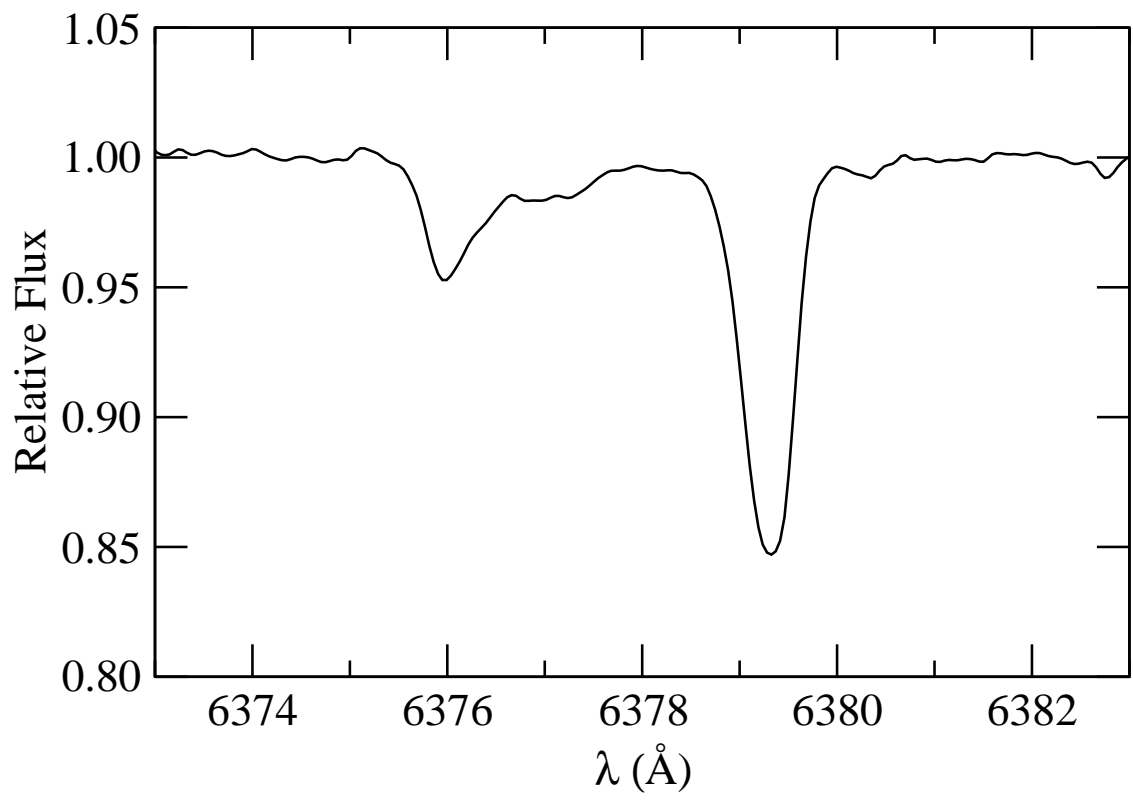

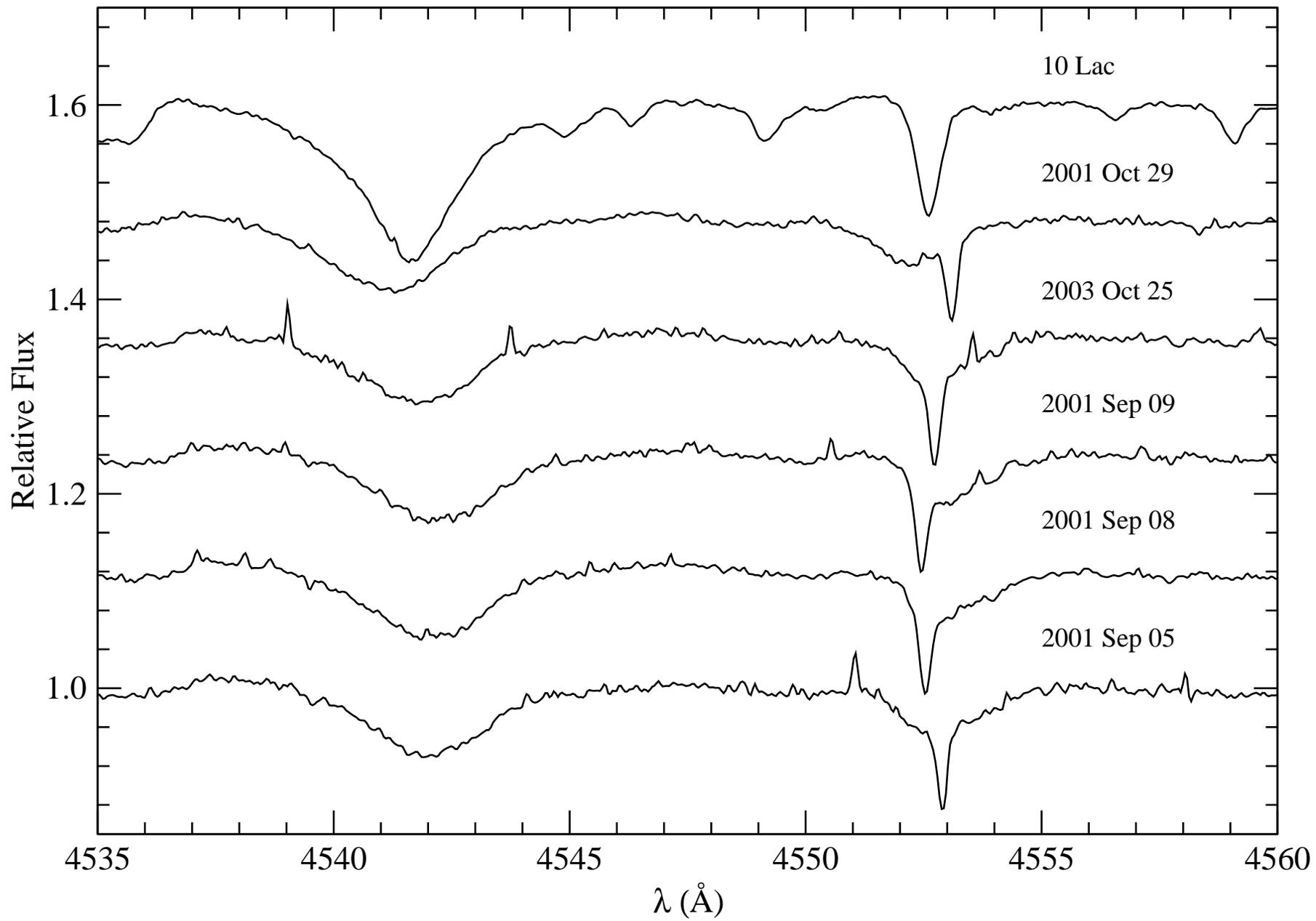

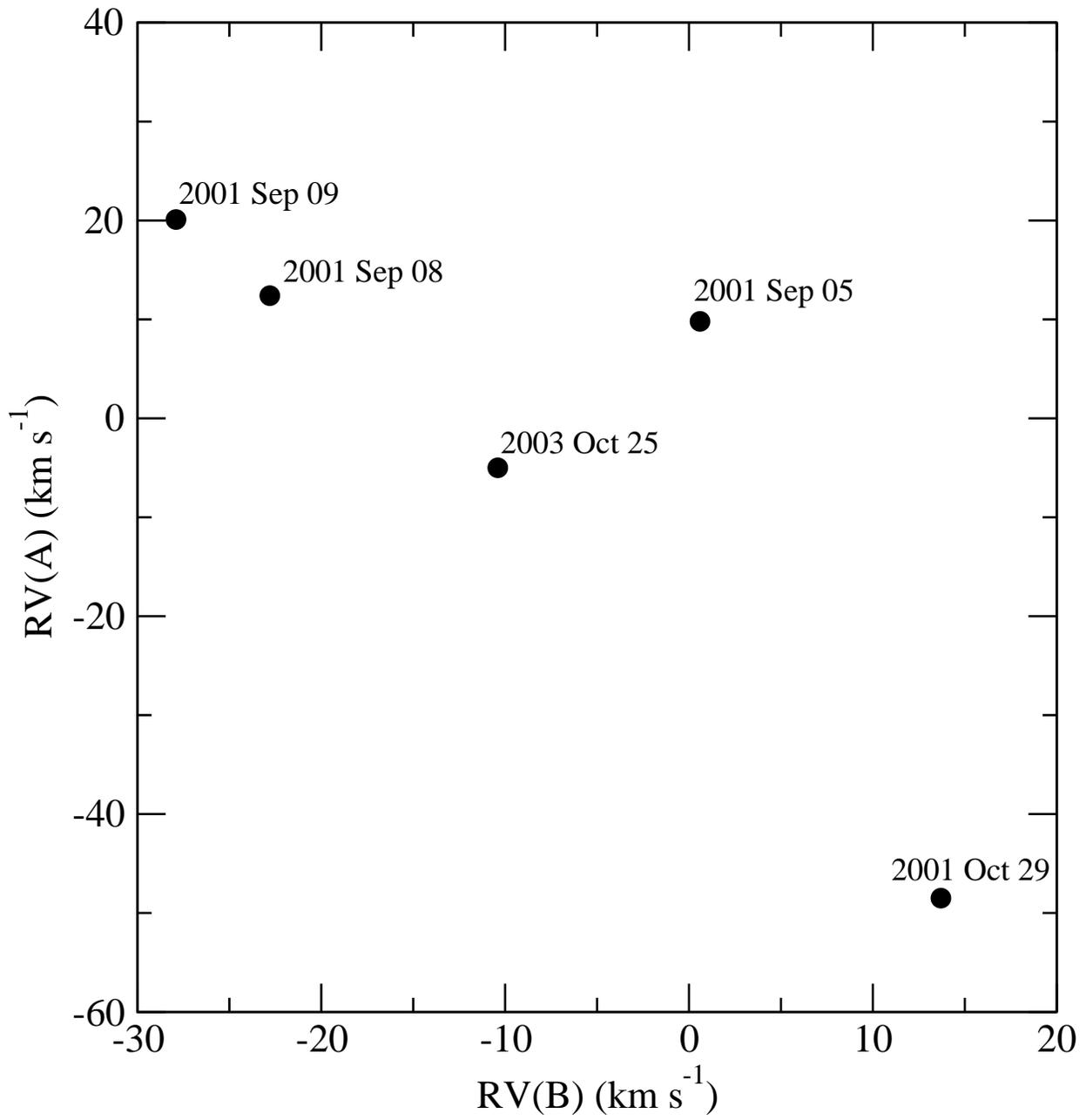

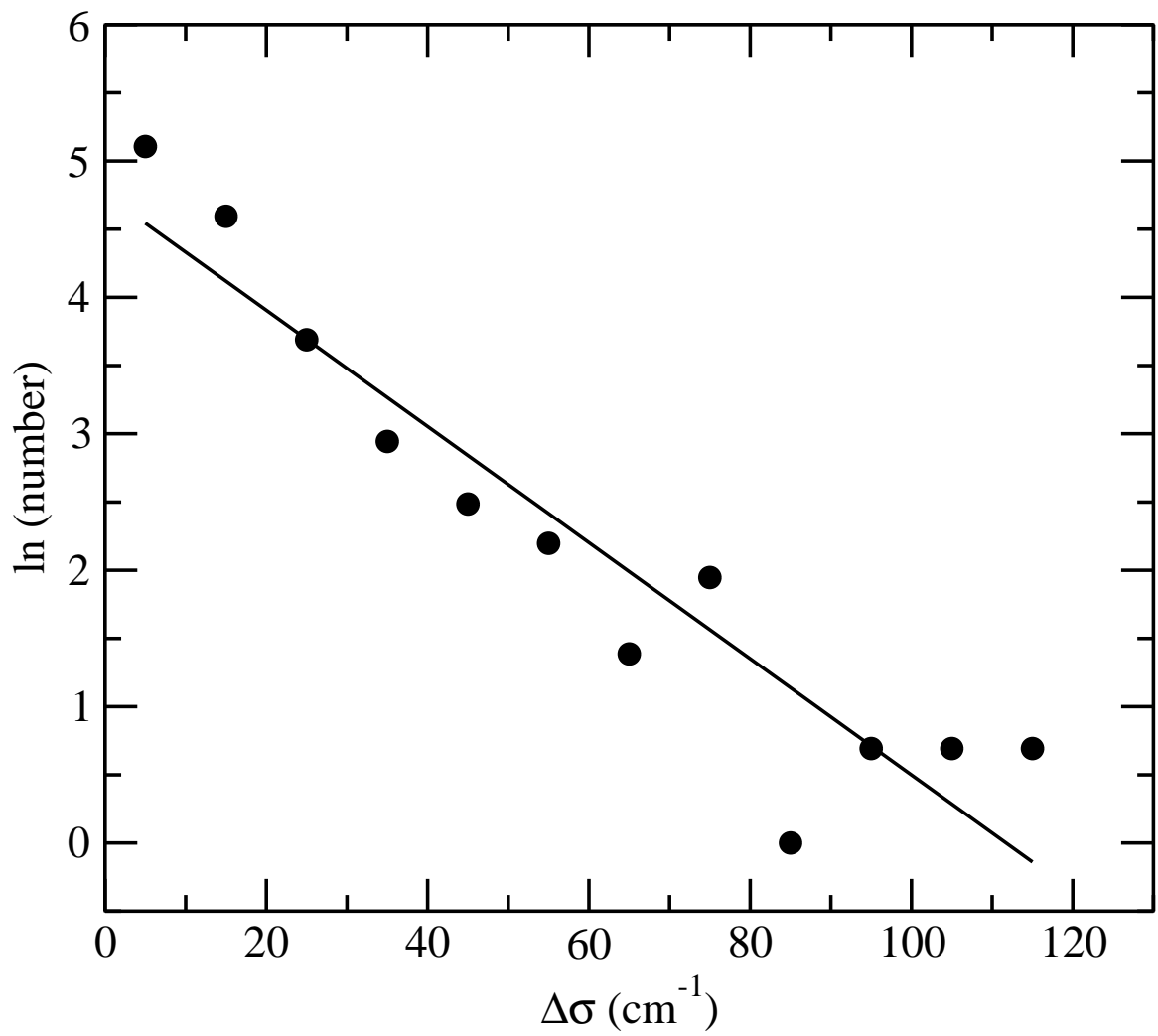

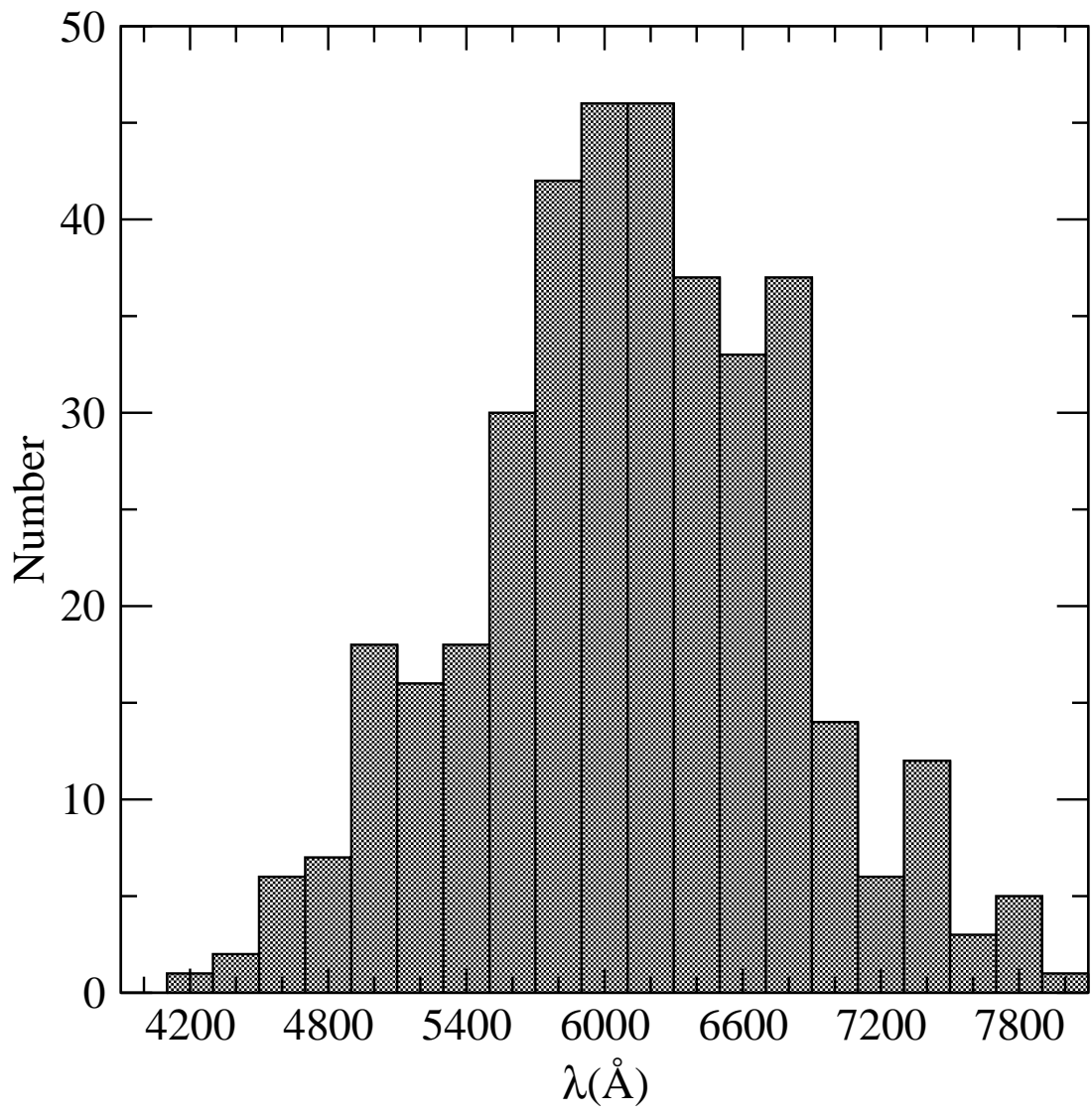

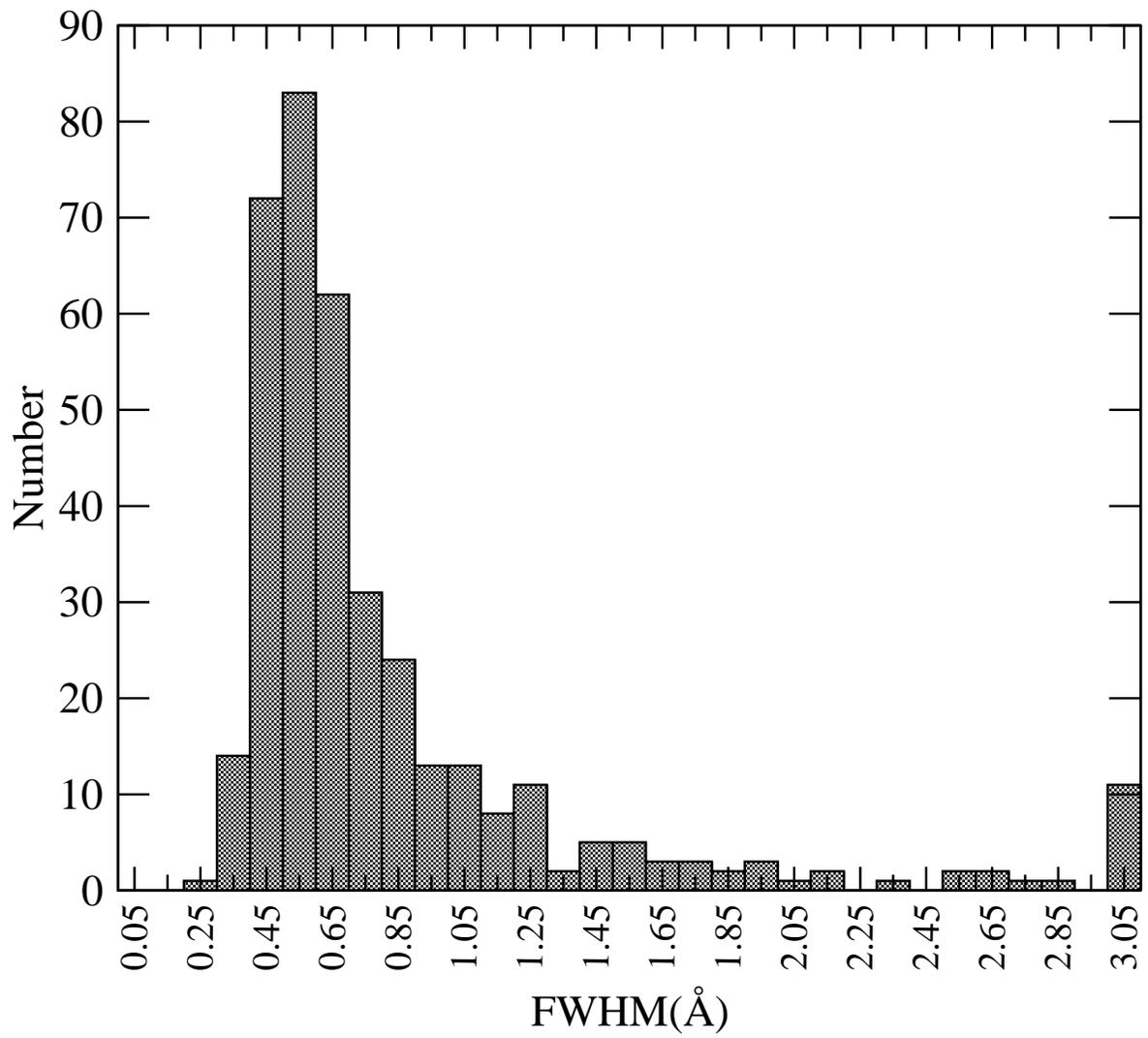

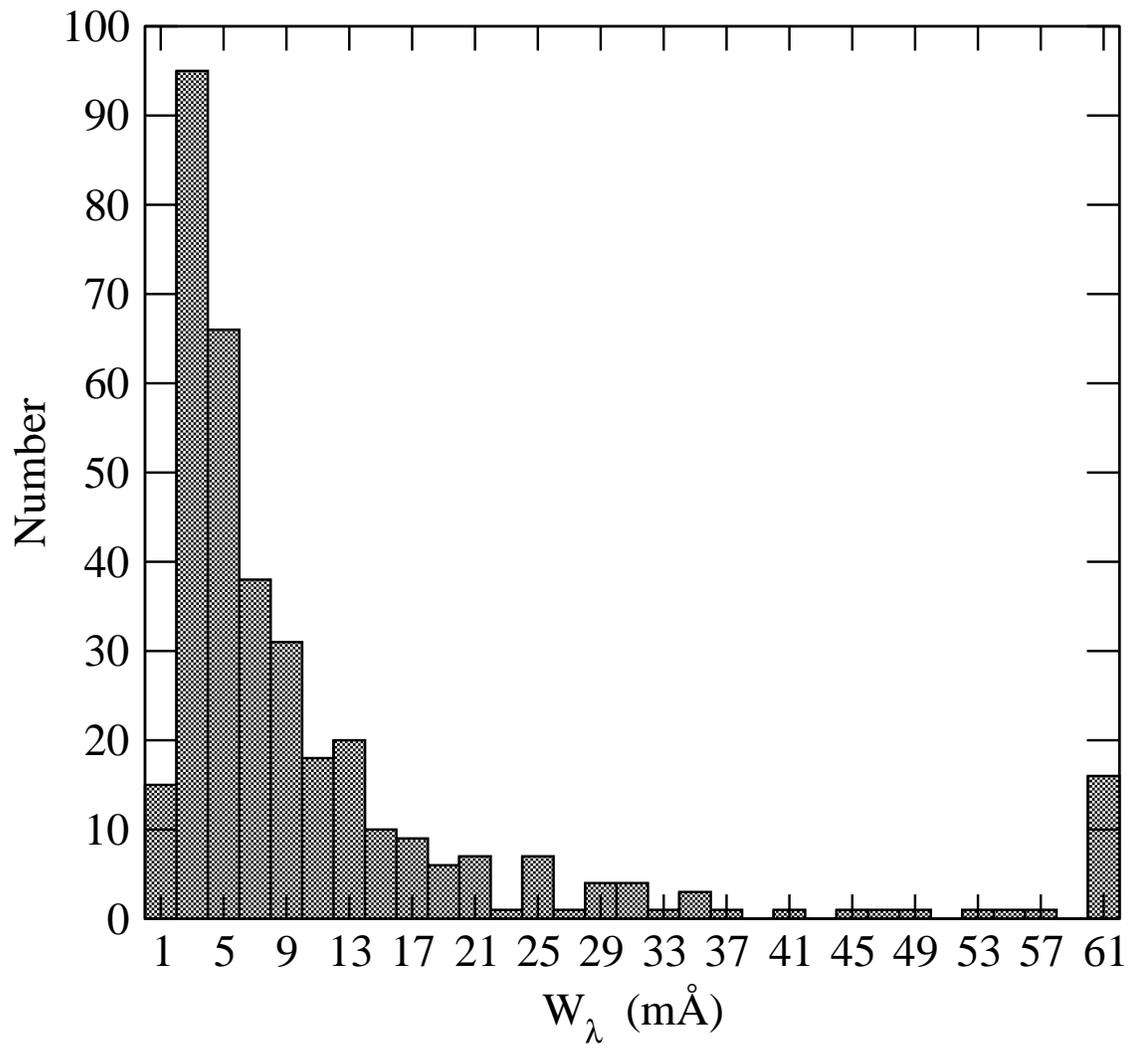

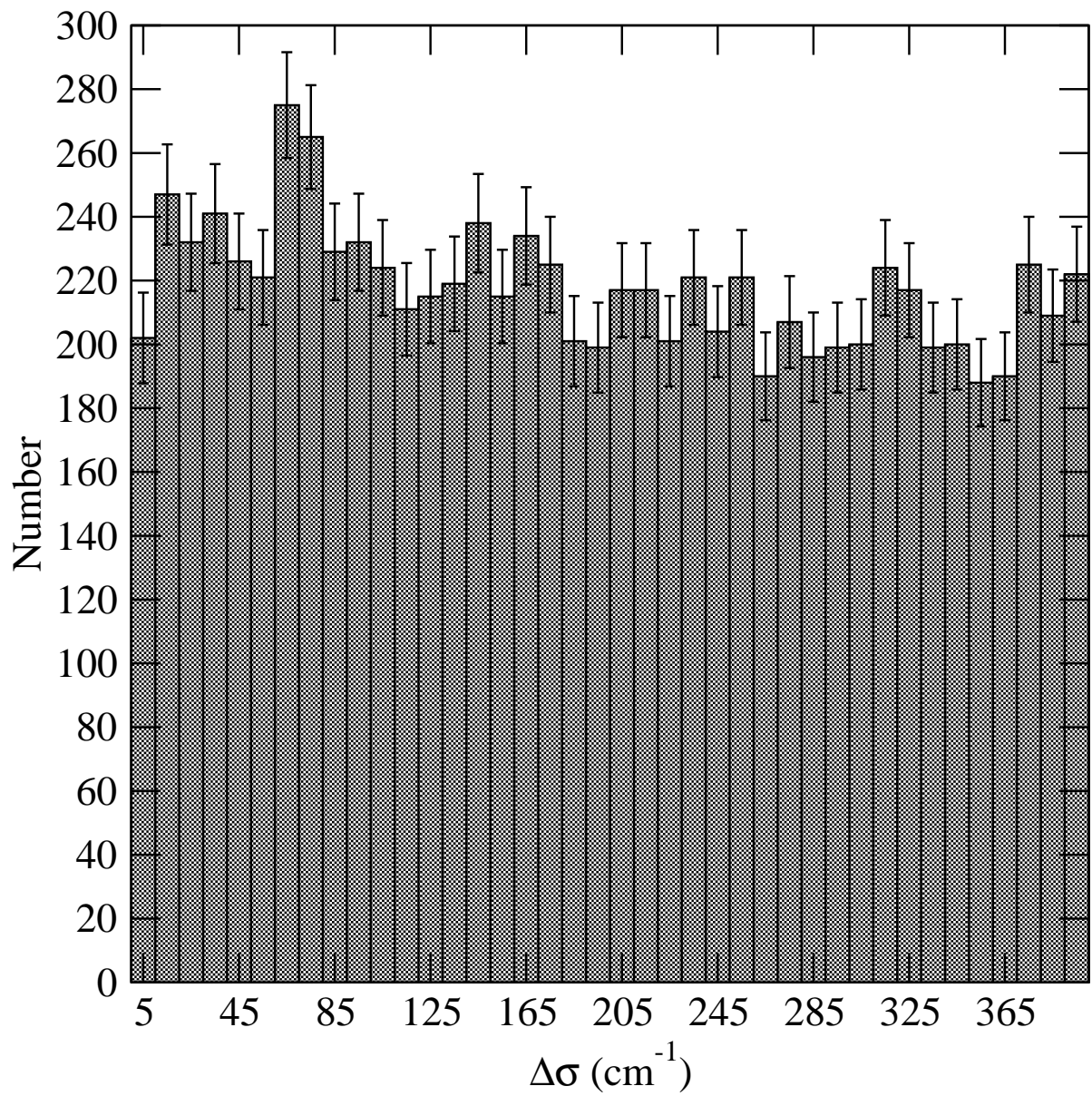

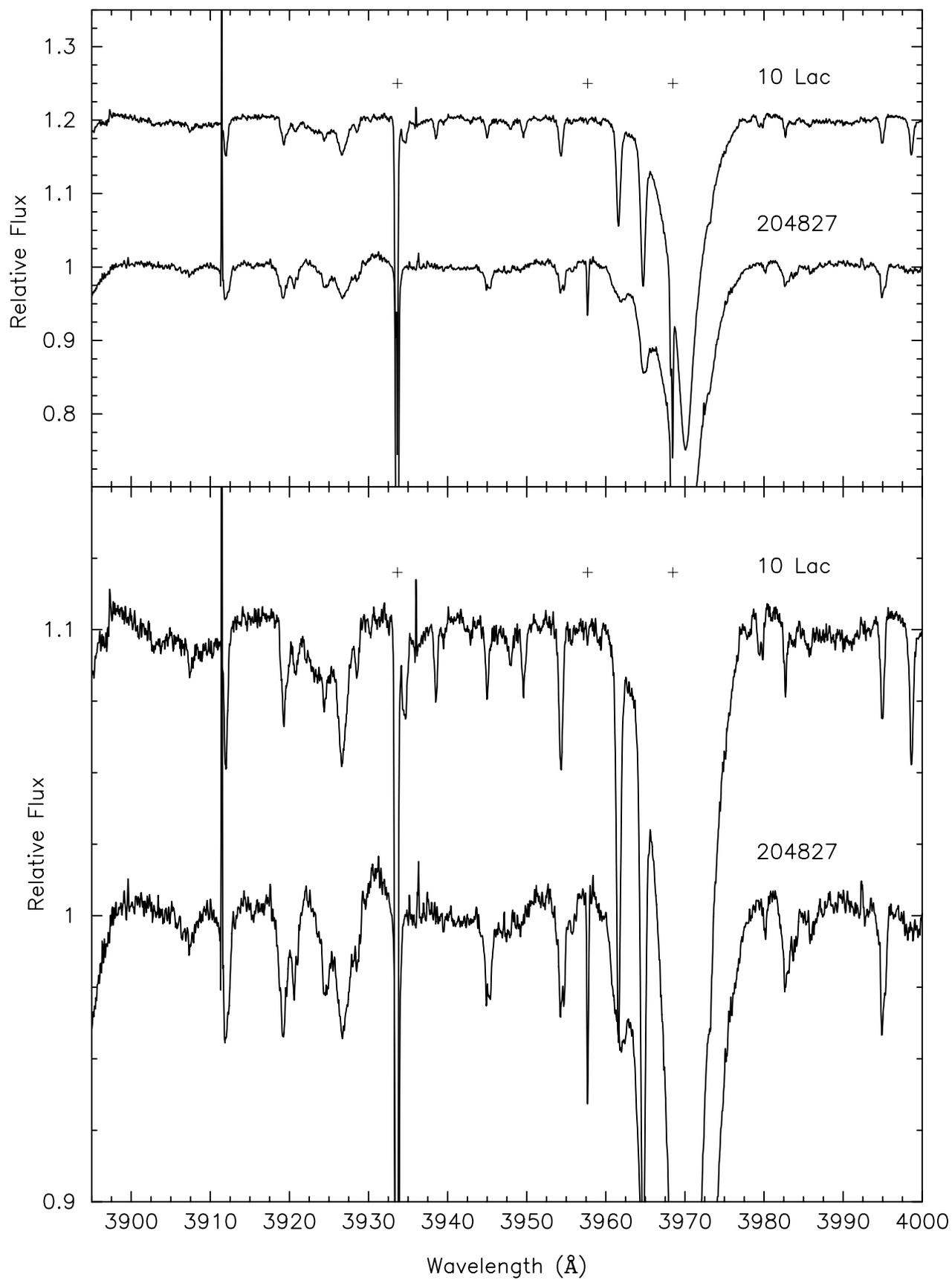

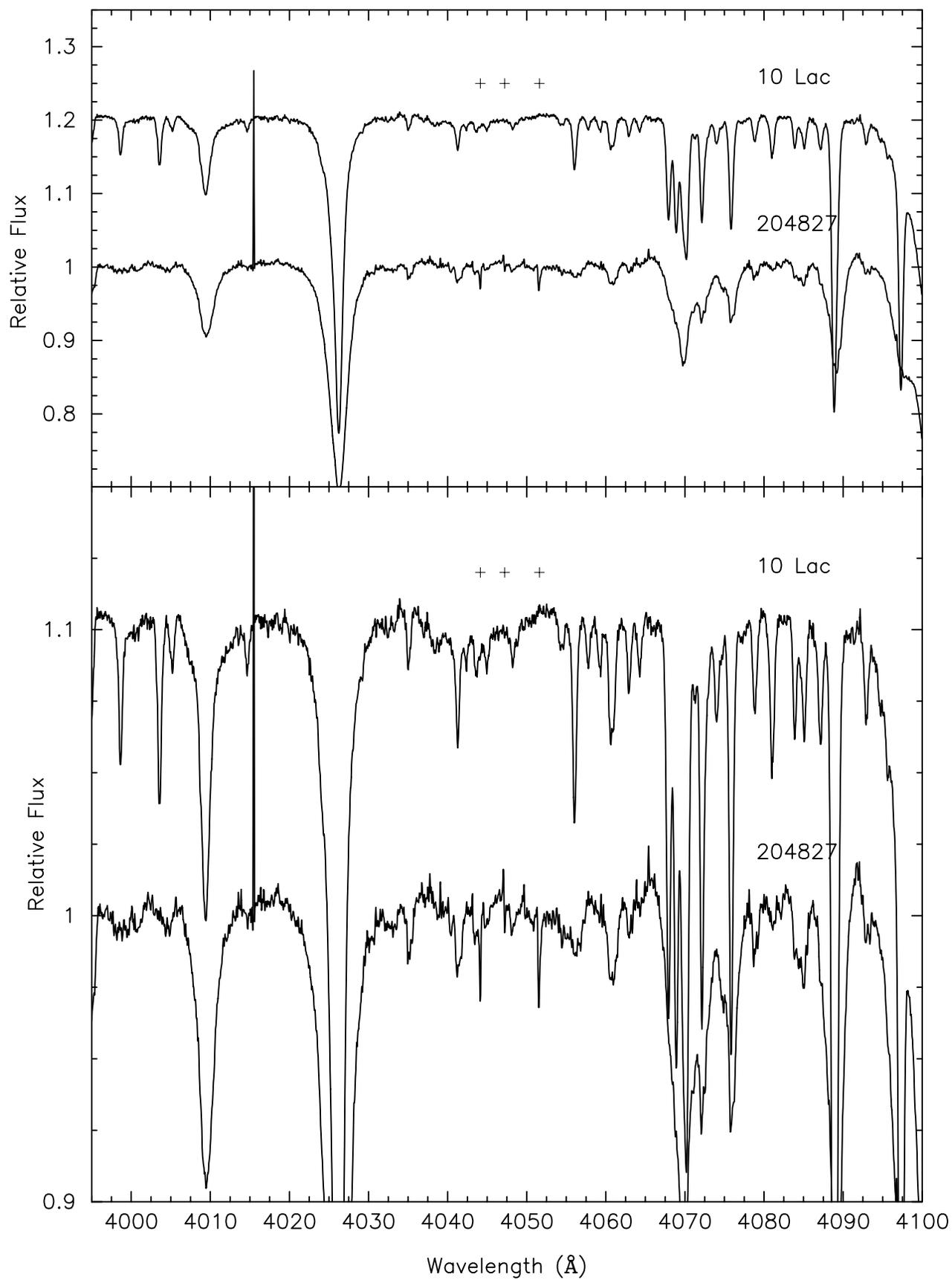

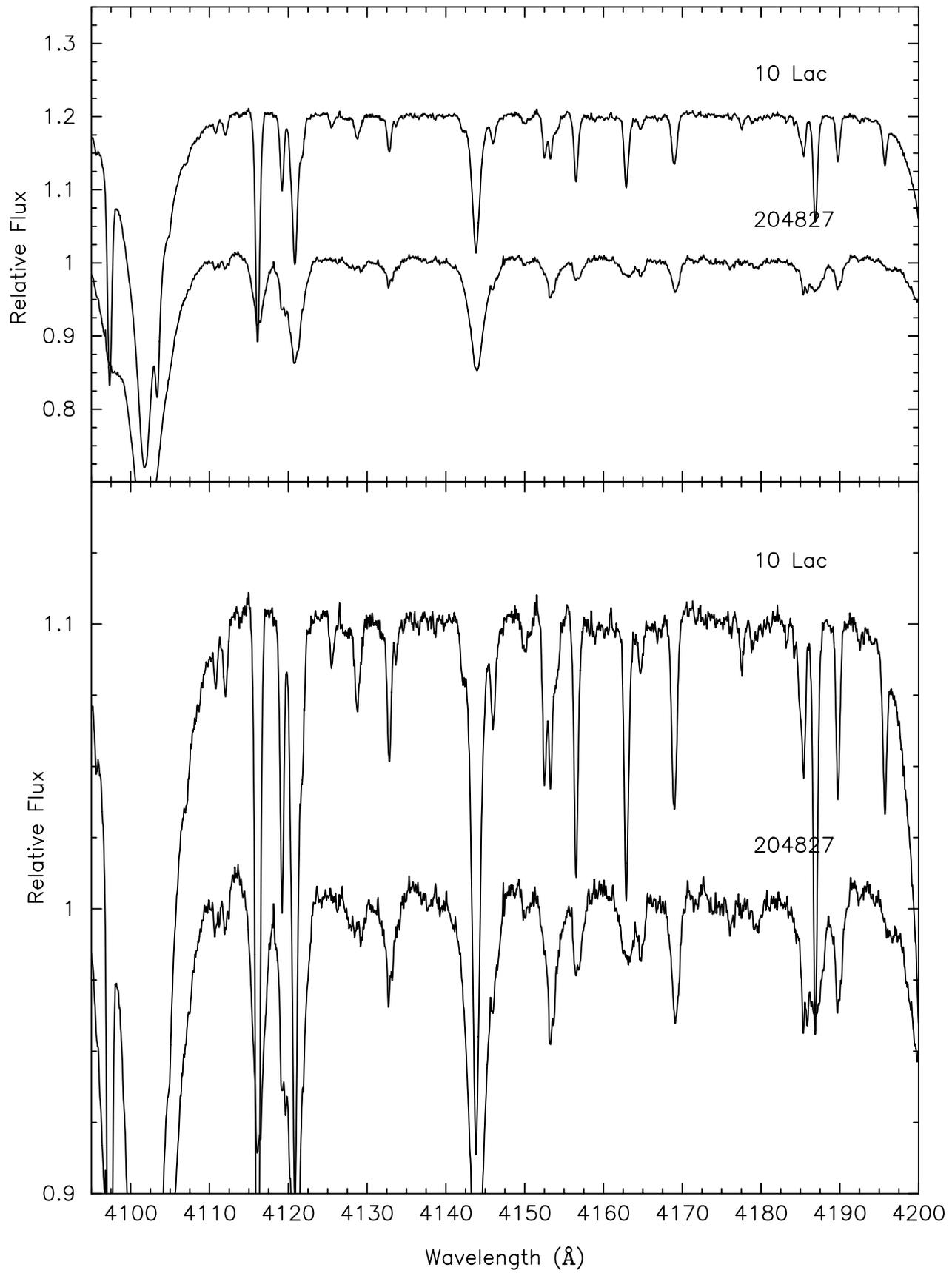

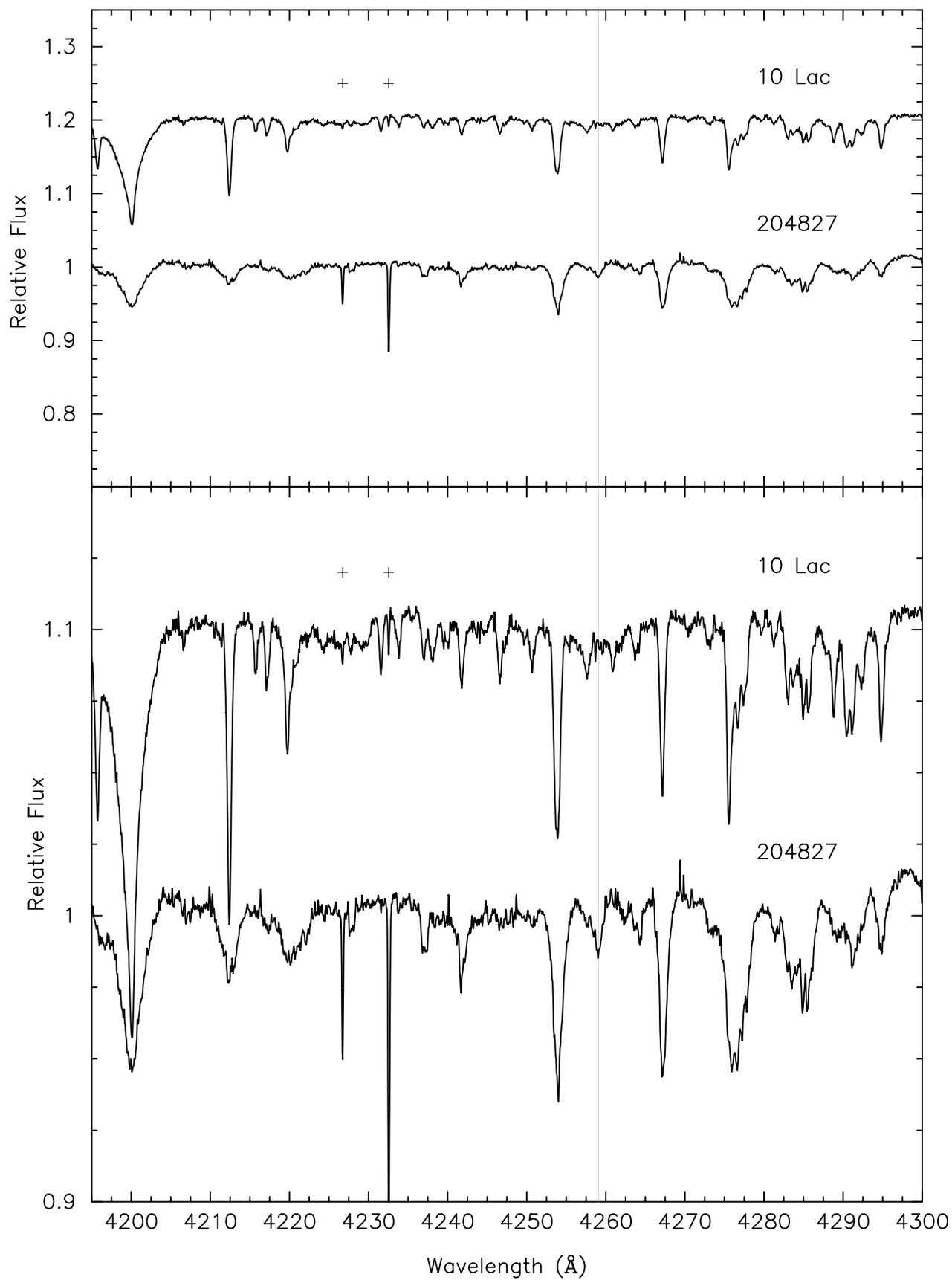

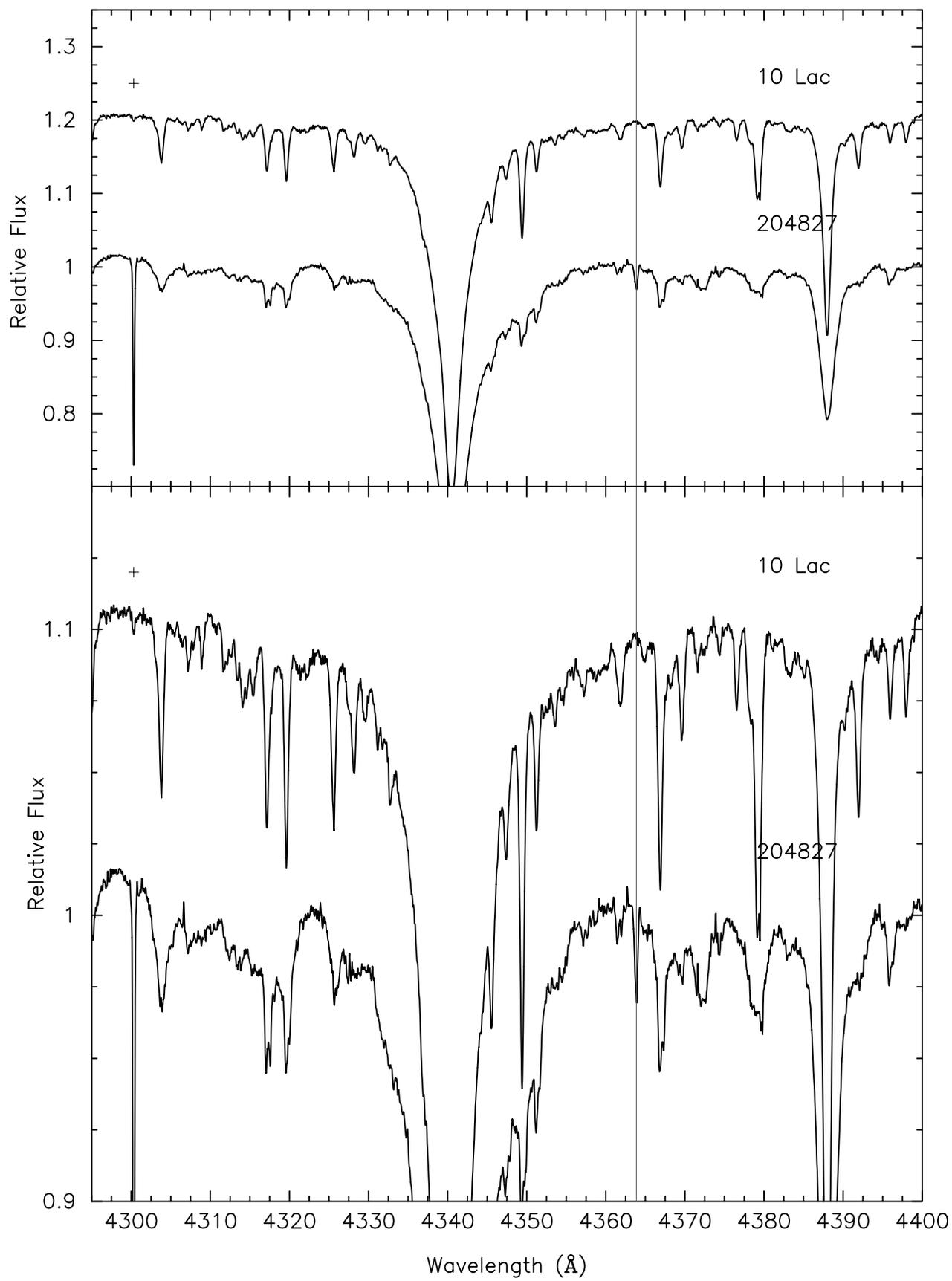

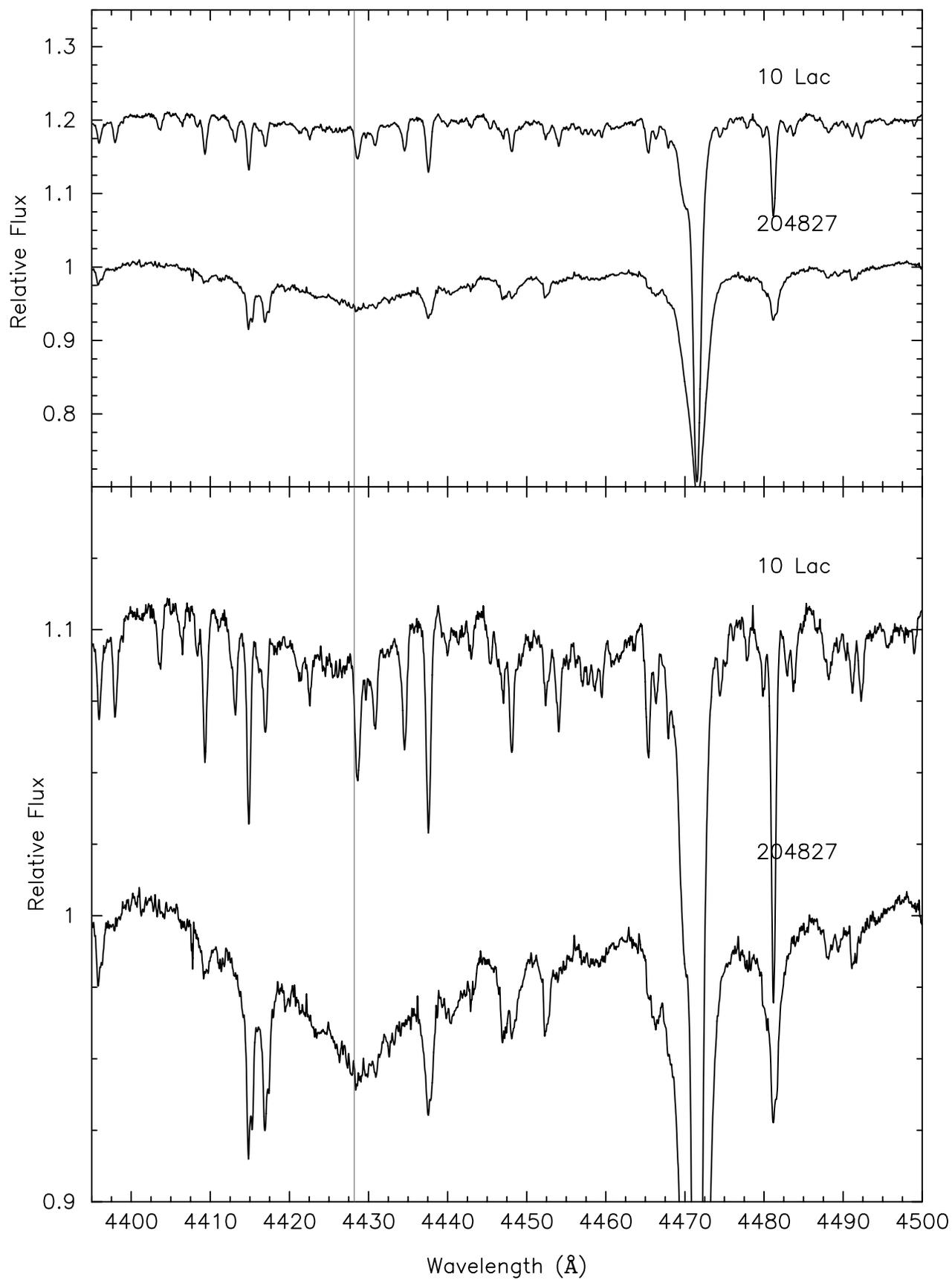

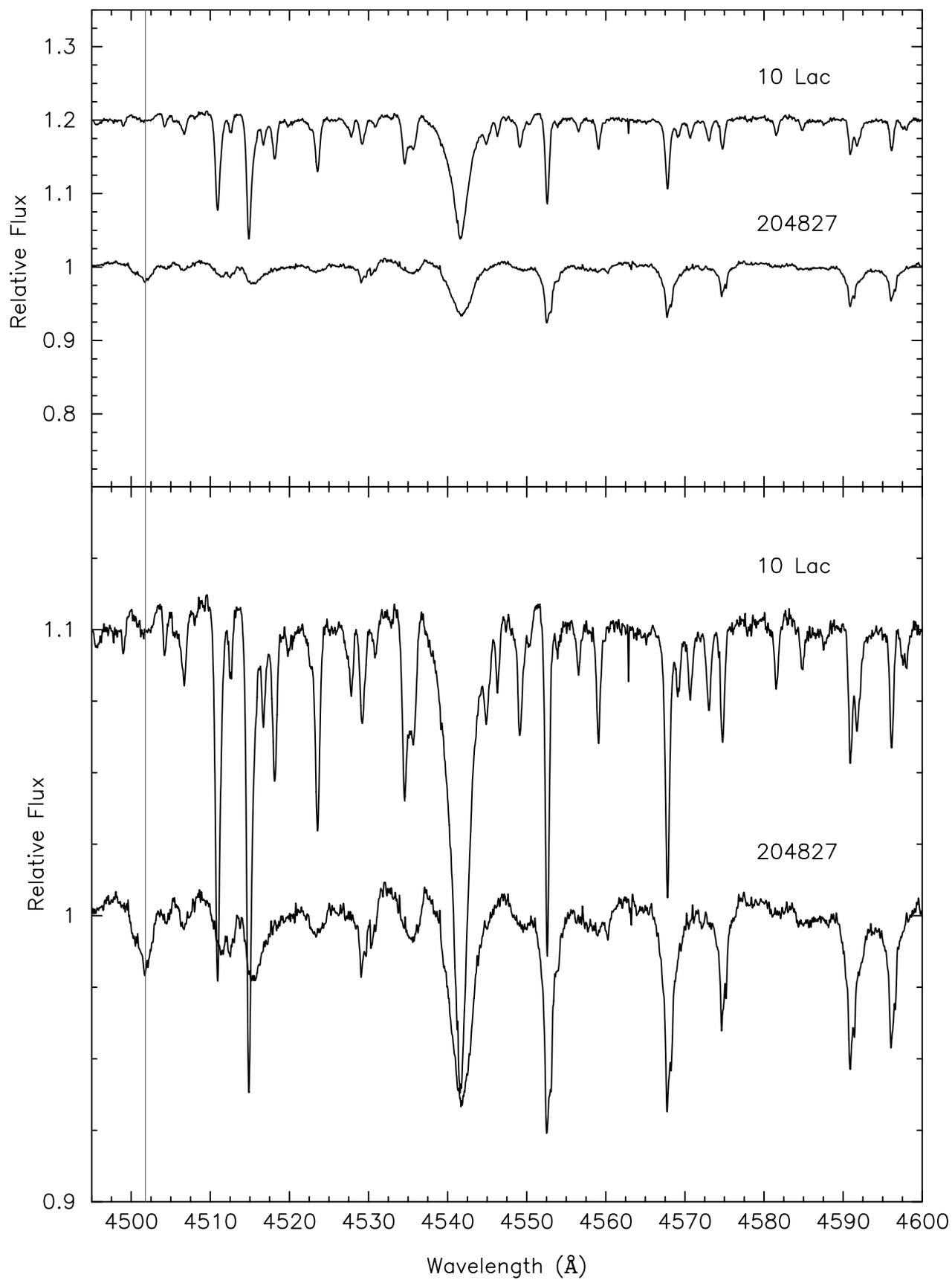

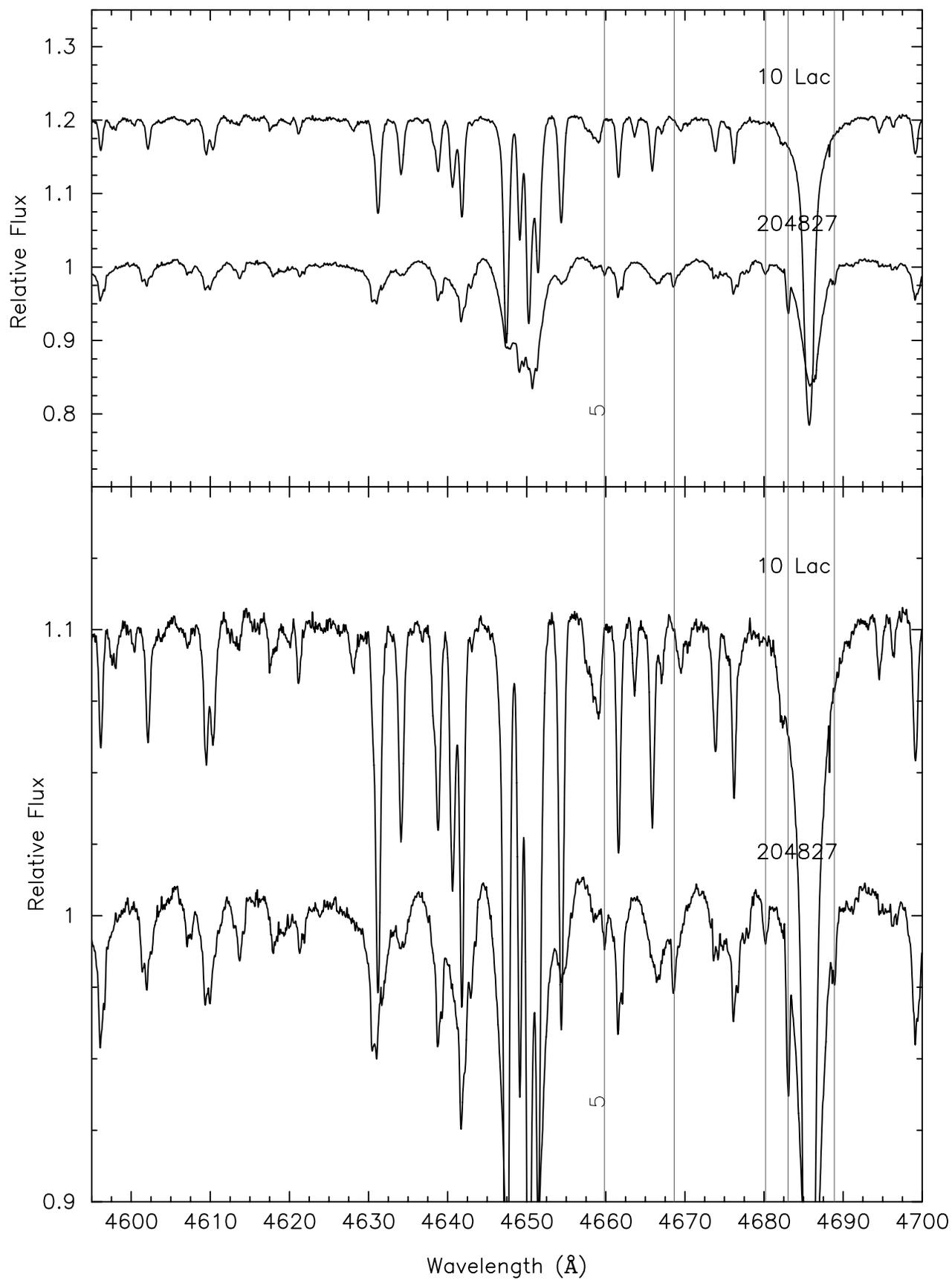

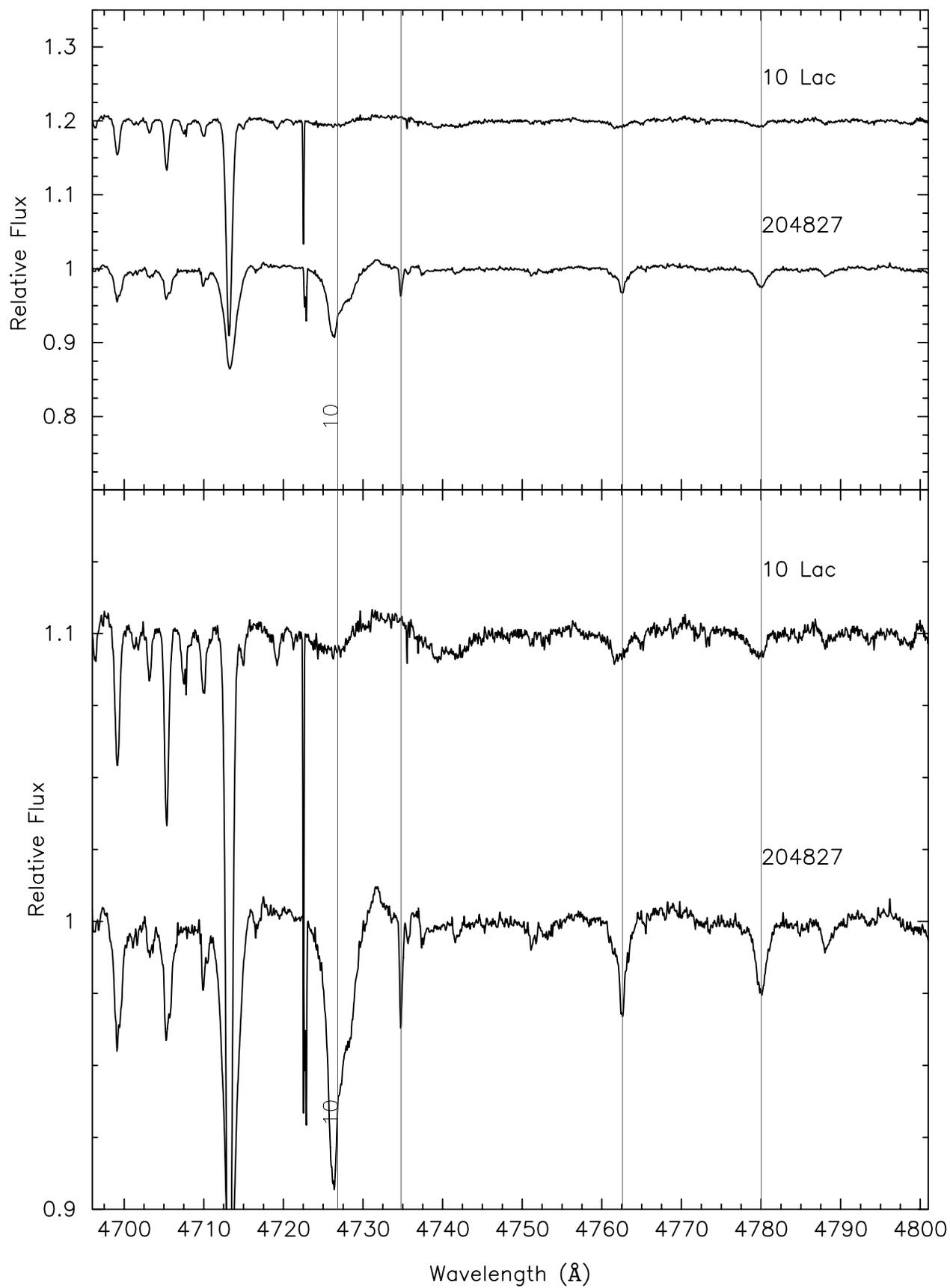

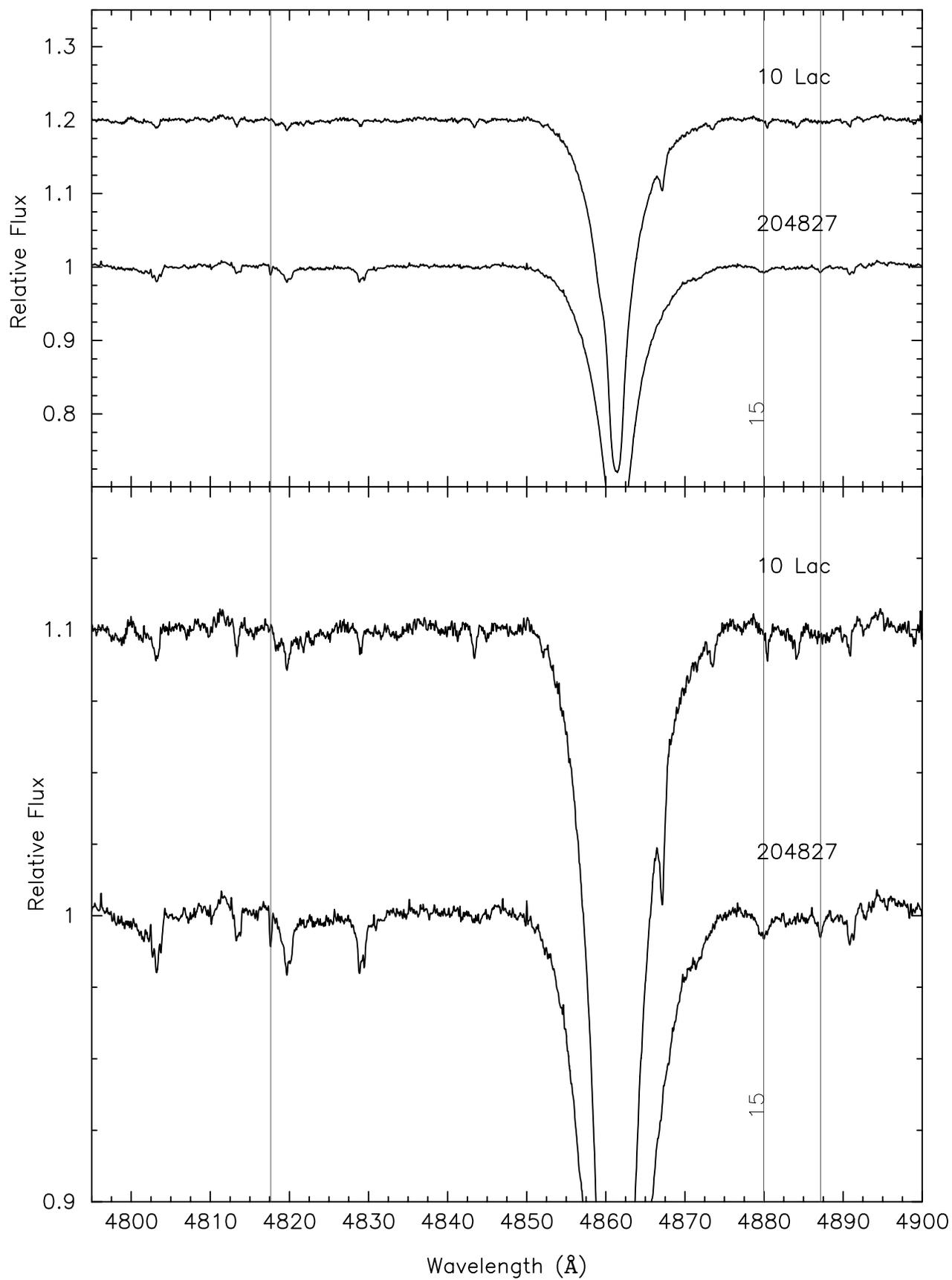

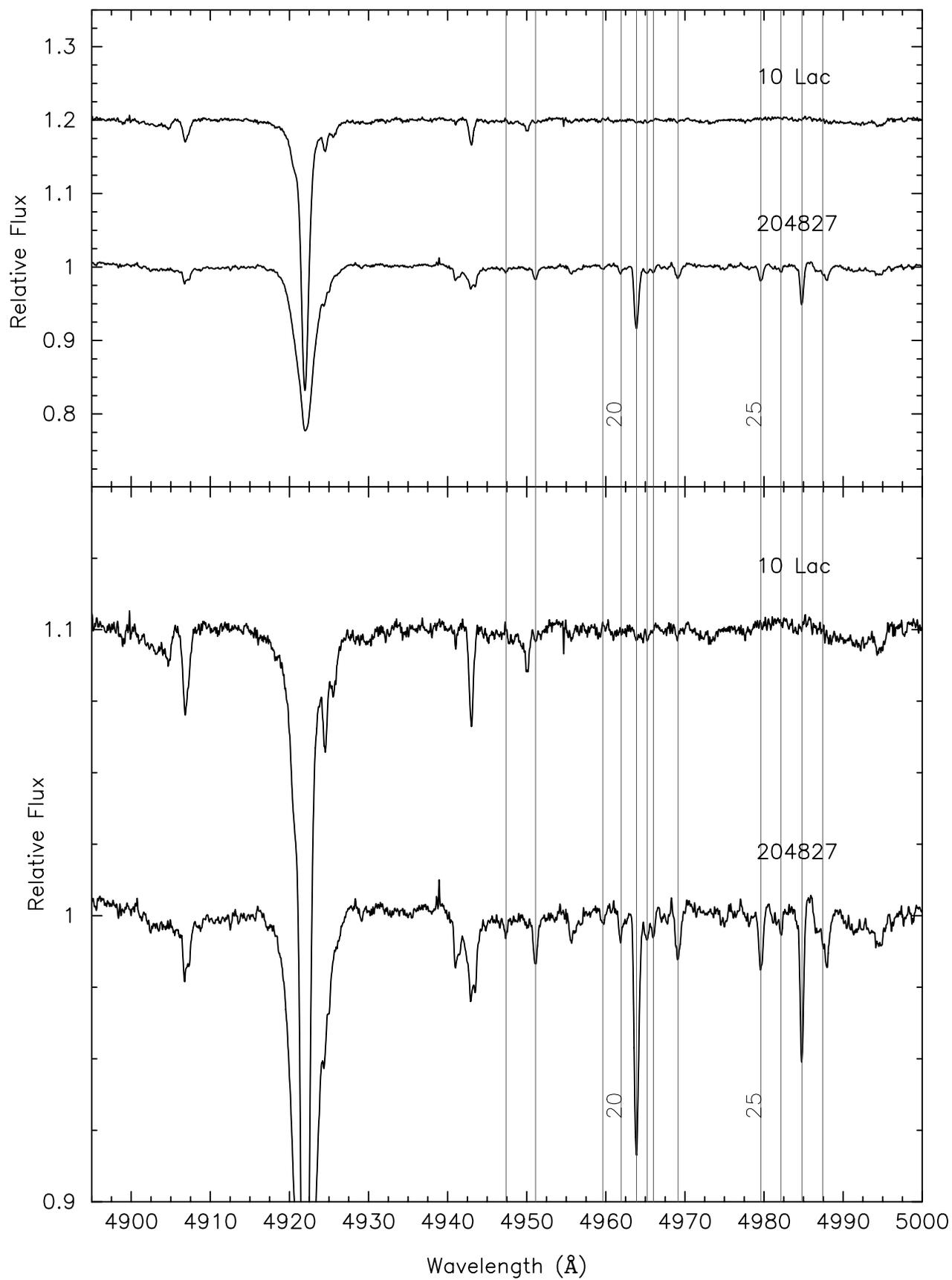

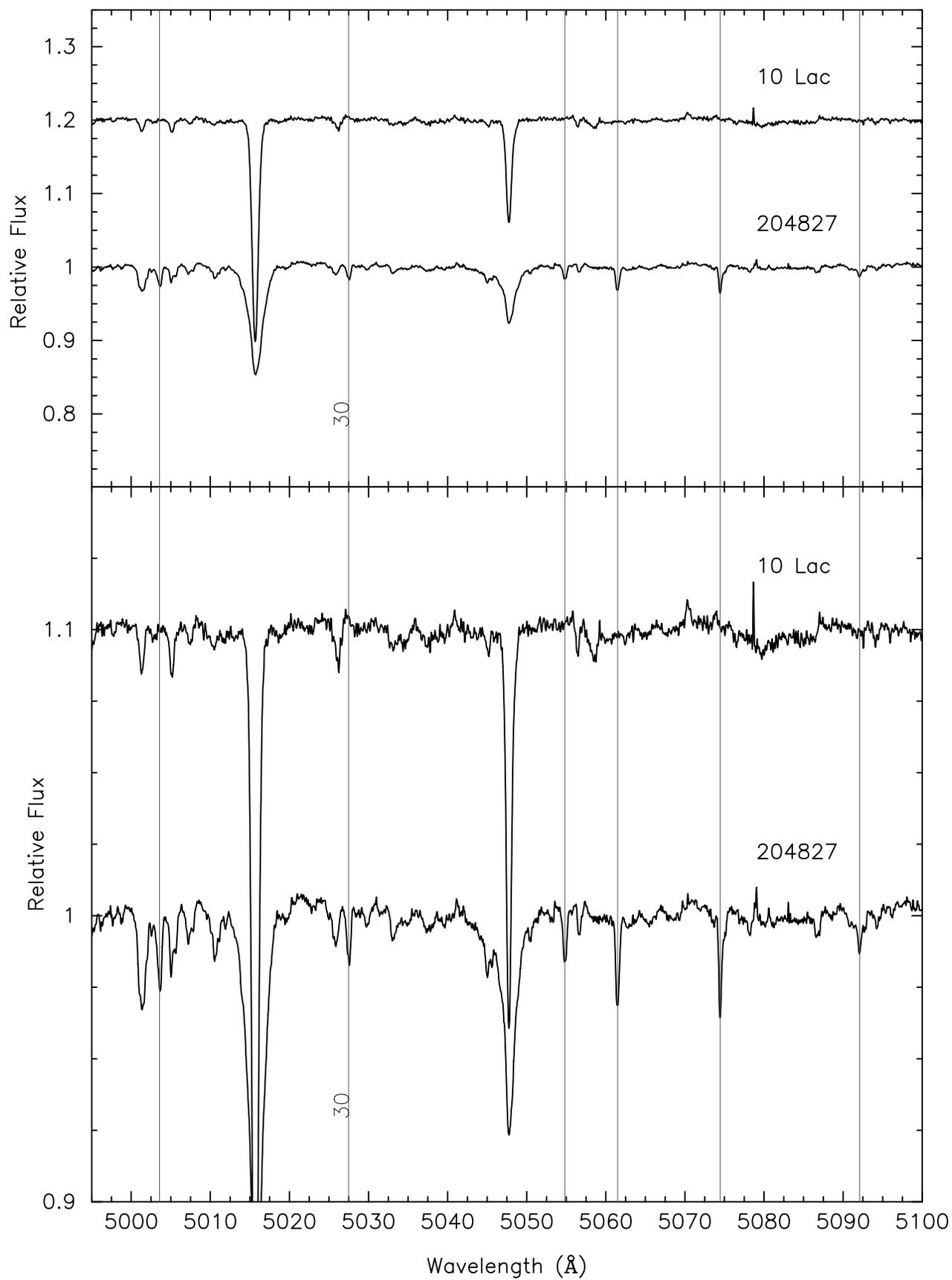

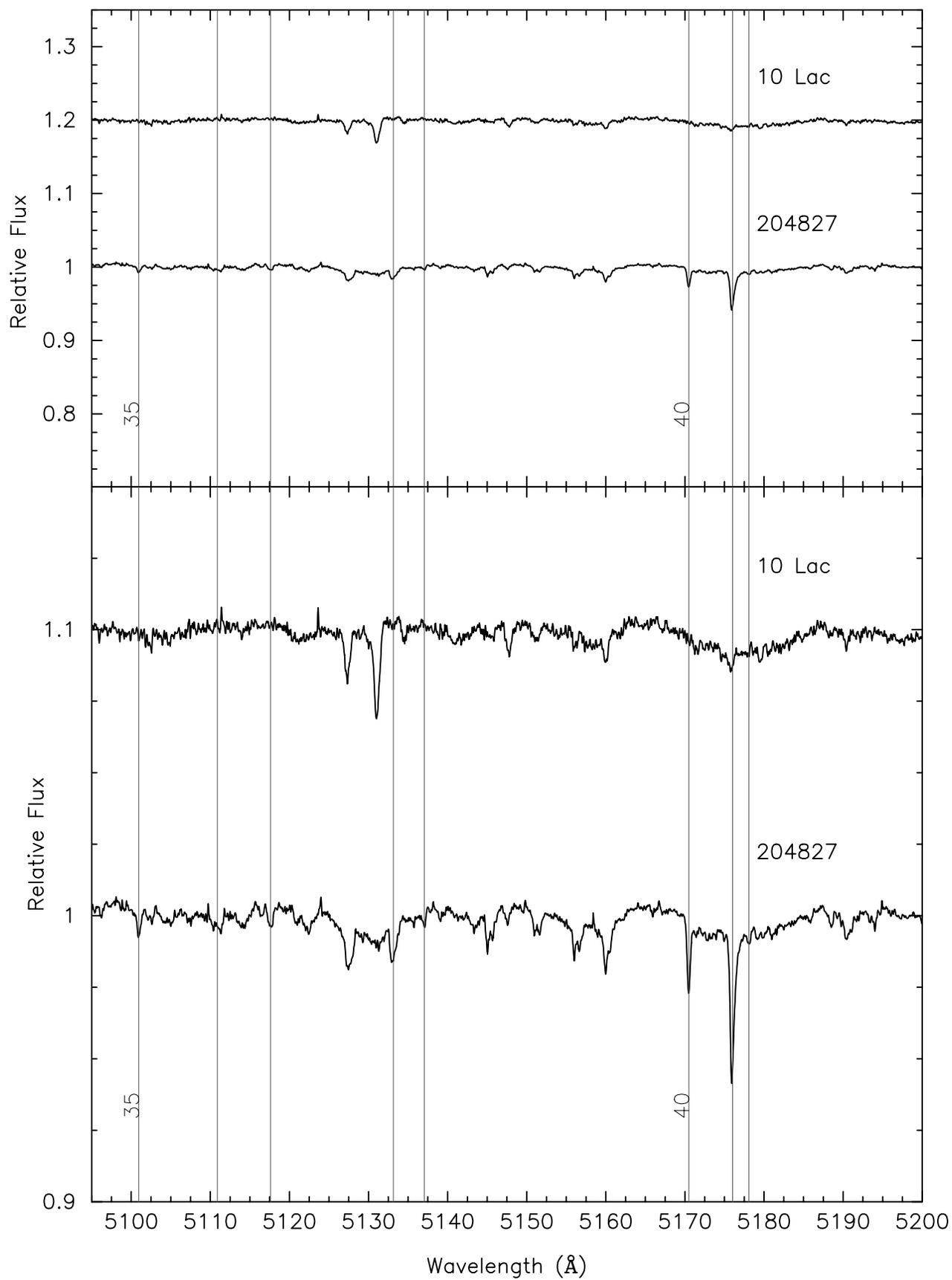

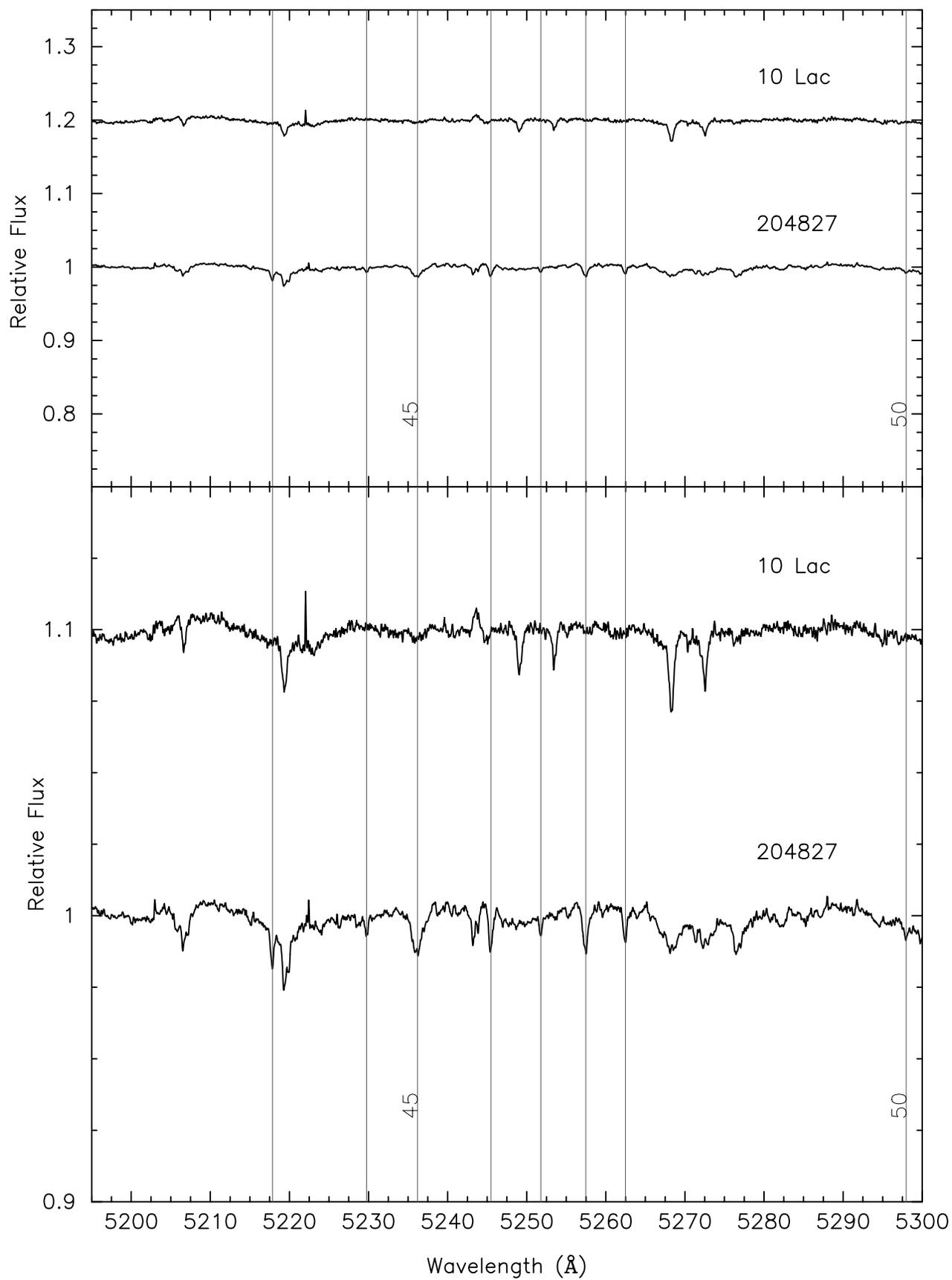

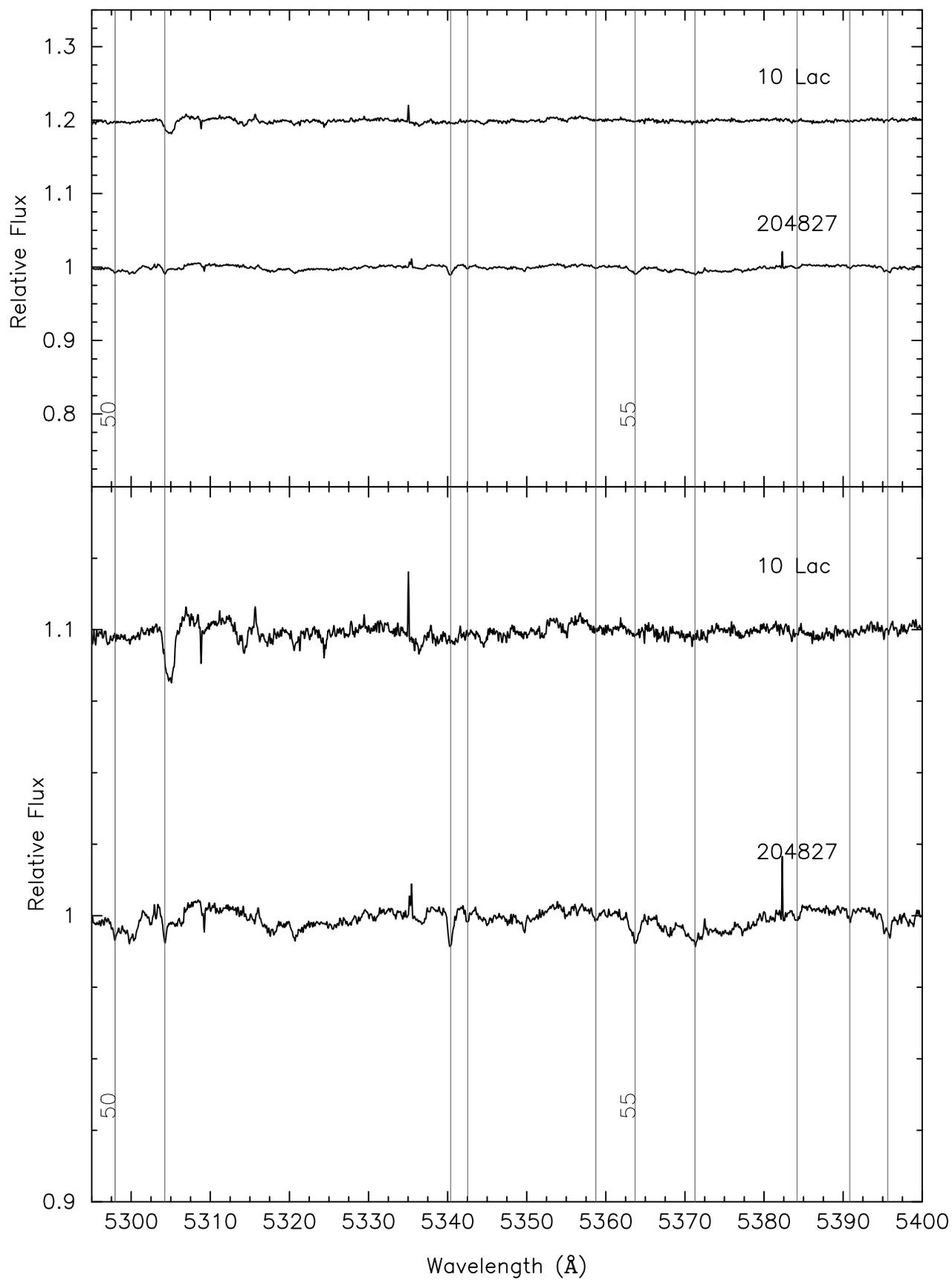

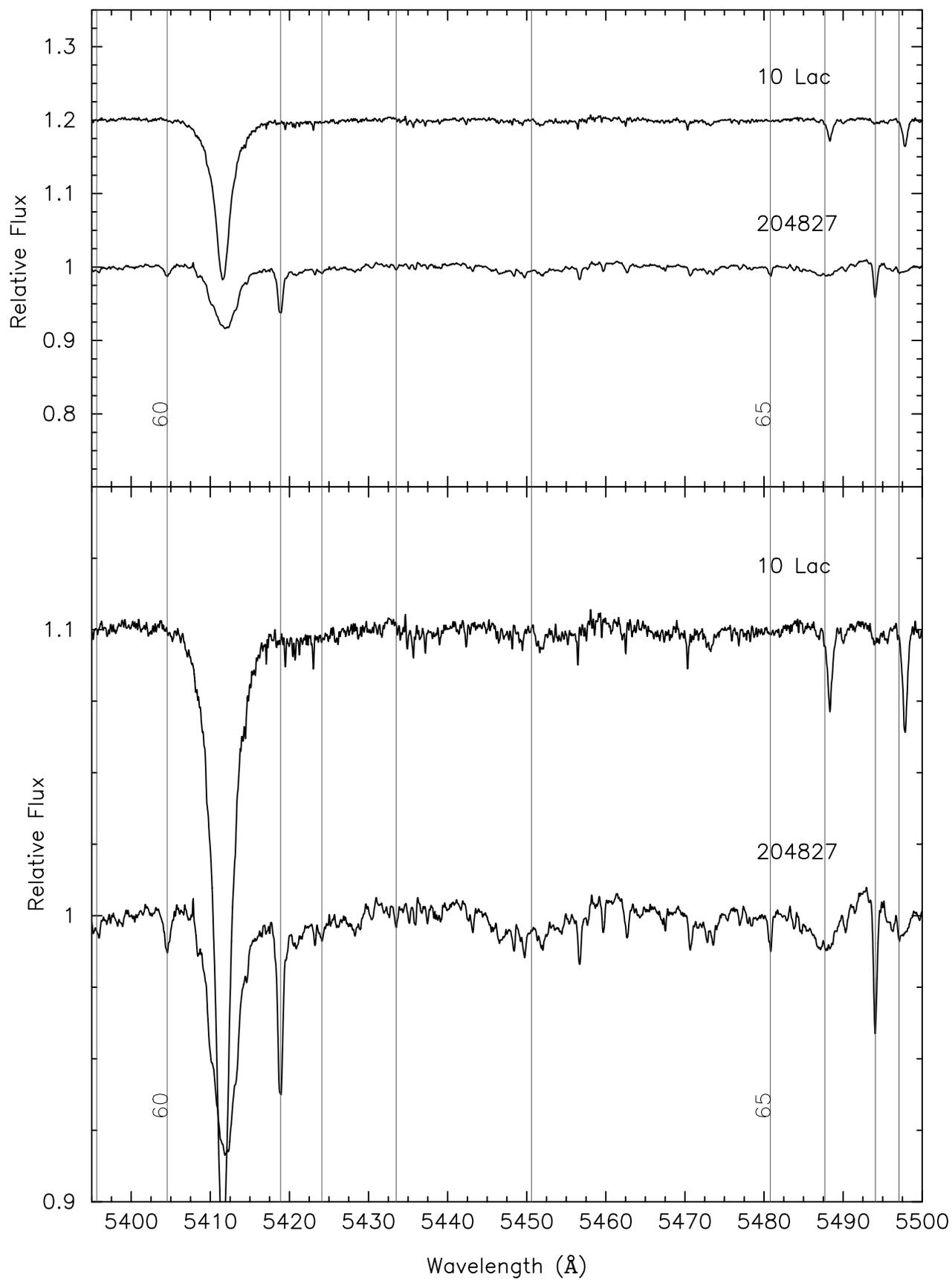

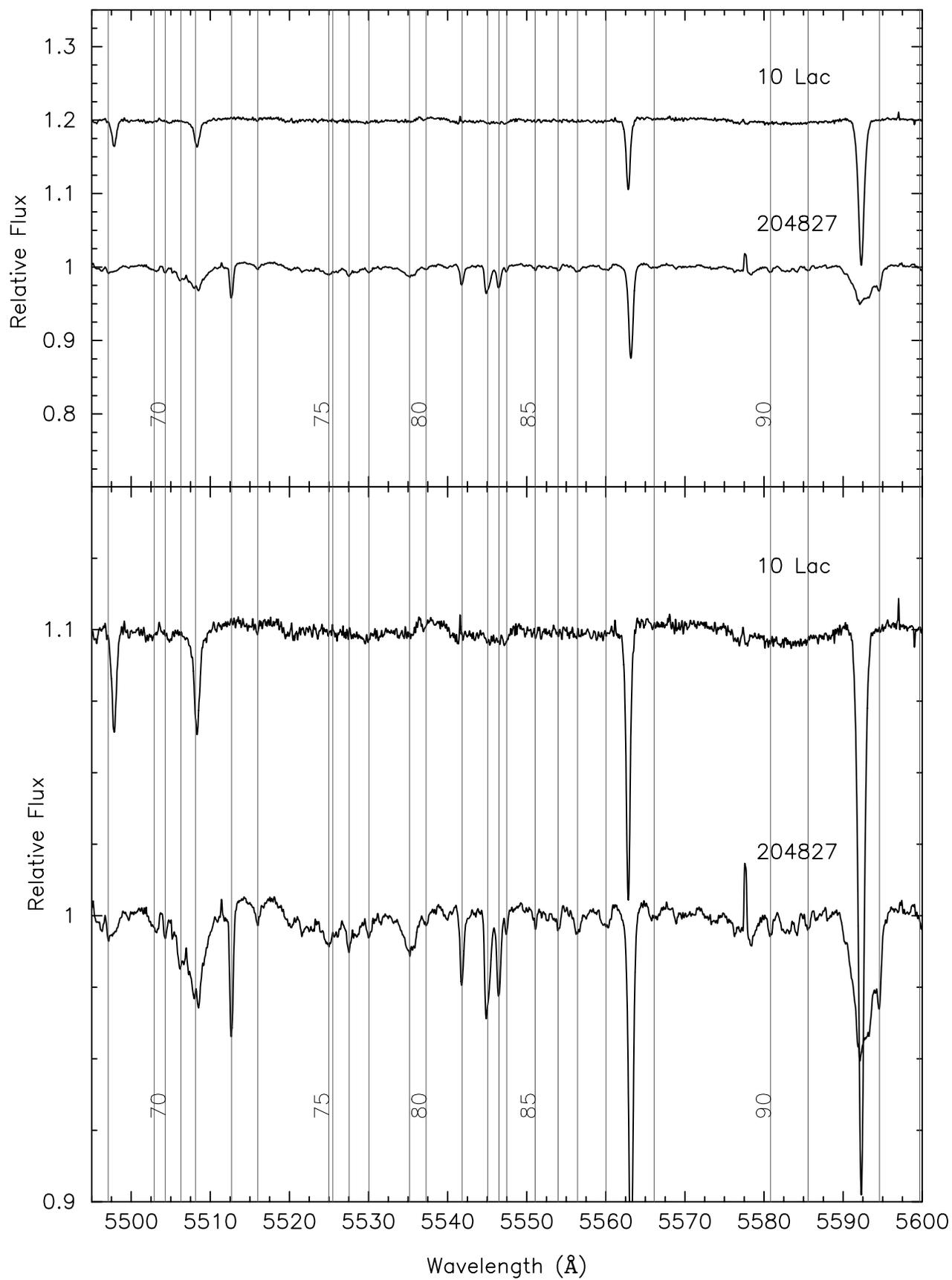

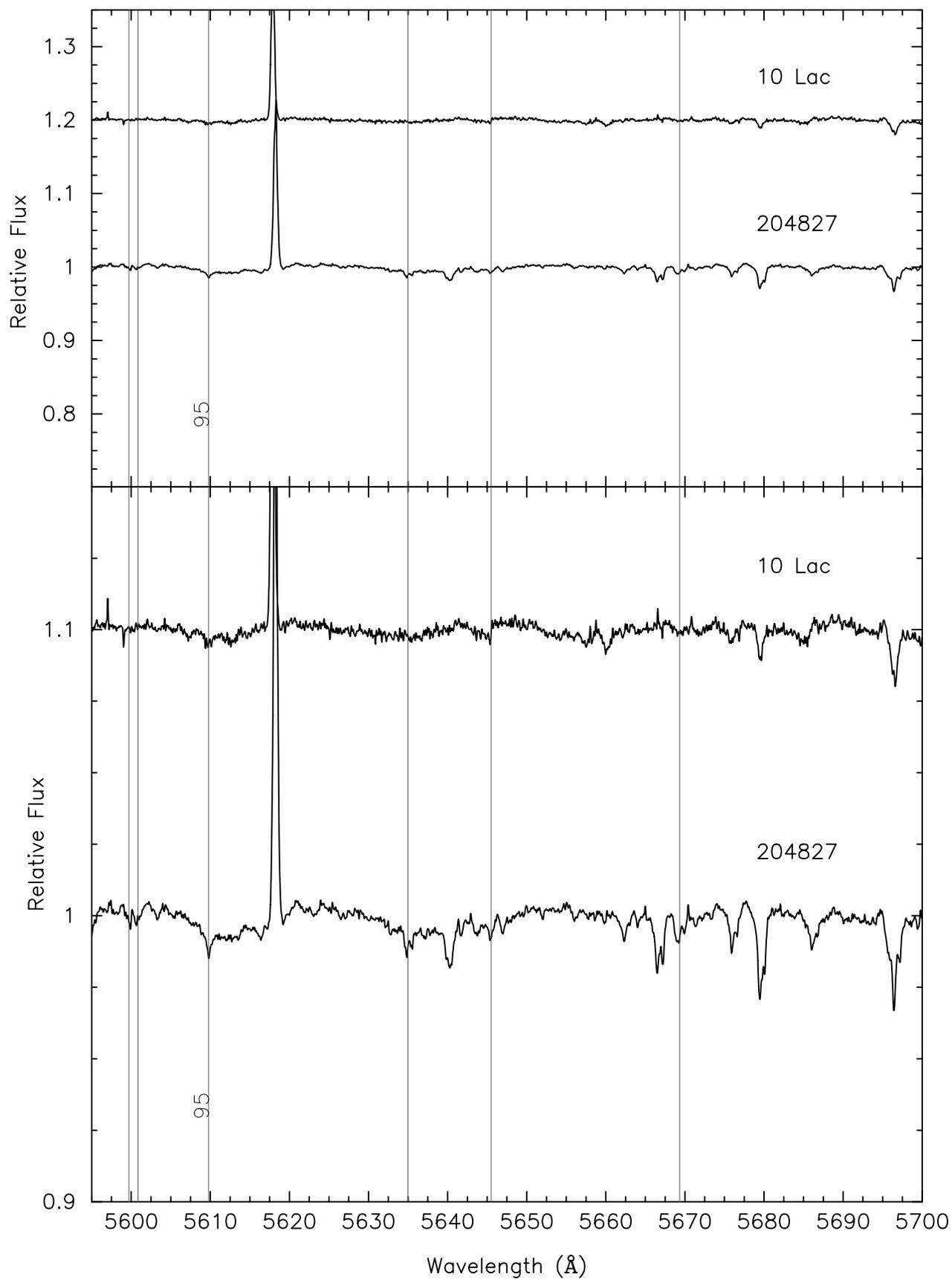

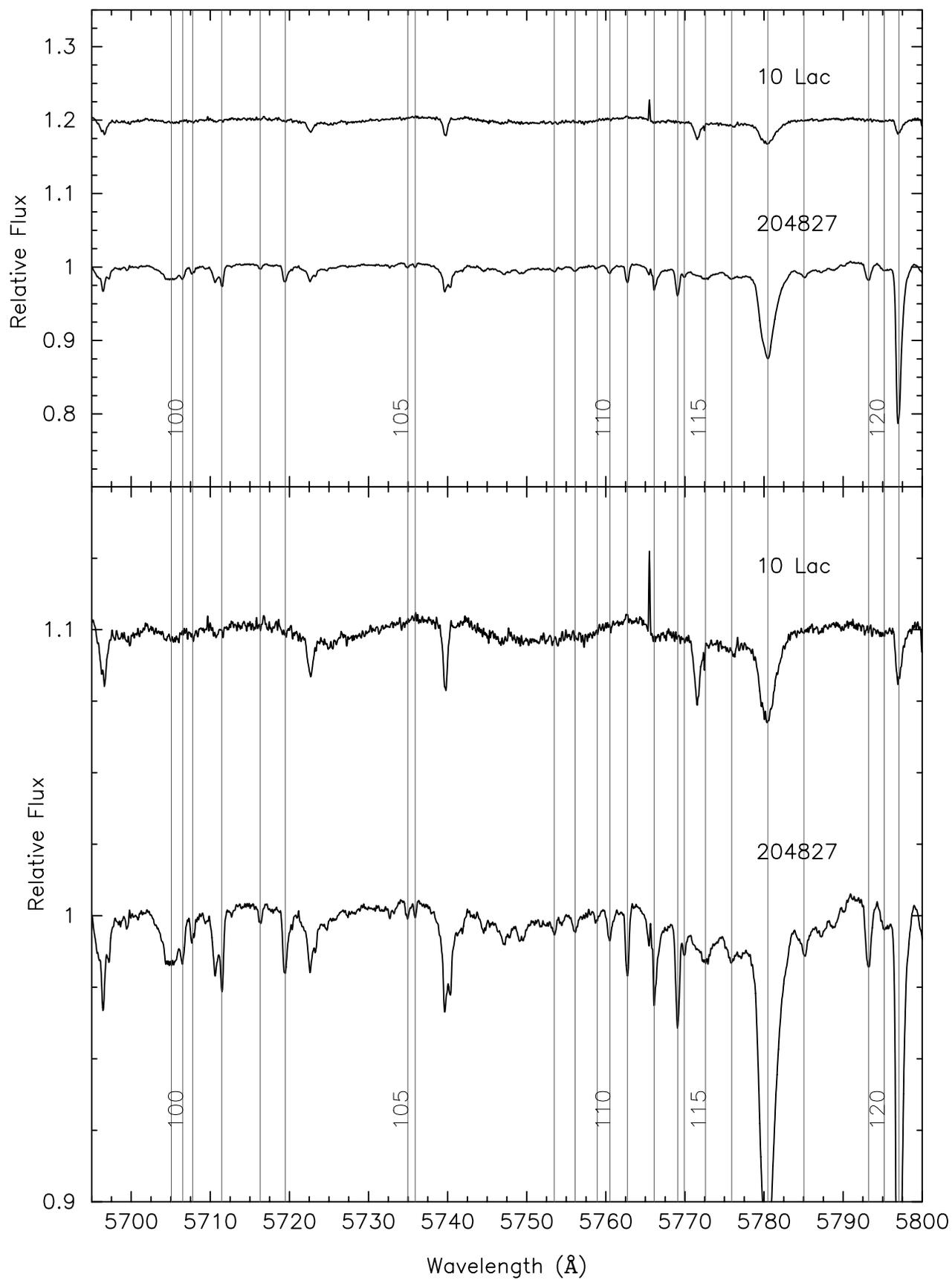

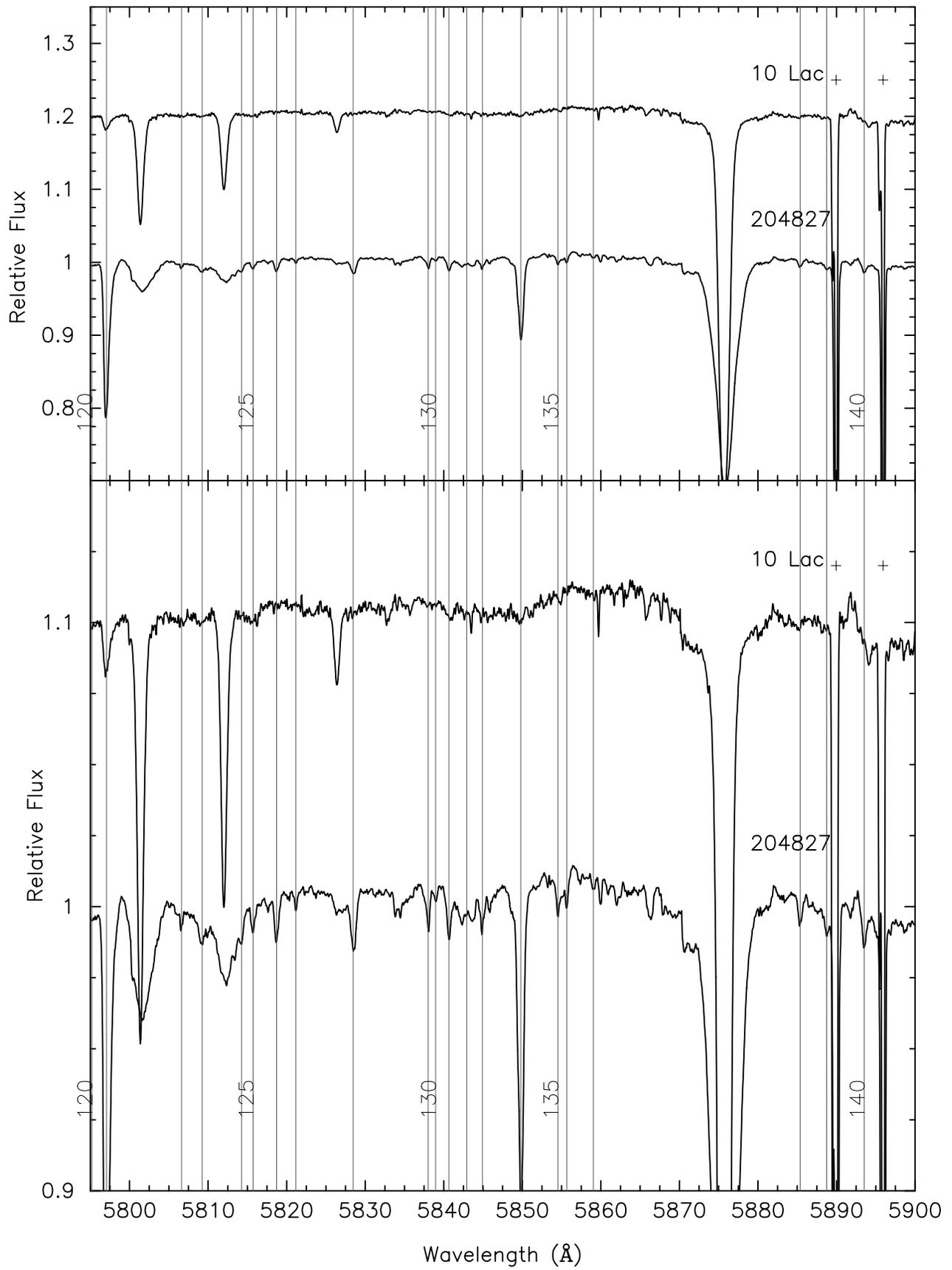

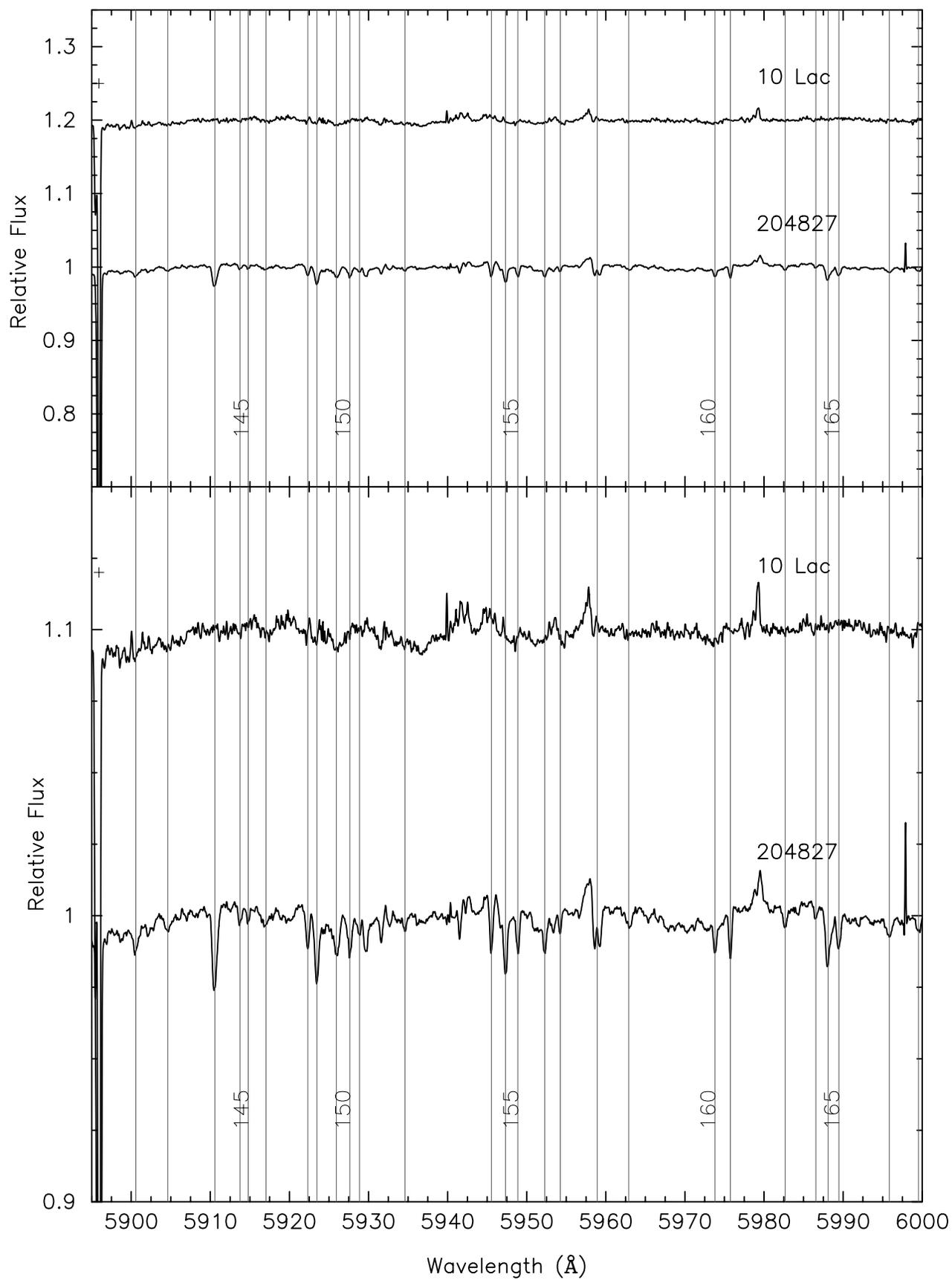

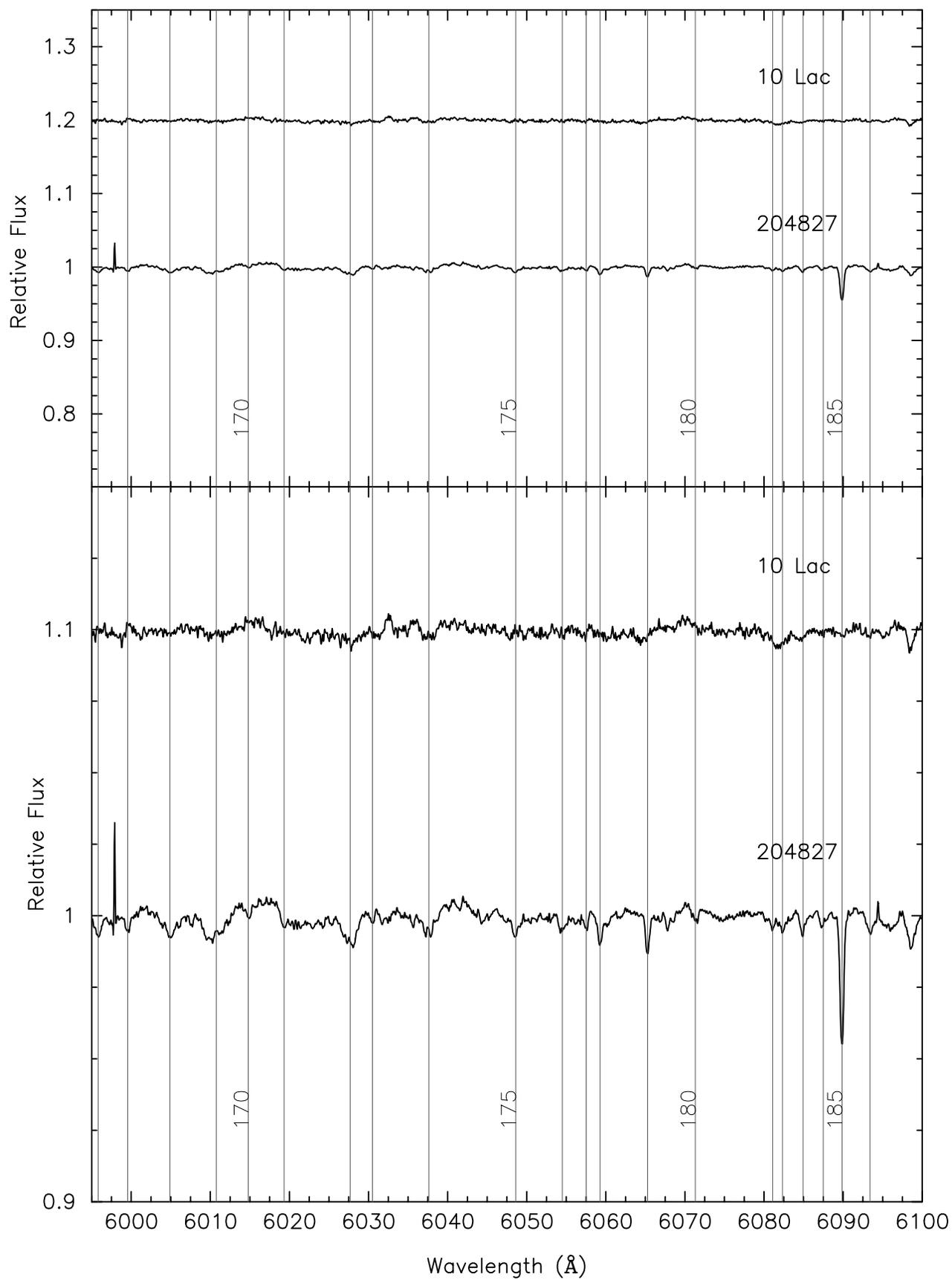

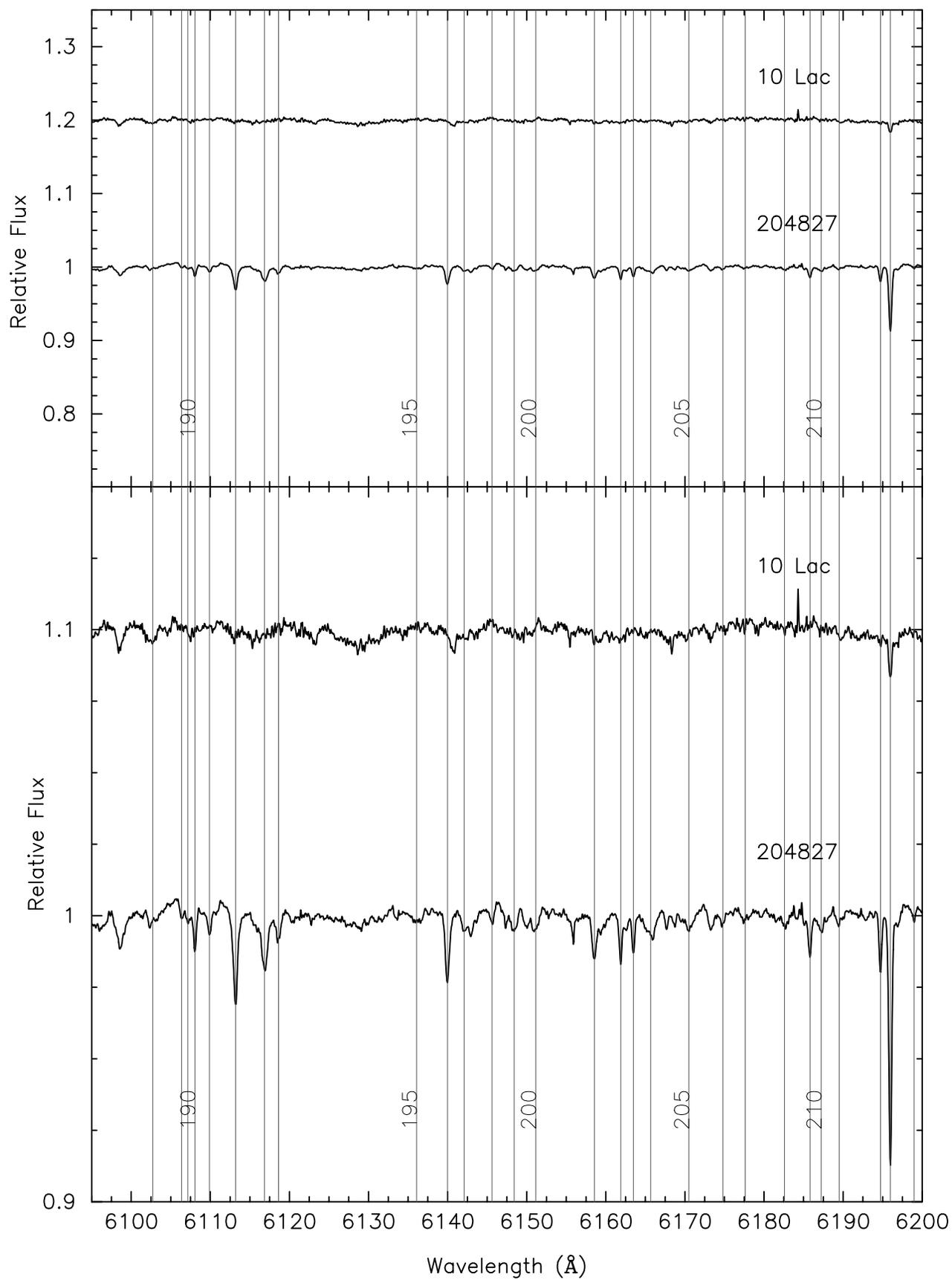

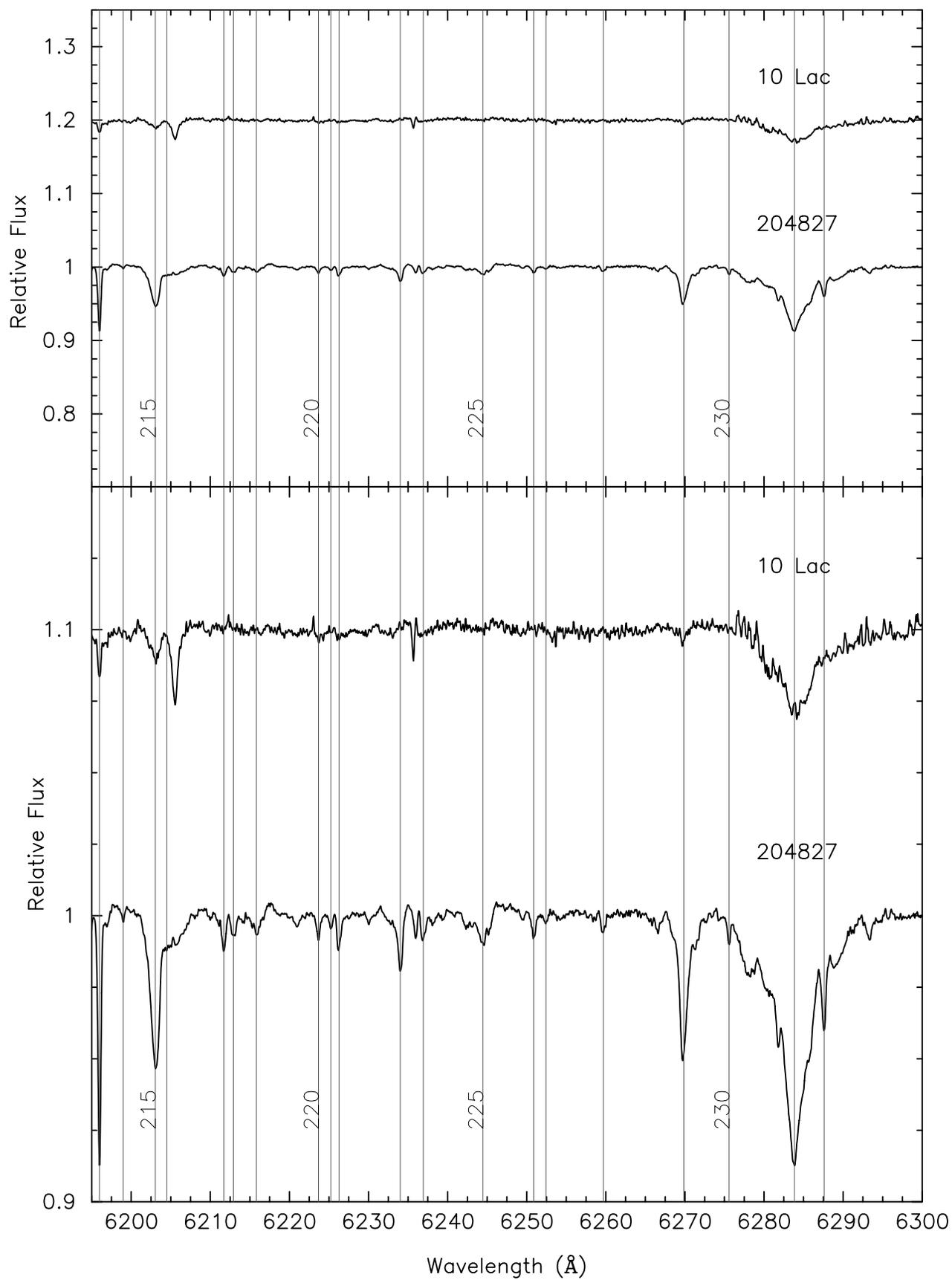

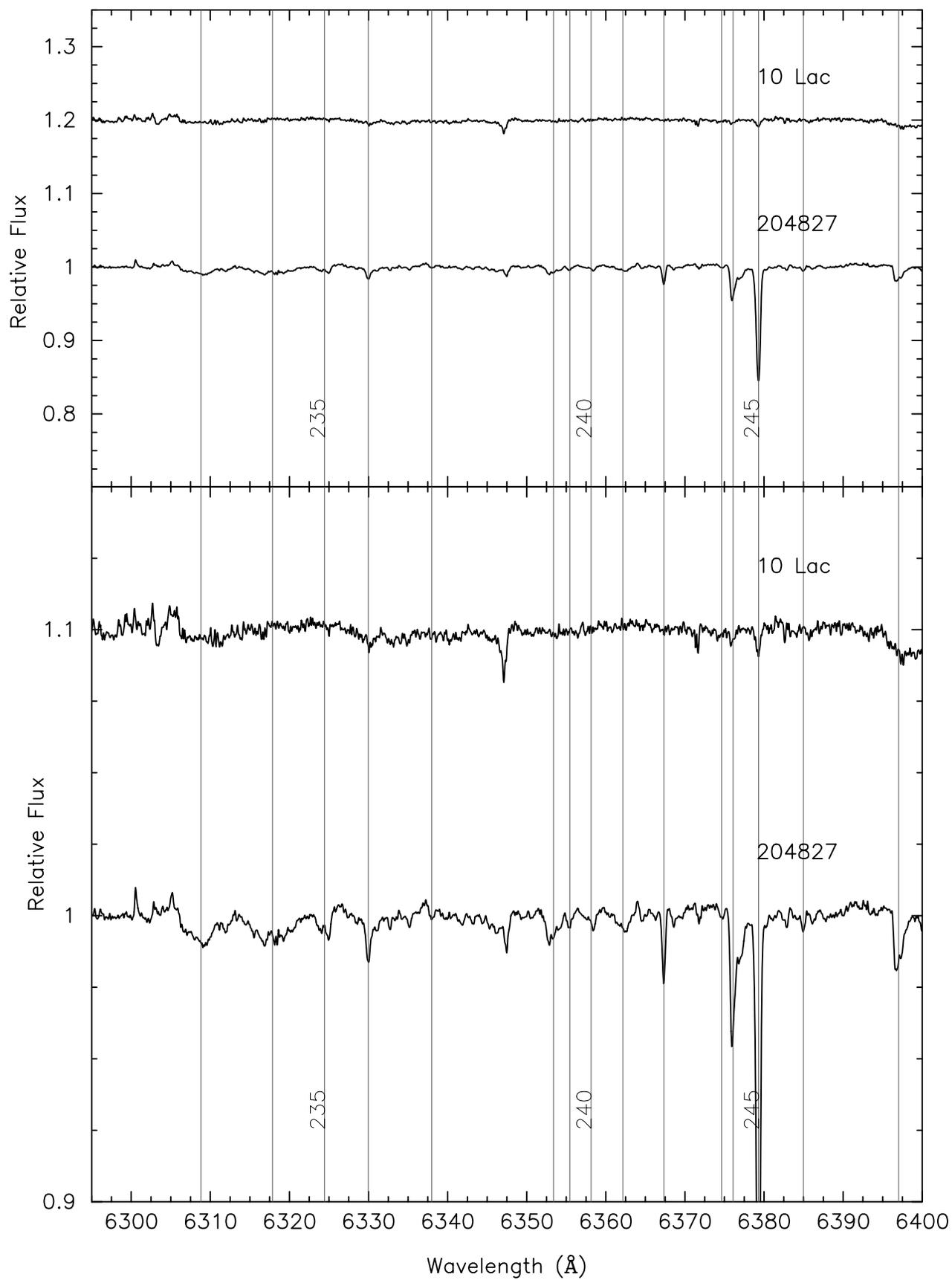

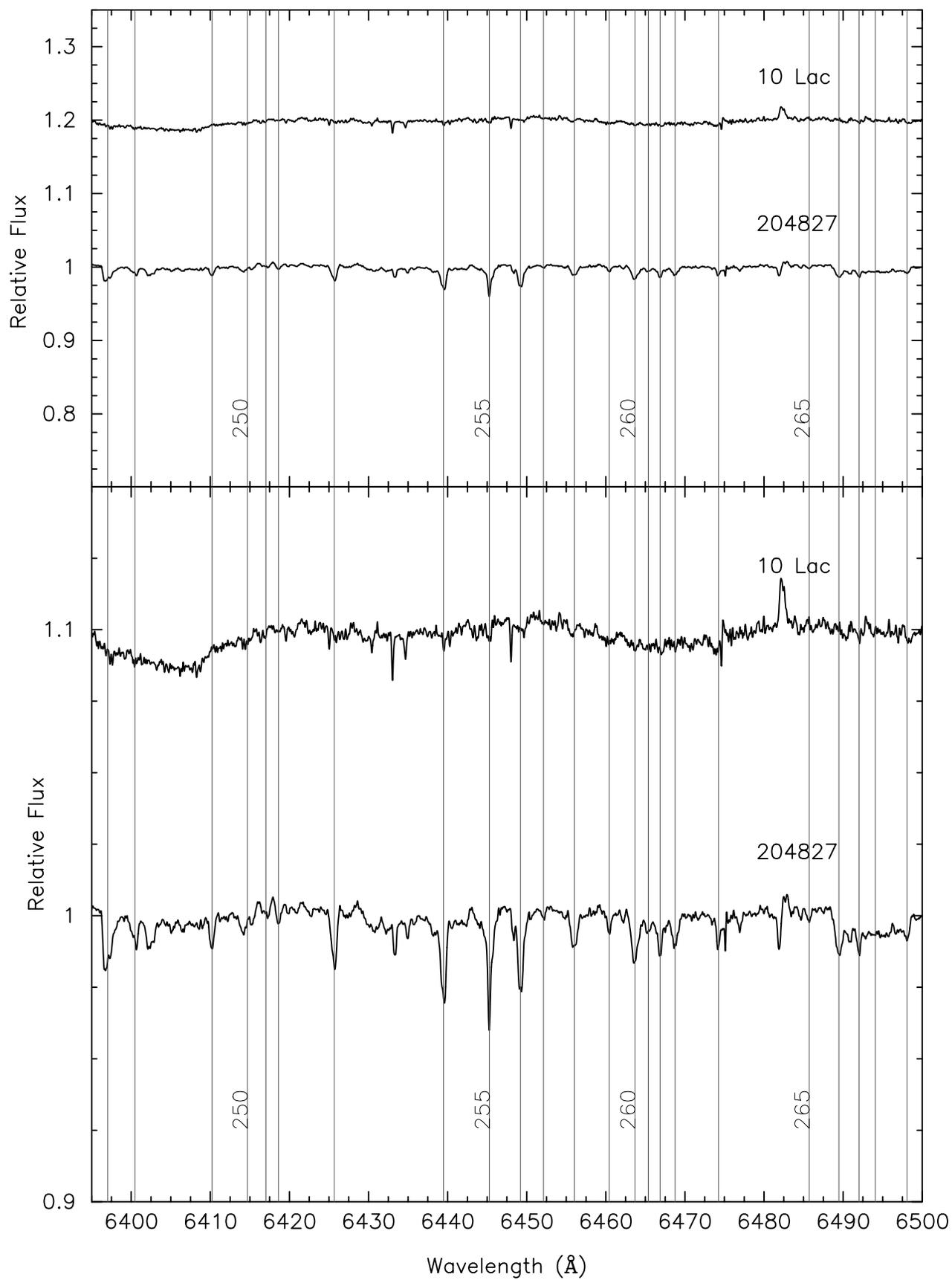

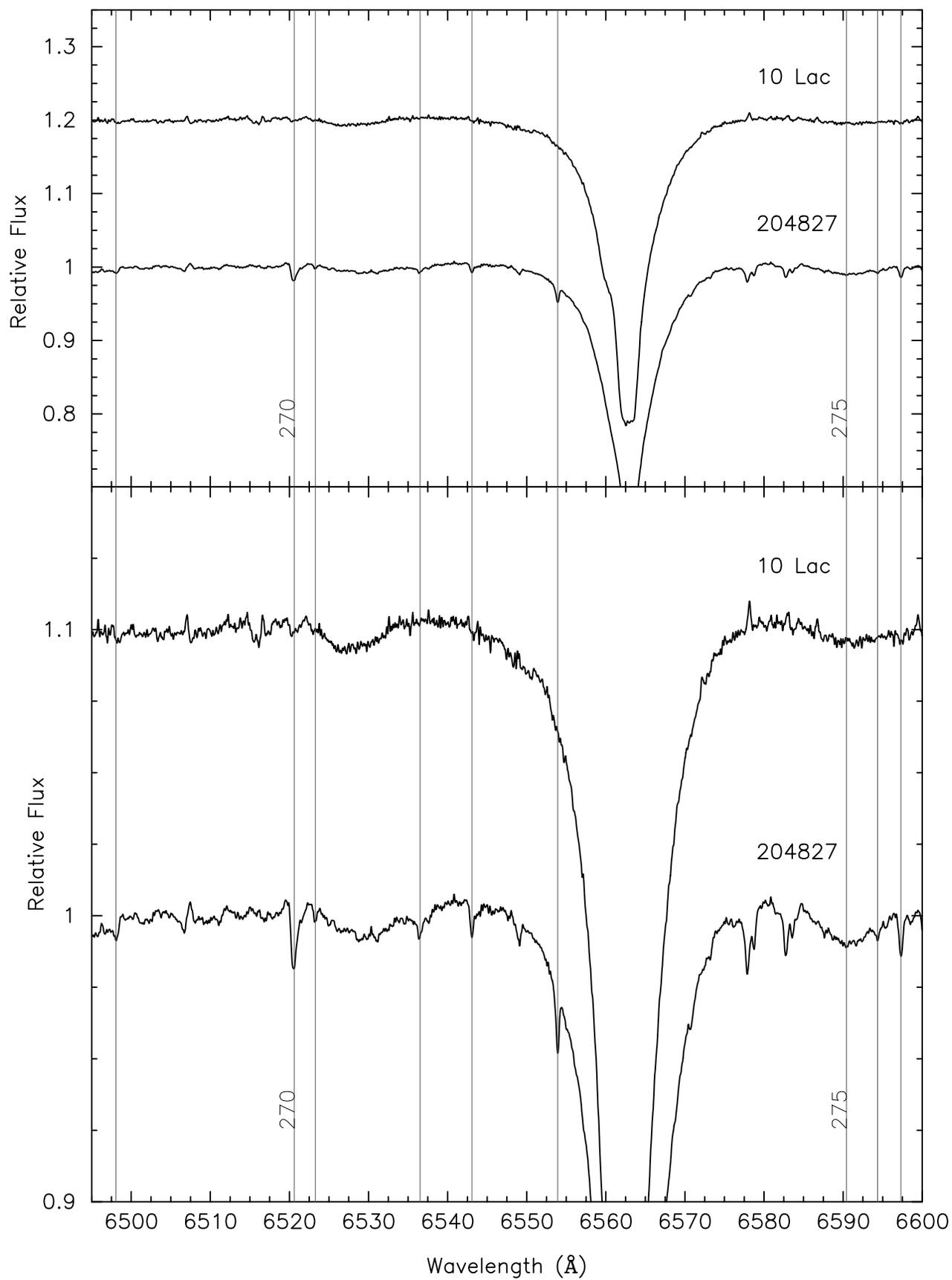

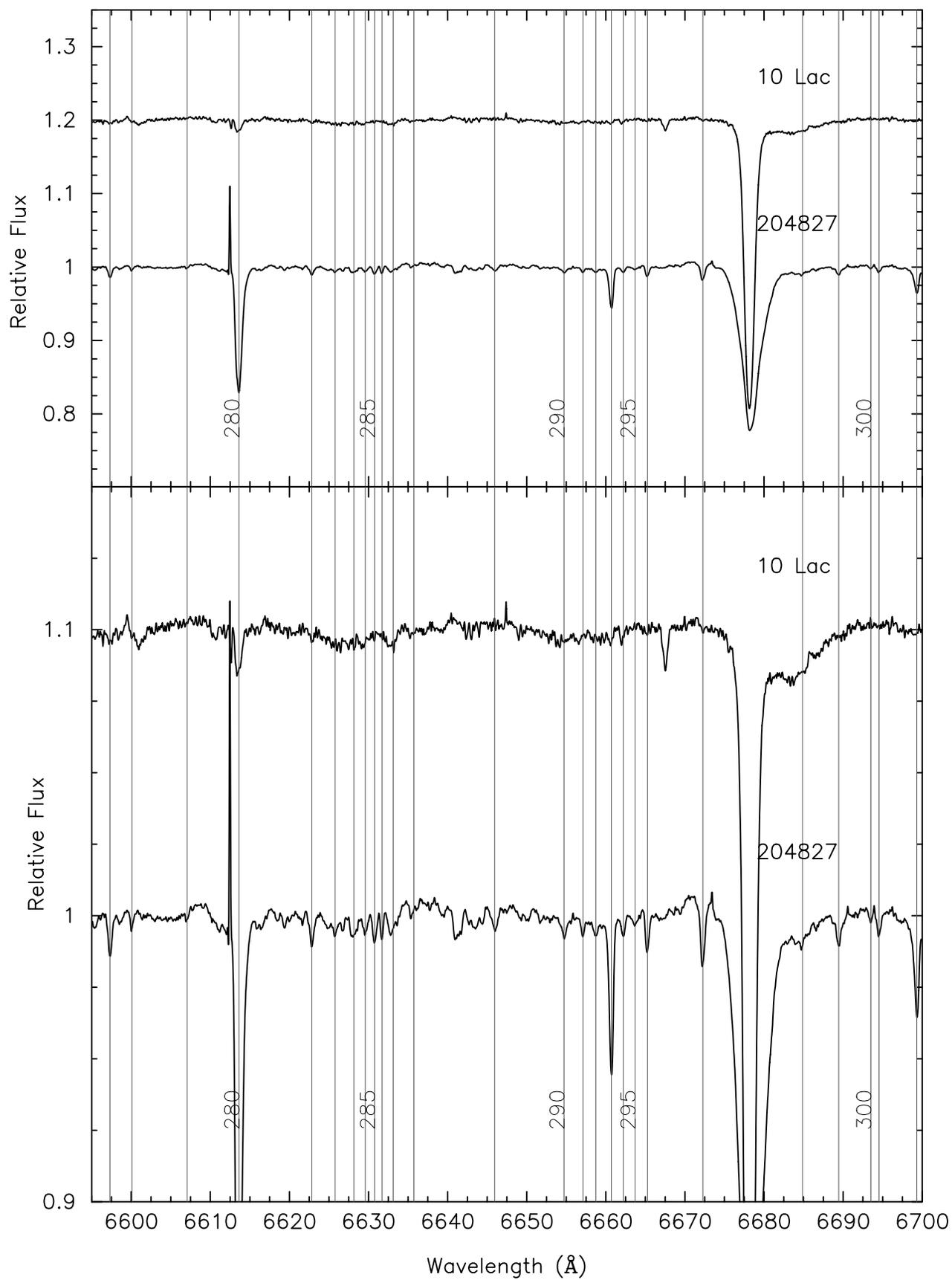

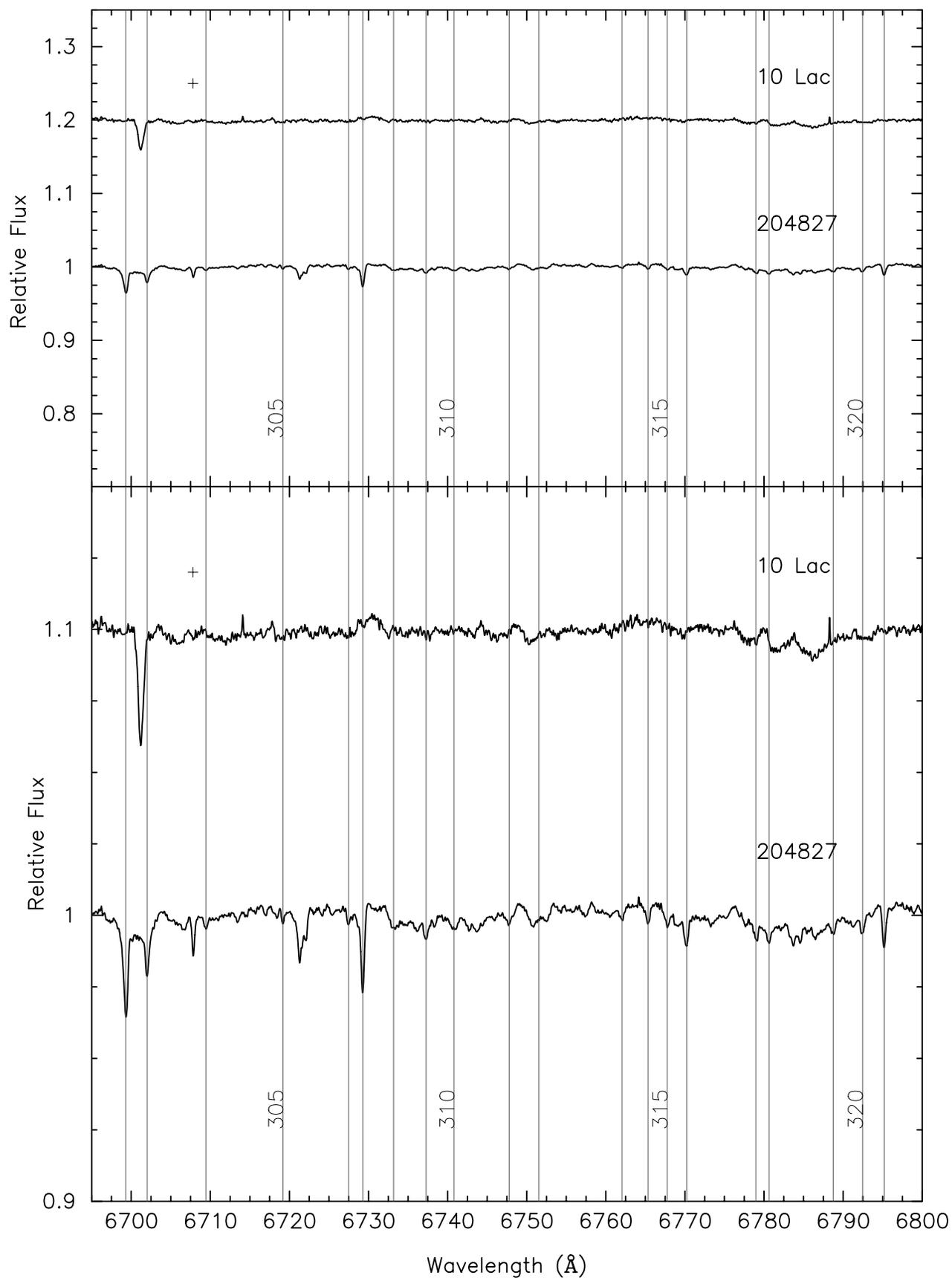

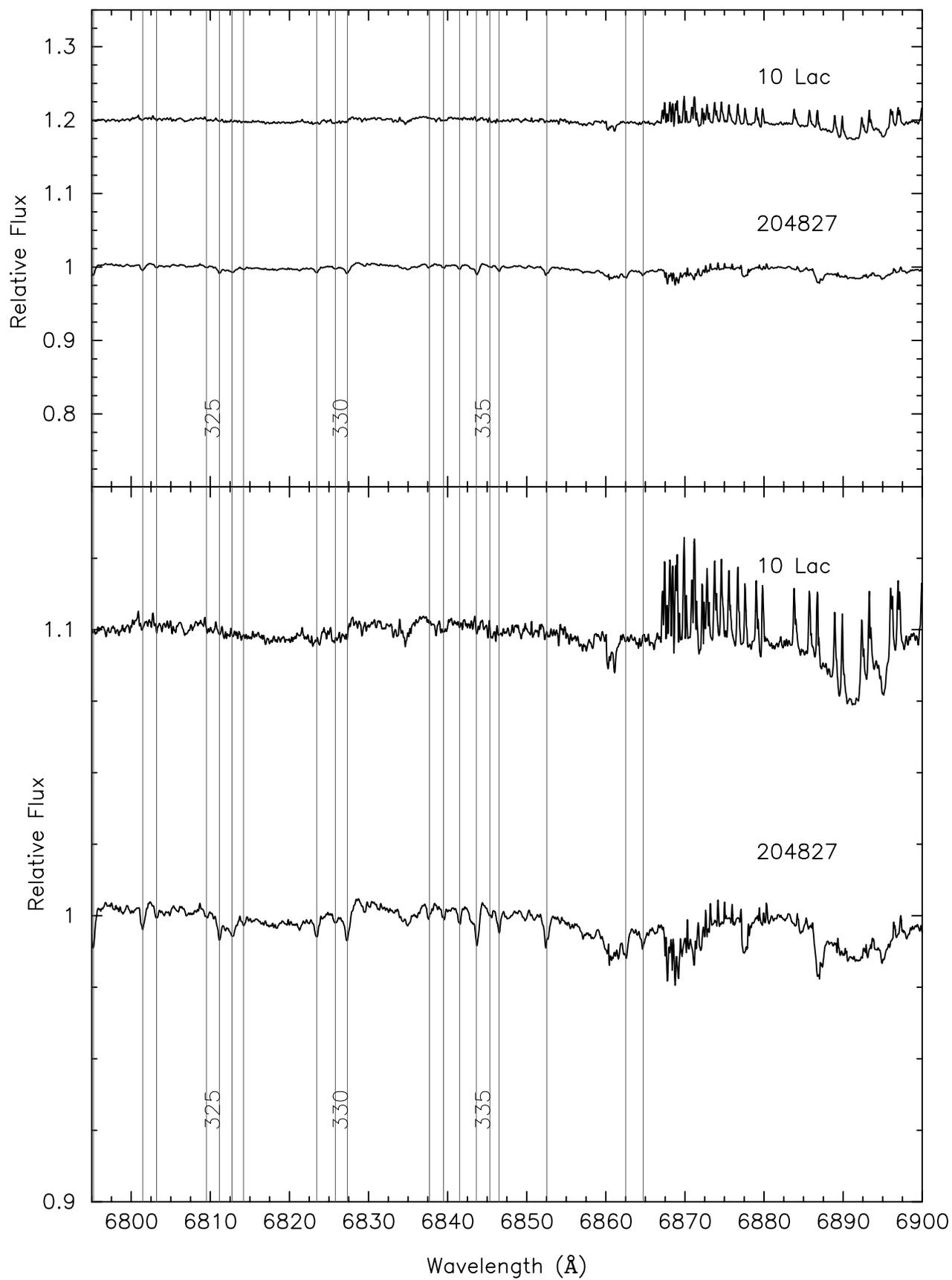

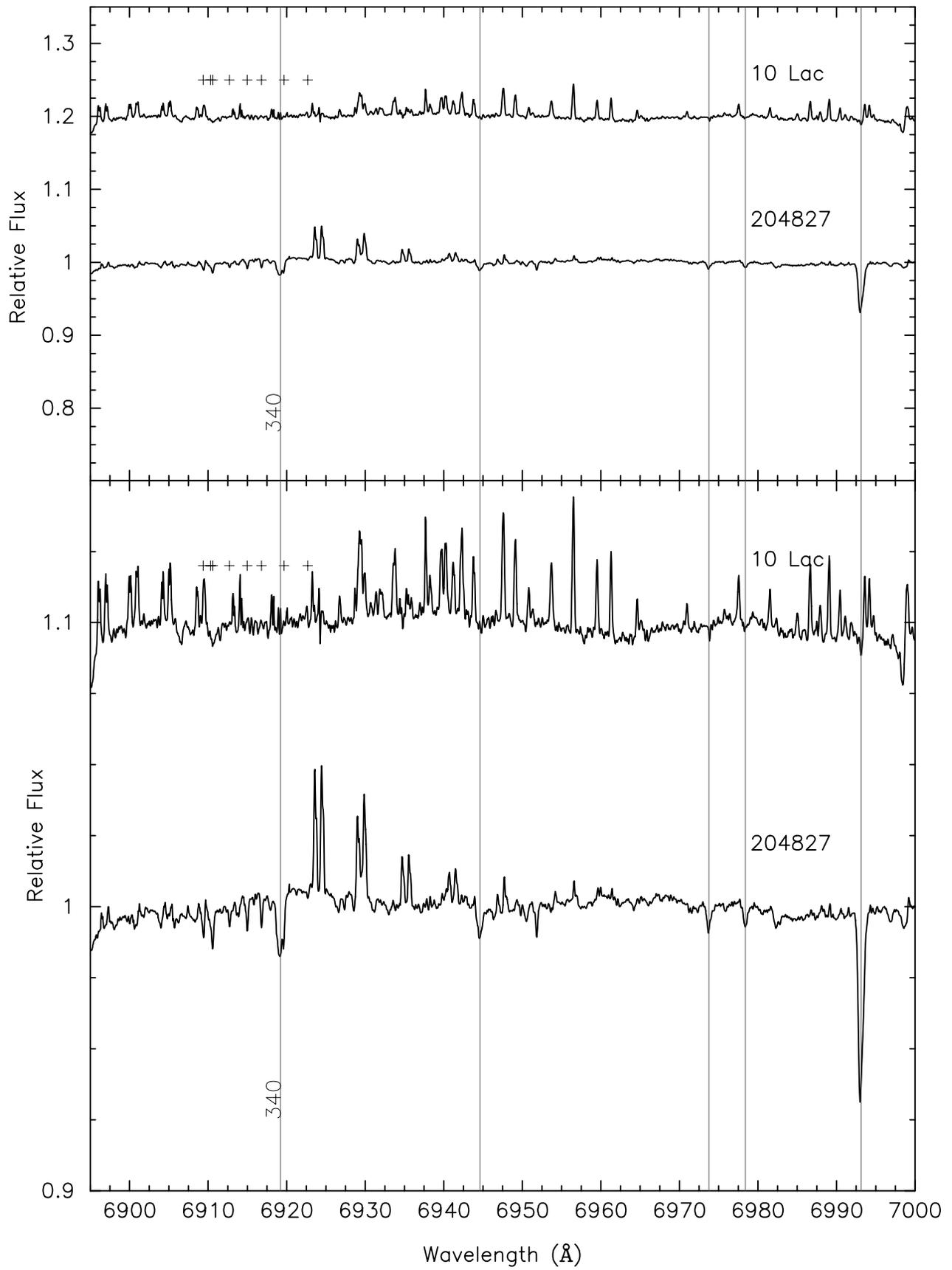

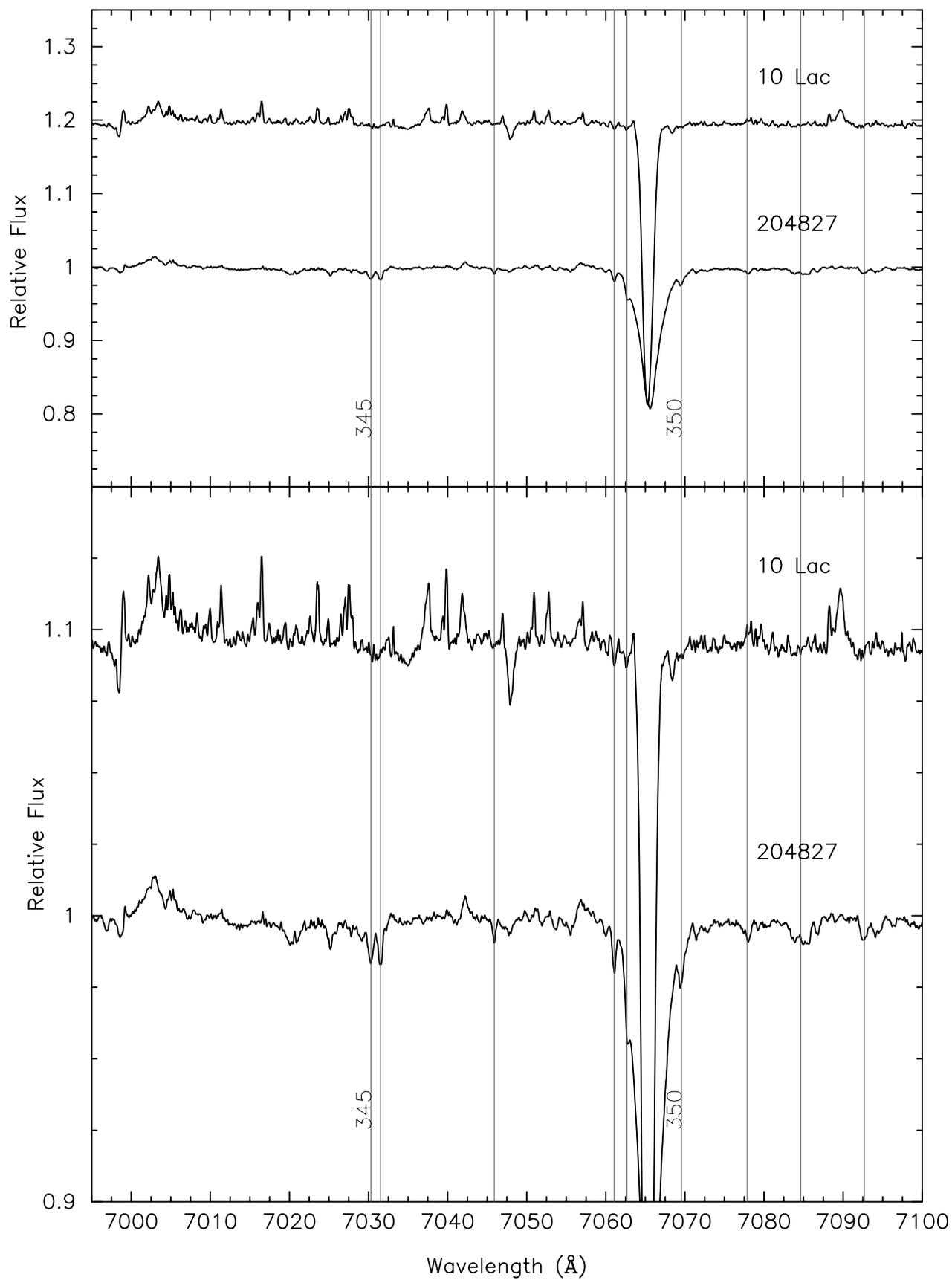

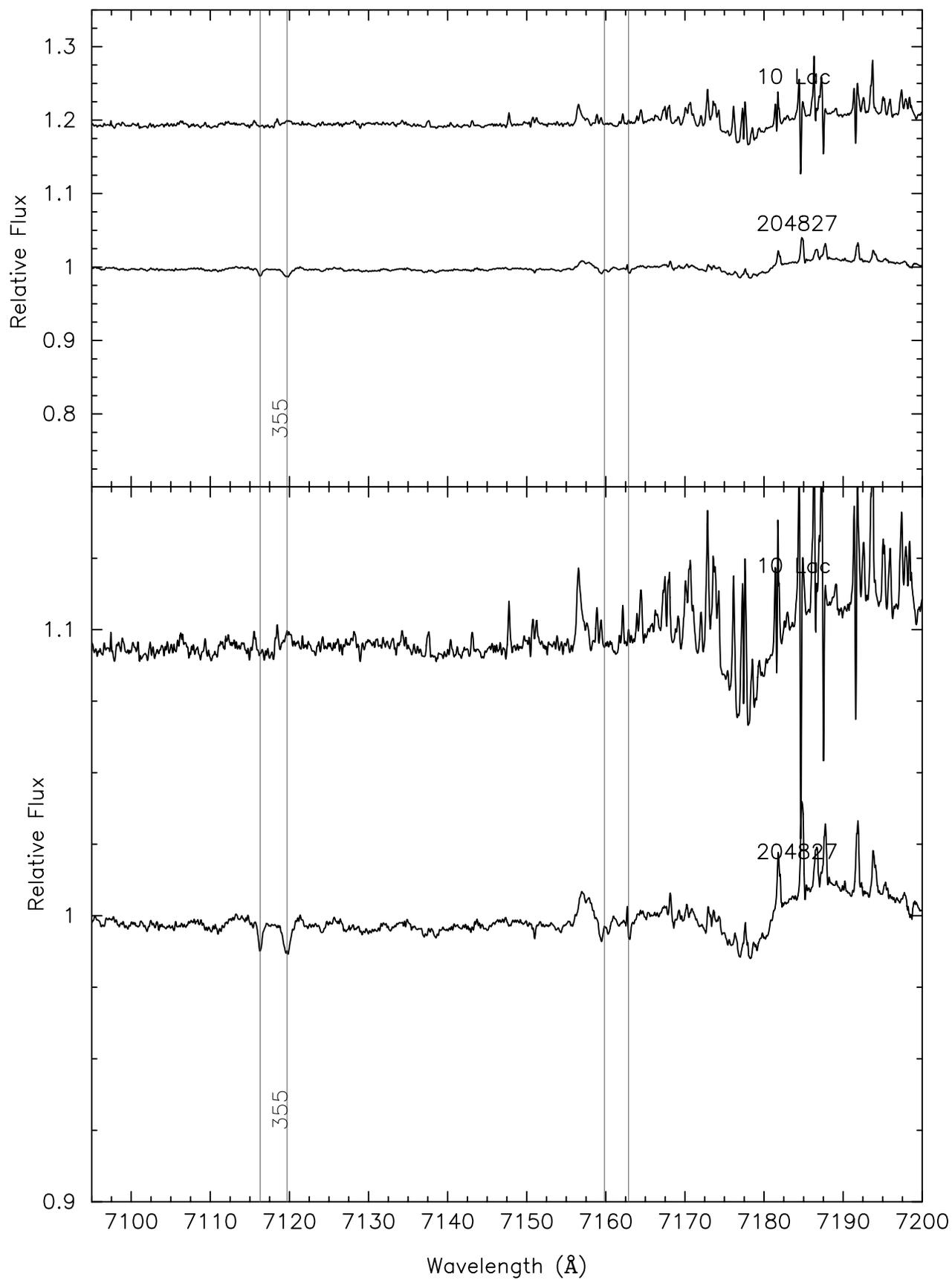

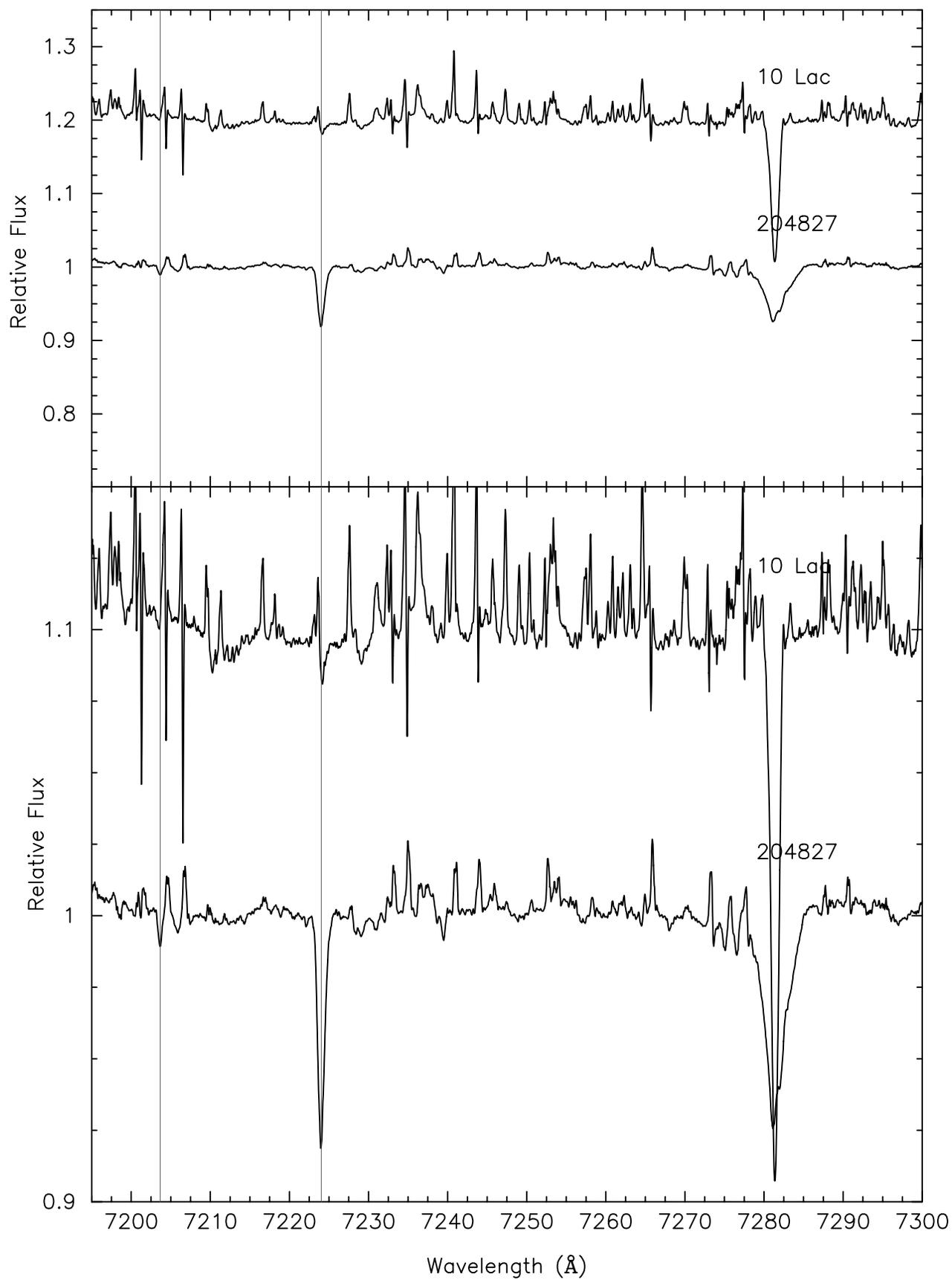

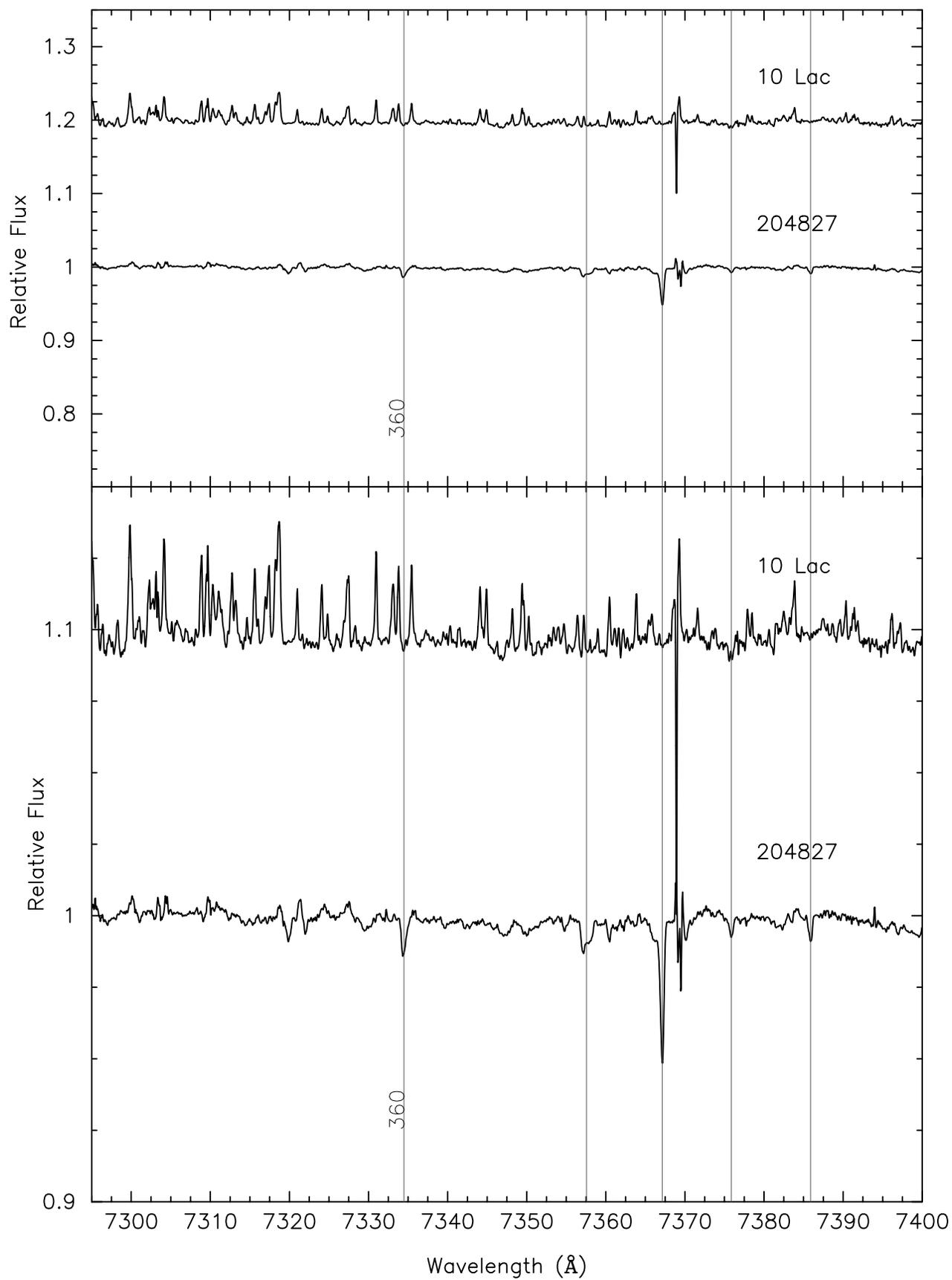

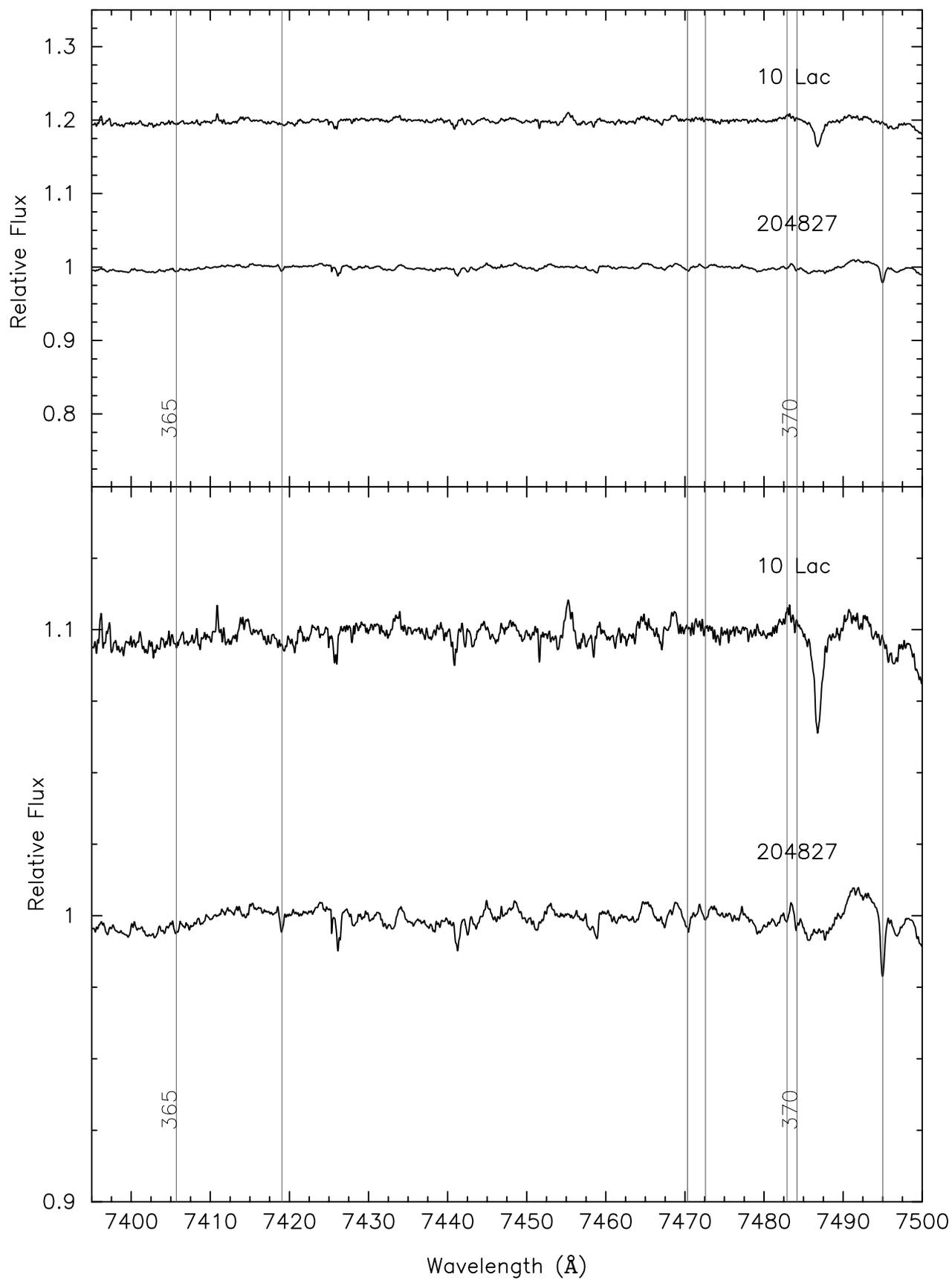

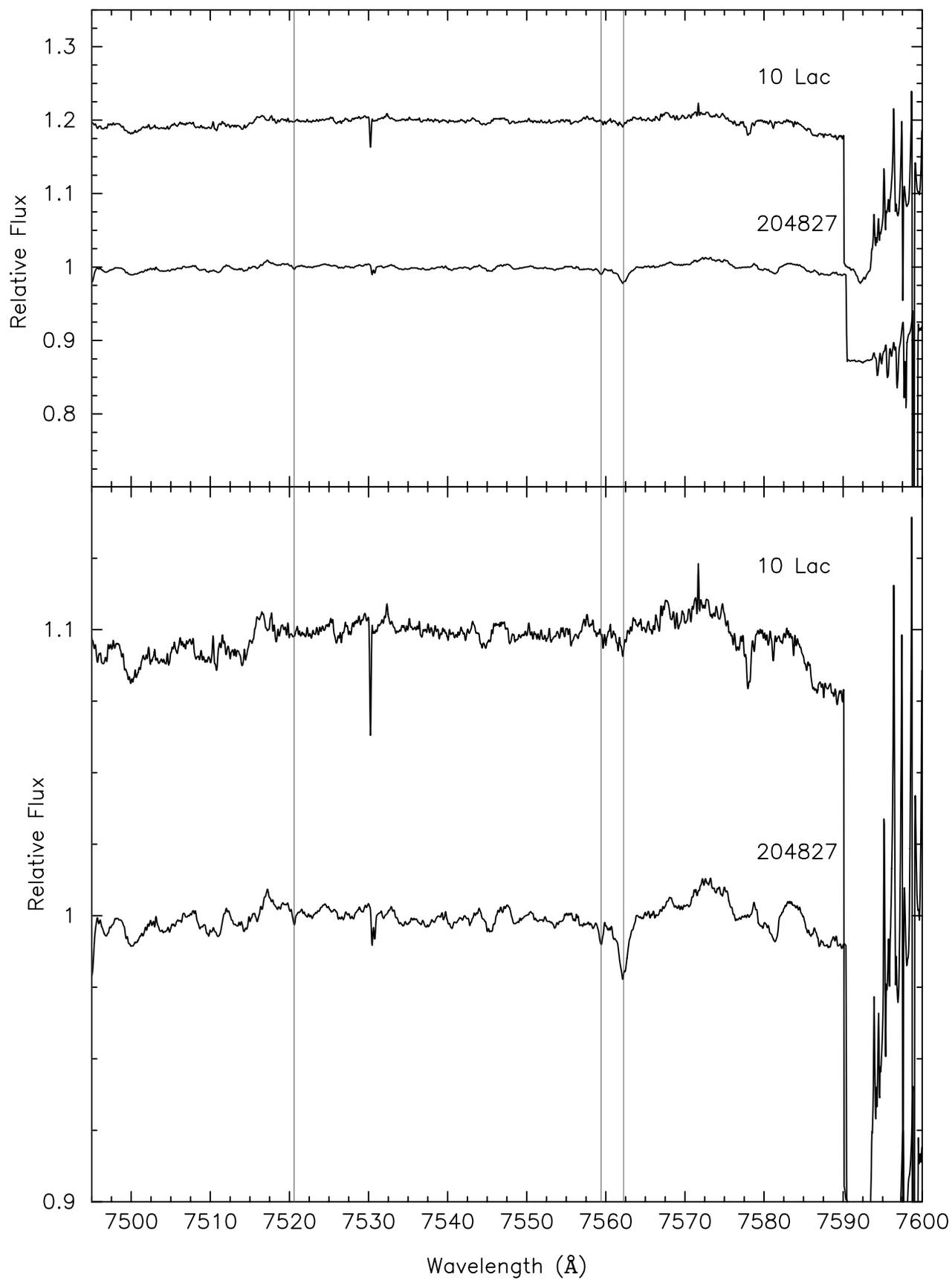

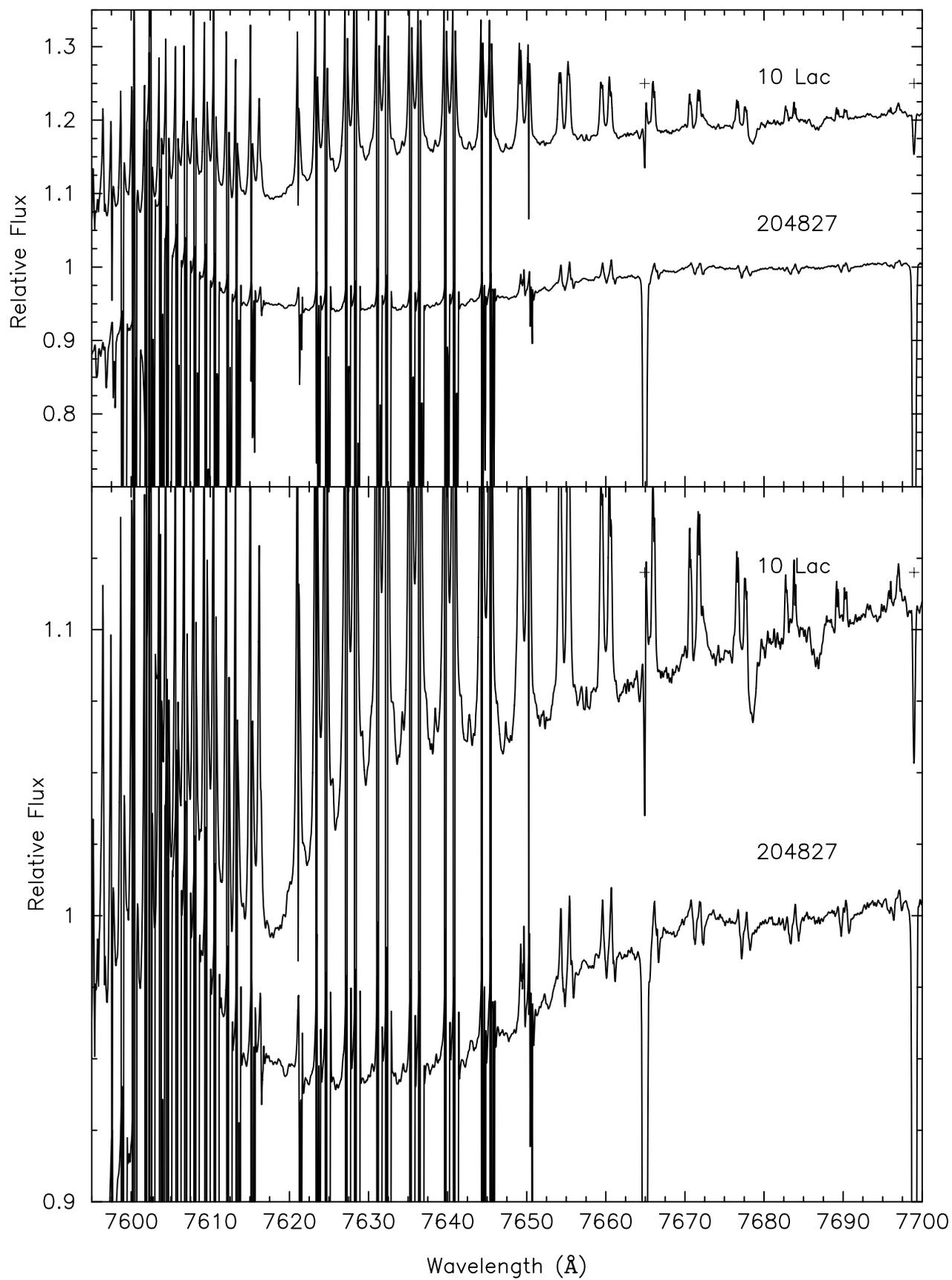

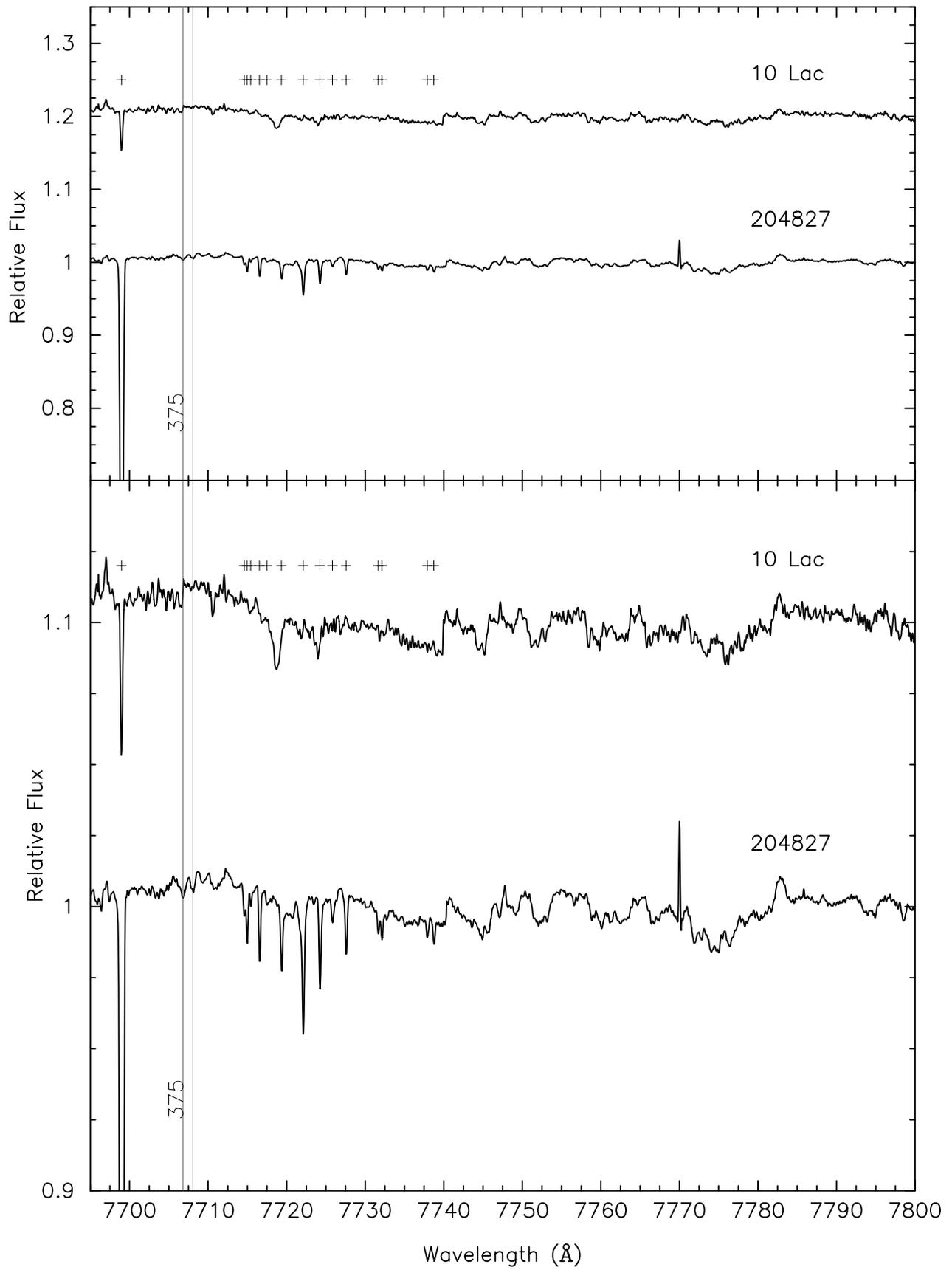

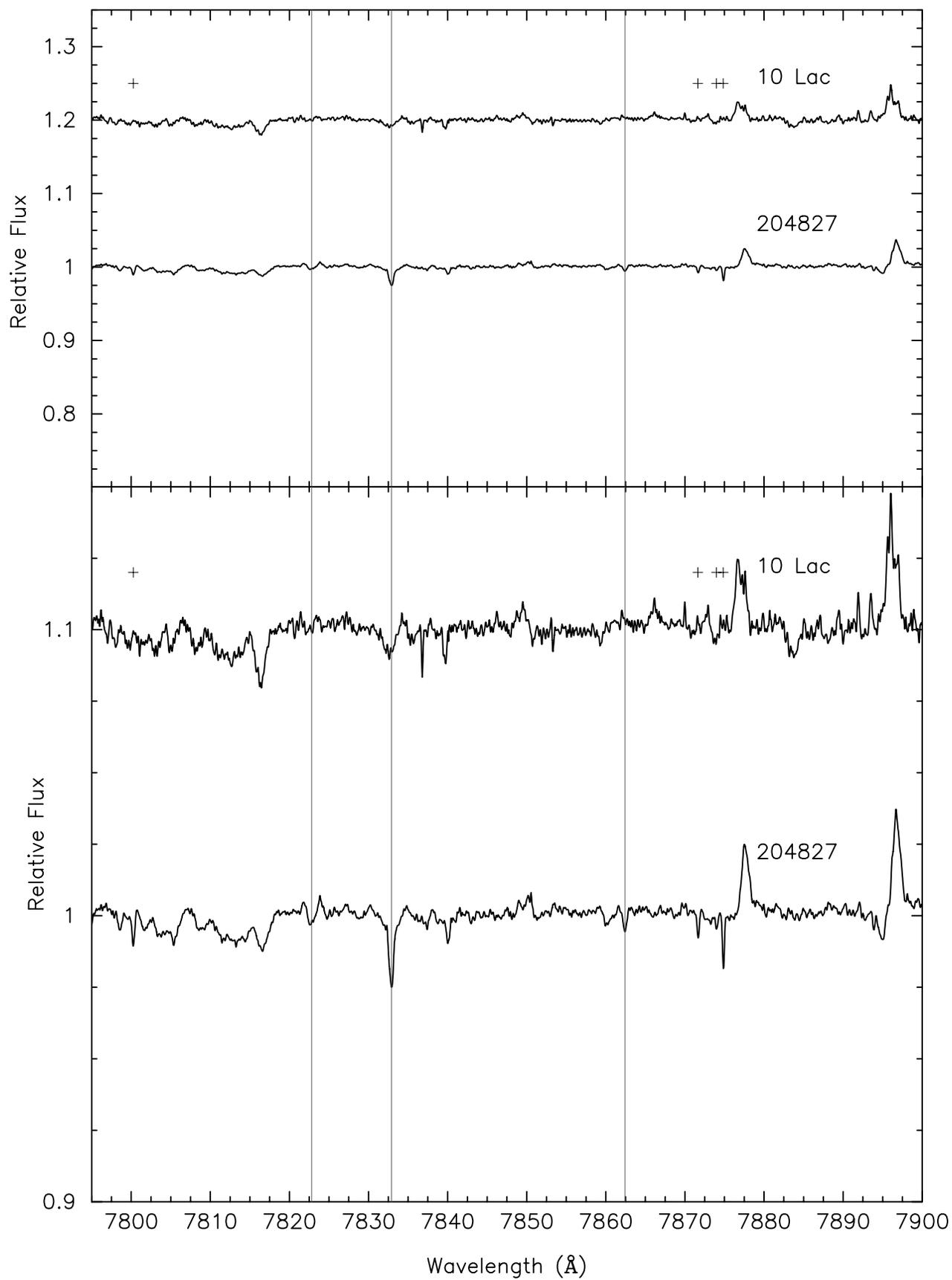

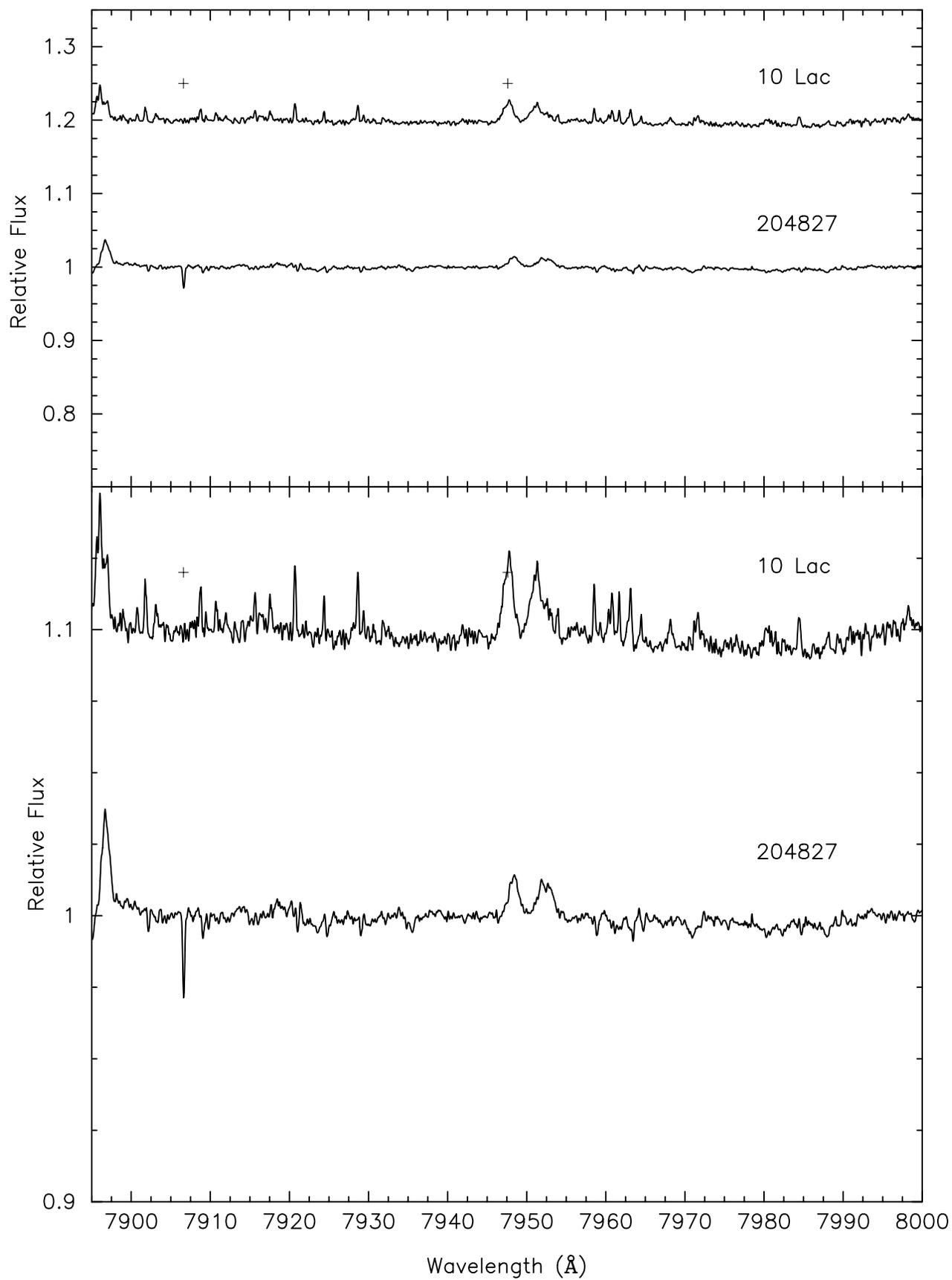

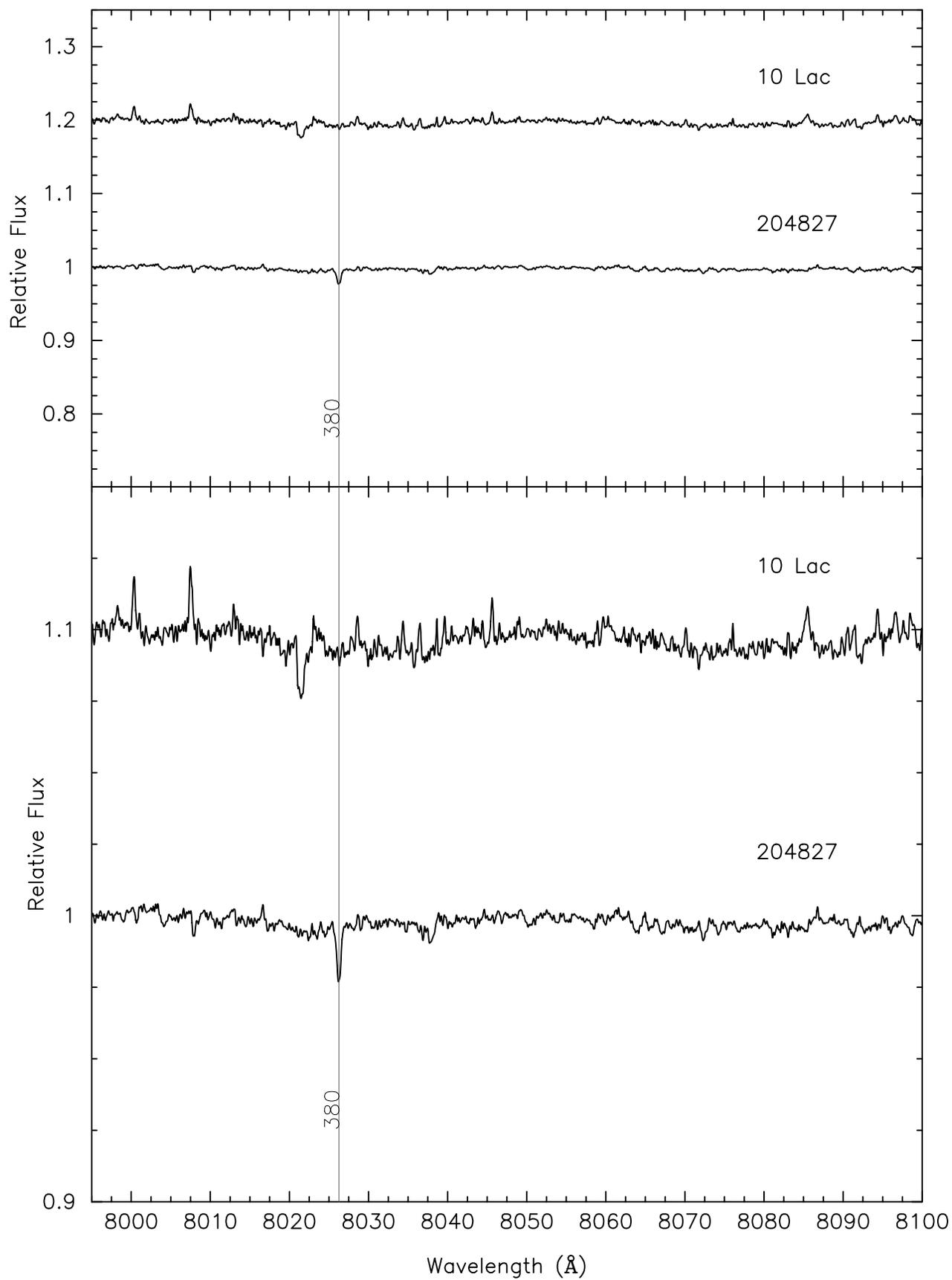